\documentclass[fleqn,10pt]{wlscirep}
\usepackage[utf8]{inputenc}
\usepackage[T1]{fontenc}
\usepackage{subcaption}
\usepackage{float}

\newcommand{\revision}[1]{\textcolor{black}{#1}} 

\title{What is a Social Media Bot? A Global Comparison of Bot and Human Characteristics}

\author[1,*]{Lynnette Hui Xian Ng}
\author[1]{Kathleen M. Carley}
\affil[1]{Center for Informed Democracy \& Social - cybersecurity (IDeaS), Societal and Software Systems
Carnegie Mellon University, Pittsburgh, PA 15213}

\affil[*]{lynnetteng@cmu.edu}


\begin{abstract}
Chatter on social media about global events comes from 20\% bots and 80\% humans. The chatter by bots and humans is consistently different: bots tend to use linguistic cues that can be easily automated (e.g., increased hashtags, and positive terms) while humans use cues that require dialogue understanding (e.g. replying to post threads). Bots use words in categories that match the identities they choose to present, while humans may send messages that are not obviously related to the identities they present. Bots and humans differ in their communication structure: sampled bots have a star interaction structure, while sampled humans have a hierarchical structure. These conclusions are based on a large-scale analysis of social media tweets across $\sim 200$ million users across 7 events.

Social media bots took the world by storm when social-cybersecurity researchers realized that social media users not only consisted of humans, but also of artificial agents called bots. These bots wreck havoc online by spreading disinformation and manipulating narratives. However, most research on bots are based on special-purposed definitions, mostly predicated on the event studied. In this article, we first begin by asking, ``What is a bot?", and we study the underlying principles of how bots are different from humans. We develop a first-principle definition of a social media bot. This definition refines existing academic and industry definitions: ``A Social Media Bot is An automated account that carries out a series of mechanics on social media platforms, for content creation, distribution and collection, and/or for relationship formation and dissolutions." With this definition as a premise, we systematically compare the characteristics between bots and humans across global events, and reflect on how the software-programmed bot is an Artificial Intelligent algorithm, and its potential for evolution as technology advances. Based on our results, we provide recommendations for the use of bots and for the regulation of bots.
Finally, we discuss three open challenges and future directions of the study of bots: Detect, to systematically identify these automated and potentially evolving bots; Differentiate, to evaluate the goodness of the bot in terms of their content postings and relationship interactions; Disrupt, to moderate the impact of malicious bots, while not unsettling human conversations.
\end{abstract}
\begin{document}

\flushbottom
\maketitle
%
%

\section*{Introduction}
The notion of ``bots'' on social media is ubiquitous across many scholarship. These studies captured a range of different social phenomena where bots operate: politics, hate speech, toxicity and so forth. Bots were used to boost the follower count of politicians in the 2011 Arab Springs uprising, generating false impressions of popularity \cite{woolley2016automating,lotan2011arab}. In the same uprising, bots flooded news streams to interrupt efforts of political dissidents \cite{woolley2016automating,lotan2011arab}. In the US 2020 elections, bots augmented human users in strategic communications, and actively distorted or fabricated narratives to create a polarized society \cite{ng2024cyborgs,ng2024assembling,chang2021social}. 
More recently, bots aggressively pushed anti-vaccine narratives and conspiracy theories on social media during the 2021 coronavirus pandemic \cite{seckin2024mechanisms,ferrara2020types}. Bots applied social pressure to influence humans to favor the anti-vaccine ideology \cite{ng2022pro,ng2024cyborgs}.
When the tension of the online ideologies are sufficiently strong, and the spread sufficiently wide, these ideologies spillover to the offline world, resulting in protests, riots and targeted hate-speech \cite{magelinski2021synchronized,broniatowski2018weaponized,shao2018spread,ng2022cross}. Social media bots gained further media attention in 2022 when Elon Musk proclaimed that at least 20\% of the Twitter users were bots, which were influencing content quality \cite{cjrMusksTwitter}. Musk later bought the platform, and took steps to curtail the bot population in a ``global bot purge", which includes removing huge amounts of bots, and charging newly created accounts to post and interact on the platform \cite{latimesElonMusk}.

Much research on social media bots involve constructing bot detection algorithms and applying bot detection algorithms to analyze bot activity during an event. Bot detection algorithms typically extract a series of features from user and post data, then build a supervised machine learning model which classifies the likelihood of a user being a bot or a human \cite{ellaky2023systematic}. These machine learning models range from logistic regression \cite{ng2024tiny}, to random forests \cite{beskow2018bot}, to ensemble of classifiers \cite{sayyadiharikandeh2020detection,ng2024assembling}, to deep learning methods \cite{ng2023botbuster,orabi2020detection}. Another technique of bot detection is graph-based methods, which infers the probability of a user being a bot by its connections, i.e. friends \cite{kolomeets2021bot,li2023botfinder}. Most recently, Large Language Models (LLMs) are incorporated in bot detection algorithms to handle the diverse user information and content modalities \cite{feng2024does}. These bot detection classifiers have been used to study bot behavior in many events, ranging from political events \cite{jacobs2023tracking,ng2024assembling,uyheng2020bot,bessi2016social} to natural disasters \cite{khaund2018analyzing,khaund2018analyzing} to the spread of information and opinions on social media \cite{ng2022pro,ng2021does,hajli2022social}.

Although researchers have built automatic bot detection classifiers, behavioral studies show that humans are unable to differentiate between the bot and human user \cite{kenny2024duped}. In fact, the identification of bots by security students are akin to random guesses \cite{kolomeets2024experimental}. Consequently, it is important to study bots and their characteristic nature and activity patterns. Our study is positioned within the social cybersecurity realm, which studies how the digital environment, particularly bots, can be exploited to alter the content and community relationships \cite{carley2020social}.

This article is being driven by looking at a first principles approach to bots. We ask the following research questions: 
\begin{itemize}
    \item \textbf{RQ1: What is a social media bot?} Many studies are predicated on a general purpose understanding of a bot, or a specific definition particular to the event of study. Instead, we pare down the definition of a bot into its treatment of the core elements of a social media platform (users, content, relationships). 
    \item \textbf{RQ2: How does the nature of a bot differ from a human?} Systematically, we look at the difference between bots and humans. We use a large scale dataset from Twitter/X that spans over 200 million users, and analyzed four aspects: volume of bot/human user types, use of linguistic cues, use of identity terms, and social interactions. 
\end{itemize}

After an examination of the social media bot, we discuss how a bot is also an Artificial Intelligent (AI) agent, and its potential evolution alongside technological advancements. We finally follow with a discussion of the open research challenges of the study of bots to encourage future studies in this field. The challenges we identify reflect the definition of a bot: Detect, to systematically identify these automated and evolving bots; Differentiate, to evaluate the goodness of the bot in terms of their content postings and relationship interactions; and Disrupt, to moderate the impact of malicious bots, while not unsettling digital human communities.

\section*{What is a Social Media Bot?}
The term ``bot" has become a pervasive metaphor for inauthentic online users \cite{ng2022pro,ferrara2016rise}. Most social media users have an implicit understanding of a bot, as do most researchers \cite{kenny2024duped}. \autoref{tab:definitions} summarizes some of the recent definitions of bots retrieved from academic literature. The security industry also watches social media bot accounts, and \autoref{tab:definitions_industry} summarizes definitions from industry sources. Each of the definition grasps one or more relevant properties (highlighted in bold) of a social media bot, yet are not sufficiently comprehensive to describe the bot. Some of these relevant properties are: ``automated", ``interacts with humans", ``artificial agents". 

One of the problems with existing definitions is that they often define bots as being malicious and they highlight the nefarious use of bots \cite{uyheng2021active,himelein2021bots,chang2021social,ng2024exploring,cloudflarebot}: ``malicious actors", ``public opinion manipulation", ``malicious tasks" \cite{assenmacher2020demystifying,sayyadiharikandeh2020detection,cisa}. Most often, the study of bots is established upon nefarious tasks: election manipulation, information operations, even promoting extremism \cite{ferrara2016rise,danaditya2022curious,uyheng2021active}. The exact same technology can be used in both good and bad ways. There are plenty of good bots \cite{hayawi2023social,tsvetkova2017even,stieglitz2017socialsheep}. Bots provide notifications and entertainment \cite{cisainfographic}, such as the @CoronaUpdateBot found in our dataset which posts critical public health information. Bots support crisis management efforts by gathering the needs and combined locations of people after a disaster, for authorities and community volunteers to identify crucial areas and providing help \cite{hofeditz2019meaningful}. Chat bots provide emotional support during stress \cite{10.1145/3461778.3462114} and continue bonds in times of grieve \cite{krueger2022communing}. Celebrities and organizations use bots to facilitate strategic communications with their fans and clients \cite{ng2024cyborgs,boshmaf2011socialbot}. 

In essence, a bot is a computer algorithm. As an algorithm, a bot is neither bad or good. It is the persona \revision{, or public facade, that the bot} is afforded to that determines the goodness of its use. We develop a generic definition of a bot. The definition is independent of the use of the bot. The determination of the use of the bot warrants separate treatment beyond this paper. Regardless of whether a bot is used for good or ill, the behavioral characteristics of bots remain the same.

To better describe the social media bot, we first need to characterize the environment in which it lives: the social media platform. Within a social media platform, there are three main elements: users, content and relationships \cite{howard2012social}. Users are represented by their virtual accounts, and are the key actors driving the system, creating and distributing information. Content is the information generated by users on the platform. Relationships are formed from the user-user, user-content and content-content interactions.

After distilling a social media platform into its core building blocks, it follows that definition of a social media bot should be based on the foundations of a social media platform as first principles. The presence of these components in each of the reference definitions are broken down in \autoref{tab:characteristics}. The first principles of a bot are: 
\begin{itemize}
    \item User: \textit{An automated account that carries out a series of mechanics.} A key characteristic of bots is its programmability, which give it its artificial and inauthentic characteristic. Automation is an aspect that has been iterated in all the reference definitions. The key here is that a bot is automated. The model underlying the automation does not matter; any model can be applied equally well to humans and bot. A bot could be built to mimic humans, or it could be built to optimize other functions. For example, some studies describe bots in terms of its mimicry of humans \cite{ferrara2016rise,chavoshi2016debot}, but others observe that bots eventually influence the social space such that humans mimic bots \cite{stieglitz2017social,grimme2017social}. \revision{The automated nature of the bot gives rise to its systematic nature, which bot detection algorithms leverage through machine learning algorithms. In contrast, human users create accounts for a wide number of reasons, from personal to business to pet accounts, and have very varied and random behavior mechanics. }
    
    \item Content: \textit{for content creation, distribution, and collection and processing.} Bots often generate their content in bulk to distribute a certain ideology \cite{jacobs2023tracking,alsmadi2020many}, such as a good portrayal of their affiliated country \cite{ng2023deflating}. Instances where bots perform content distribution is where the spread fake news and disinformation content \cite{ng2021does,shao2017spread,vziatysheva2020fake}, or when they distribute general news information \cite{lokot2016news}. Bots in the form of web crawlers and scrapers download and index data from social media in bulk \cite{gorwa2020unpacking}, and sometimes process the data to perform functions like analyzing sentiment of opinions \cite{arora2023developing}. \revision{Humans create and distribute content to create their online personalities, to update their friend groups of their lives, or to promote their businesses \cite{liu2016analyzing,ng2023recruitment}. However, humans seldom generate or collect content in bulk; if they do so, they are likely to use a bot to aid them, rendering their account a cyborg \cite{ng2024cyborgs}. }
    
    \item Relationships: \textit{for relationship formation and dissolution.} In other words, forming a relationship online means to connect with other users via post-based (i.e., retweet, mention) or user-based (i.e., following, friend) mechanisms. Dissolving a relationship means to destroy a connection by forcing users to leave a community. Bots are an actively form and destroy relationships on social media platforms. An example of the formation of post-based relationship is the active amplification of narratives. This technique is mostly employed in the political realm where the bots retweet political ideology in an organized fashion \cite{jacobs2023tracking,mckelvey2017computational,woolley2018computational}. User-based relationships can grow through coordinated fake-follower bots, that are used to boost online popularity \cite{zouzou2024unsupervised}, or can be dissolved through toxic bots that spread hate and directed religious ideologies \cite{albadi2019hateful,danaditya2022curious}, causing users to break away from the community \cite{blane2023social}. \revision{In general, bots form and dissolve automated relationships towards some single-purpose goal, while humans form and dissolve relationships organically, sometimes based on real-life encounters. }
    
\end{itemize}

\begin{figure}[h]
    \centering
    \includegraphics[width=\linewidth]{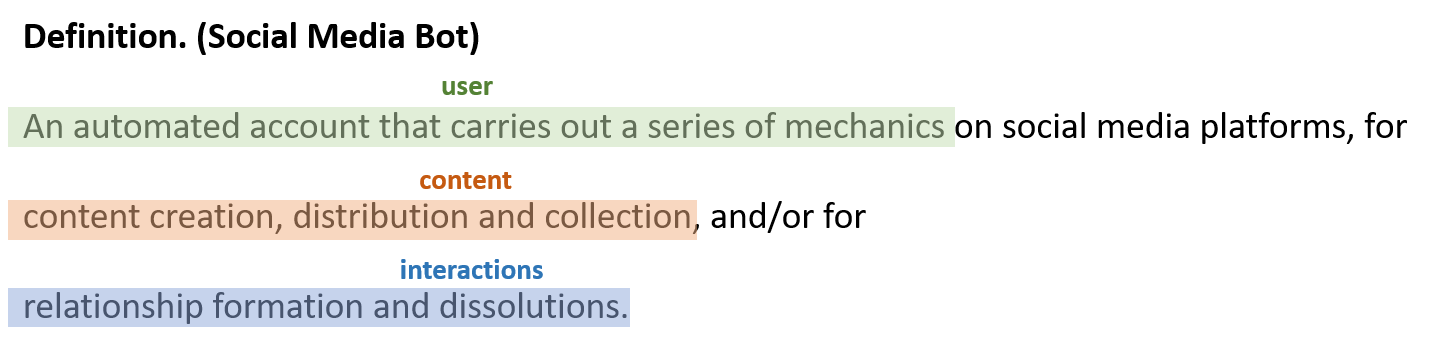}
    \caption{Definition of Social Media Bot. This definition displays the possibilities of mechanics that the bot account can carry out. A bot does not necessarily carry out all the mechanics.}
    \label{fig:what_bot}
\end{figure}

\autoref{fig:what_bot} reflects a first principles definition of a social media bot. A Social Media Bot is ``An automated account that carries out a series of mechanics on social media platforms, for content creation, distribution, and collection and processing, and/or for relationship formation and dissolutions." This definition displays the possibilities of mechanics that the bot account can carry out. A bot does not necessarily carry out all the mechanics. The combinations of mechanics that a bot carries out thus affects the type of bot it is and the role it plays within the social media space, and as shown in \autoref{tab:bot_type_illustration}, those mechanics can be used for either good or bad. Bot types can be named for their actions or for their content. For example, a bot that carries out relationship formation between two different communities, and does not do any content mechanics can be classified as a bridging bot \cite{ng2024tiny}. We illustrate a few types of bots and their use for good and bad in \autoref{tab:bot_type_illustration}. Note that this list is not meant to be an exhaustive list but an illustrative list of the variety of bots in the social media space.

\begin{table}[h]
    \centering
    \begin{tabular}{p{2cm}p{7cm}p{7cm}}
    \toprule
        \textbf{Type of Bot} & \textbf{Use for Good} & \textbf{Use for bad} \\ \midrule
        General Bot & search engine optimization, data collection \cite{gorwa2020unpacking} & spread disinformation \cite{hajli2022social}, manipulate opinion \cite{ng2022pro} \\
        
        Bridging Bot & political commentators that aggregate information \cite{ng2024tiny} & disseminate information \revision{to instigate anger across social, cultural, } community differences \cite{ng2023deflating} \\
        
        Political Bot & ``establishing brand and amplifying messages" \cite{ng2024cyborgs,uyheng2020bot}, digital campaigning\cite{ng2024tiny} & political manipulation \cite{badawy2018analyzing} \\ 

        Chat Bot & emotional support during stress \cite{10.1145/3461778.3462114} and grieve \cite{krueger2022communing} & Post offensive and inflammatory comments \cite{schwartz20192016} \\ 
        
        Activist Bot & crisis management \cite{hofeditz2019meaningful} & trigger and initiate activism \cite{lotan2011arab,magelinski2021synchronized}  \\
    \bottomrule
    \end{tabular}
    \caption{Illustration of type of bots and their role in the social media space. Note that this list is not exhaustive but illustrative.}
    \label{tab:bot_type_illustration}
\end{table}

\begin{table}
    \centering
    \begin{tabular}{p{2cm}p{2cm}p{8cm}}
    \toprule
        \textbf{Year} & \textbf{Reference} & \textbf{Definition} \\ \midrule
         2016 & \cite{ferrara2016rise} & A social bot is a computer algorithm that \textbf{automatically produces content} and \textbf{interacts with humans} on social media, trying to emulate and possibly alter their behavior. \\ 
         2016 &   \cite{bessi2016social} & [...] social bots, \textbf{algorithmically controlled accounts} that \textbf{emulate the activity of human users} but operate at much higher pace (e.g., automatically producing content or engaging in social interactions), while successfully keeping their artificial identity undisclosed \\
         2016 & \cite{chavoshi2016debot} & \textbf{Automated accounts}, called bots, [...] \\ 
         2018 & \cite{gorwa2020unpacking} & Bots are have been generally defined as \textbf{automated agents} that function on an online platform [..]. As some put it, these are programs that run continuously, formulate decisions, act upon those decisions without human intervention, and are able adapt to the context they operate in. \\
         2018 & \cite{brachten2018threat} & The term “social bot” describes accounts on social media sites that are \textbf{controlled by bots} and \textbf{imitate human users} to a high degree but differ regarding their intent. \\
         2018 & \cite{beskow2018bot} & [...] malicious \textbf{automated} agents \\
         2020 &   \cite{orabi2020detection} & Social Media Bots (SMB) are \textbf{computer algorithms} that \textbf{produce content} and \textbf{interacts with users} \\ 
         2020 & \cite{assenmacher2020demystifying} &[...] social bots, \textbf{(semi-) automatized} accounts in social media, gained global attention in the context of \textbf{public opinion manipulation}. \\
         2020 &  \cite{sayyadiharikandeh2020detection} & Malicious actors create \textbf{inauthentic social media accounts} controlled in part by \textbf{algorithms}, known as social bots, to disseminate misinformation and agitate online discussion. \\
         2021 & \cite{martini2021bot} & Social bots – partially or fully \textbf{automated accounts} on social media platforms [...] \\ 
         2022 & \cite{ng2022stabilizing} & Social media bots are \textbf{automated accounts} controlled by software algorithms rather than human users \\
         2023 & \cite{hayawi2023social} & Social bots are \textbf{automated} social media accounts \textbf{governed by software} and controlled by humans at the backend. \\
         2023 &  \cite{ellaky2023systematic} & A bot is a software that mimics human behavior and \textbf{operates autonomously and automatically}.\\
         2023 & \cite{tan-etal-2023-botpercent} & Twitter accounts controlled by \textbf{automated programs}. \\
         2023 & \cite{yan2023exposure} & \textbf{Automated accounts} on social media that \textbf{impersonate real users}, often called “social bots,” \\
         2023 & \cite{yan2023landscape} & Social bots are social media accounts \textbf{controlled by software} that can \textbf{carry out content} and \textbf{post content} automatically. \\
         2024 & \cite{kenny2024duped} & Social bots are \textbf{artificial agents} that infiltrate social media\\
         2024 & \cite{yang2024anatomy} & Social bots are social media accounts \textbf{controlled in part by software} [...] Social media bots display profiles and \textbf{engage} with others through various means, including following, liking, and retweeting \\ 
        \bottomrule
    \end{tabular}
    \caption{Definitions of ``Social Media Bot" in academic literature.}
    \label{tab:definitions}
\end{table}

\begin{table}
    \centering
    \begin{tabular}{p{2cm}p{3cm}p{8cm}}
    \toprule
        \textbf{Year} & \textbf{Reference} & \textbf{Definition} \\ \midrule
         2018 & US Department of Homeland Security\cite{cisa} & [...] Social Media Bots as programs that vary in size depending on their function, capability, and design; and can be used on social media platforms to \textbf{do various useful and malicious tasks while simulating human behavior} \\
         2024 & Microsoft\cite{microsoftSpotBots} & Social media bots are \textbf{automated programs} designed to \textbf{interact} with account users. \\
         2024 & Meltwater\cite{meltwaterSocialMedia} & Refers to the definition by US CSIA (see below) \\
         Not Dated & CloudFlare \cite{cloudflarebot} & [...] social media bots are \textbf{automated programs} used to \textbf{engage} in social media. These bots behave in an either partially or fully \textbf{autonomous fashion}, and are often designed to \textbf{mimic human users}. \\ 
         Not Dated & Cybersecurity and Infrastructure Security Agency (CISA) \cite{cisainfographic} & Social Media Bots are \textbf{automated programs} that \textbf{simulate human engagement} on social media platforms. \\
         Note Dated & Imperva \cite{impervaWhatBots} & An Internet bot is a software application that runs \textbf{automated tasks} over the internet. \\
        \bottomrule
    \end{tabular}
    \caption{Definitions of ``Social Media Bot" in industry literature}
    \label{tab:definitions_industry}
\end{table}

\begin{table}
    \centering
    \begin{tabular}{p{3cm}ccccccc}
        \toprule
        ~ & \multicolumn{2}{c}{\textbf{User}}& \multicolumn{2}{c}{\textbf{Content}} & \multicolumn{2}{c}{\textbf{Interactions}} & \\ \midrule
        \textbf{Reference} & \textbf{Automation} & \textbf{Mimicry} & \textbf{Creation} & \textbf{Distribution} & \textbf{Communication} & \textbf{Relationship} \\ \midrule
        \cite{ferrara2016rise} & x & x & x & & & \\ 
        \cite{bessi2016social} & x & x & x &  & x & x \\ 
        \cite{chavoshi2016debot} & x & & & & & \\ 
        \cite{gorwa2020unpacking} & x & & & & & \\ 
        \cite{brachten2018threat} & x & x & & x & & \\ 
        \cite{beskow2018bot} & x & & & & & \\ 
        \cite{orabi2020detection} & x & & x & & x & x \\ 
        \cite{assenmacher2020demystifying} & x & & & & & \\ 
        \cite{sayyadiharikandeh2020detection} & x & & & x & & \\ 
        \cite{martini2021bot} & x & & & & & \\ 
        \cite{ng2022stabilizing} & x & & & & & \\ 
        \cite{ellaky2023systematic} & x & x & & & & \\ 
        \cite{hayawi2023social} & x & x & x & & & \\
        \cite{tan-etal-2023-botpercent} & x & & & & & \\ 
        \cite{yan2023exposure} & x & x & & & & \\ 
        \cite{yan2023landscape} & x & & x & x & & \\ 
        \cite{kenny2024duped} & x & & & & & \\ 
        \cite{yang2024anatomy} & x & & & & x & x \\ 
        US Department of Homeland Security & x & x & & & & \\ 
        Microsoft & x & & & & x& x\\
        CloudFlare & x & x& & & x & x\\ 
        CISA  & x & x & & & x & x \\ 
        Imperva  & x & & & & & \\ 
        \bottomrule
    \end{tabular}
    \caption{Components of definitions of ``Social media Bot"}
    \label{tab:characteristics}
\end{table}

\section*{Results}
We perform a global comparison of bot and human characteristics by combining several datasets obtained from X (previously named Twitter) using the Twitter V1 Developed API. These events are: Asian Elections \cite{uyheng2020bot,uyheng2021active}, Black Panther \cite{babcock2018beaten}, Canadian Elections 2019 \cite{king2020lying}, Captain Marvel \cite{babcock2020pretending}, Coronavirus,\cite{ng2023combined} ReOpen America \cite{magelinski2021synchronized,ng2023combined} and US Elections 2020 \cite{ng2023combined}. In total, these datasets contain $\sim 5$ billion tweets and $\sim 200$ million users. Each user in this database is labeled as bot or human using the BotHunter algorithm \cite{beskow2018bot}. 

\begin{figure}[h]
    \centering
    \includegraphics[width=0.8\linewidth]{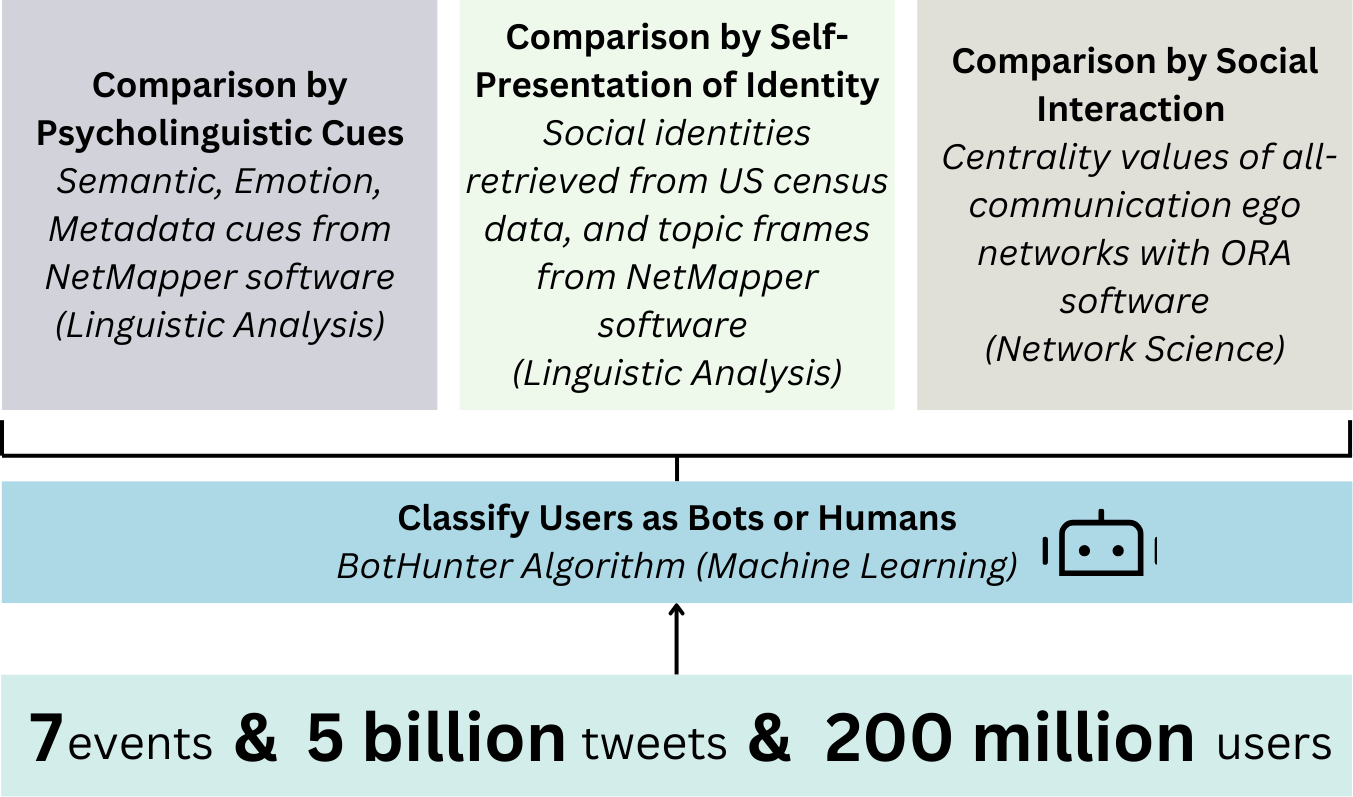}
    \caption{Illustration of multidisiplinary methods used for our analysis}
    \label{fig:methodology}
\end{figure}

\revision{We used a suite of multidisciplinary methods for our analysis. This is illustrated in \autoref{fig:methodology}. We integrated machine learning classification, social network analysis and linguistic feature extraction in a robust hybrid methodology that bridged computational, linguistic and network sciences. 
We analyzed the behavior of bots and humans along four axes. We analyzed the behavior of social media users along four axes. The first is classifying users into bot or human using a machine learning model called BotHunter. The second was linguistic feature analysis of tweets and user metadata. These information were extracted from the NetMapper software, then averaged across user type (i.e., bot vs human) and event type, and compared using statistical tests. The third is the self-presentation of social identities in the user metadata. These identities are derived from matching with an occupation census. In this third analysis, we also extracted topic frames that were written in the tweet texts using the NetMapper software. Thereafter, we correlated the social identities with the topic frames, linking together user presentation with user writing. Fourth, we construct interaction networks of users using the software ORA, and analyzed the influence of bots and humans in terms of their position and structure of their ego networks. Further details about our methodology and the implementation can be found in the Supplementary Material.}

\subsection*{How many bots are there?}
\begin{figure}[h]
\centering
\includegraphics[width=0.6\textwidth]{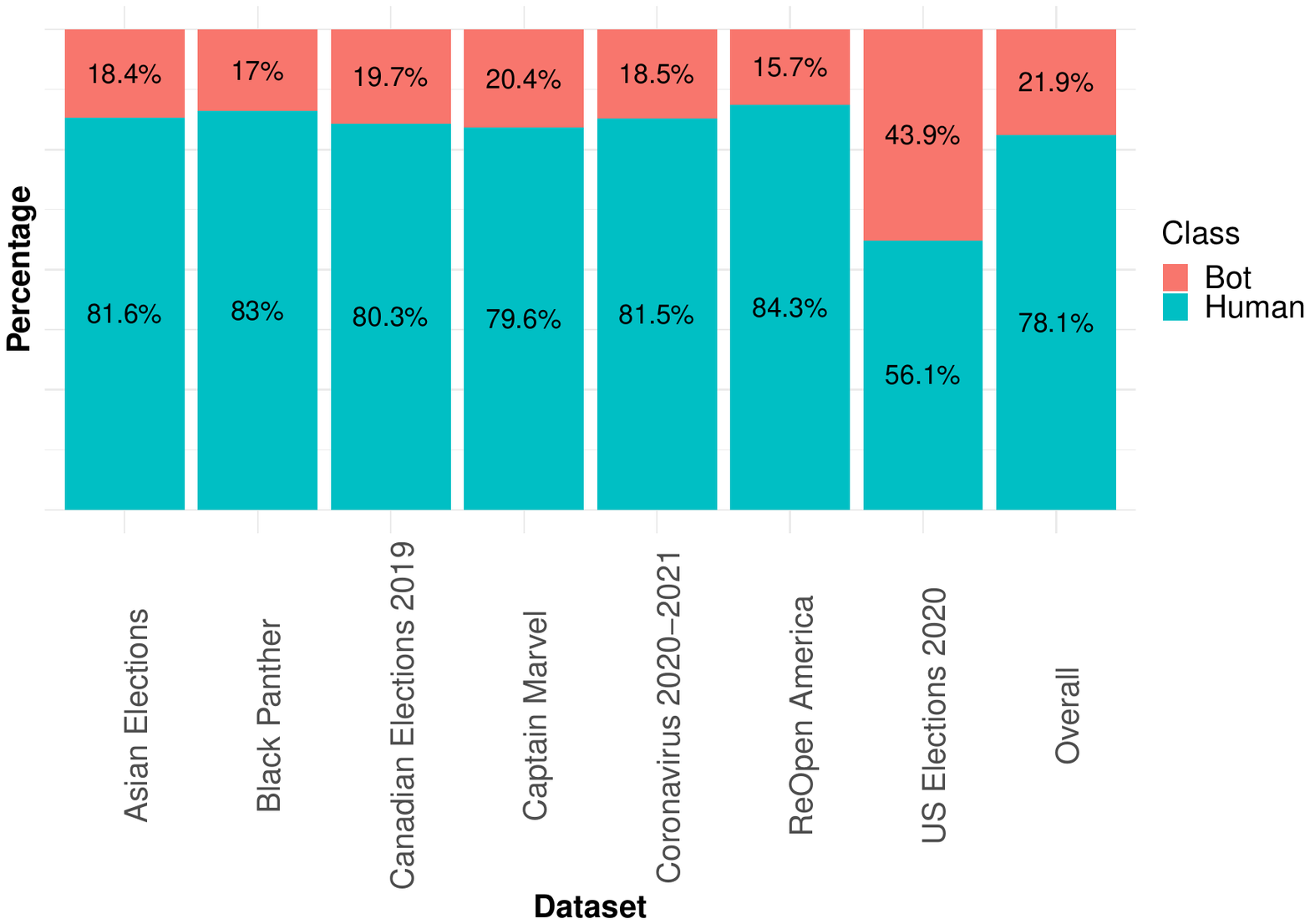}
\caption{Comparison of Bot volume across events. The percentage of bot users across the events are on average around 20\%.}
\label{fig:bot_comparison_by_volume}
\end{figure}

\autoref{fig:bot_comparison_by_volume} presents the percentage of bot users within each dataset. On average, the bot volume across the events are about 20\% with the bot percentage spiking up to 43\% during the US Elections. This is in line with past work, where a general sample of users usually reveal a bot percentage below 30\% \cite{tan-etal-2023-botpercent}, yet in a politically-charged topic (i.e. elections, tensions between countries), the bot percentage rises \cite{ng2023combined,uyheng2021active}. Our estimate is also empirically consistent with Elon Musk's estimate of 20\% \cite{cjrMusksTwitter}. This finding is important for event analysis, because it provides a comparison baseline towards the percentage of bot-like users within an event. Spikes in bot user percentage beyond 20\% suggest that the event and conversation has caught the interest of bot operators, and the analyst can monitor for signs of conversation manipulation.

\subsection*{How do bots differ from humans linguistically?}
We extract psycholinguistic cues from the tweets using the NetMapper software\cite{carley2018ora}. The software returns the count of each cue in the sentence, i.e., the number of words belonging to the cue in the tweet. There are three categories of cues: semantic, emotion and metadata. Semantic and emotion cues are derived from the tweet text, while metadata cues are derived from the metadata of the user. Semantic cues include: first person pronouns, second person pronouns, third person pronouns and reading difficulty. Emotion cues include: abusive terms, expletives, negative sentiment, positive sentiment. Metadata cues include: the use of mentions, media, URLs, hashtags, retweets, favorites, replies, quotes, and the number of followers, friends, tweets, tweets per hour, time between tweets and friends:followers ratio.

\begin{figure}[H]
\centering
\begin{subfigure}[t]{0.8\textwidth}
    \centering
    \includegraphics[width=\textwidth]{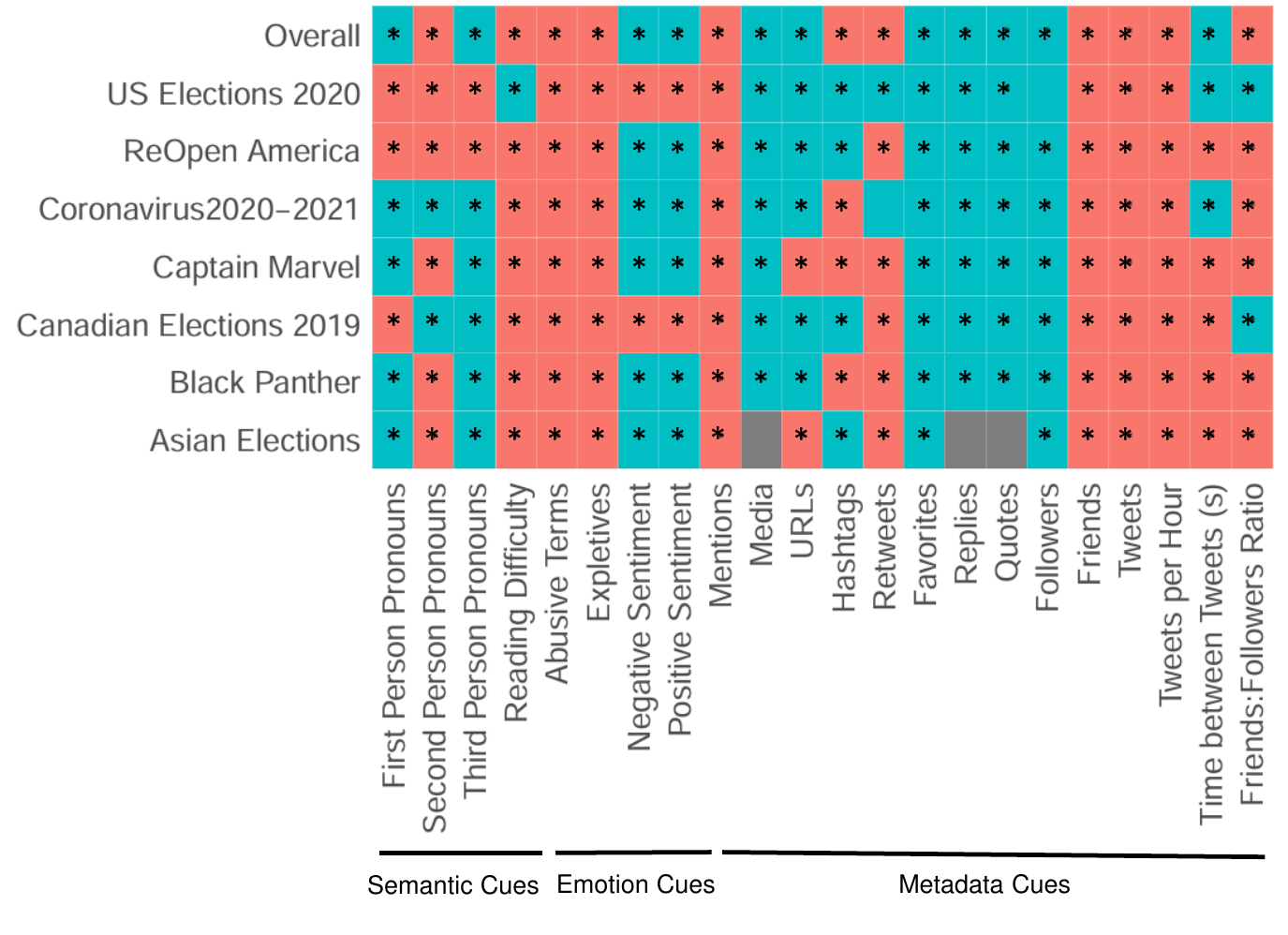}
    \caption{Differences in the use of psycholinguistic cues between bots and humans.}
    \label{fig:bot_comparison_by_linguistic}
\end{subfigure}


\begin{subfigure}[t]{0.8\textwidth}
    \centering
    \includegraphics[width=\textwidth]{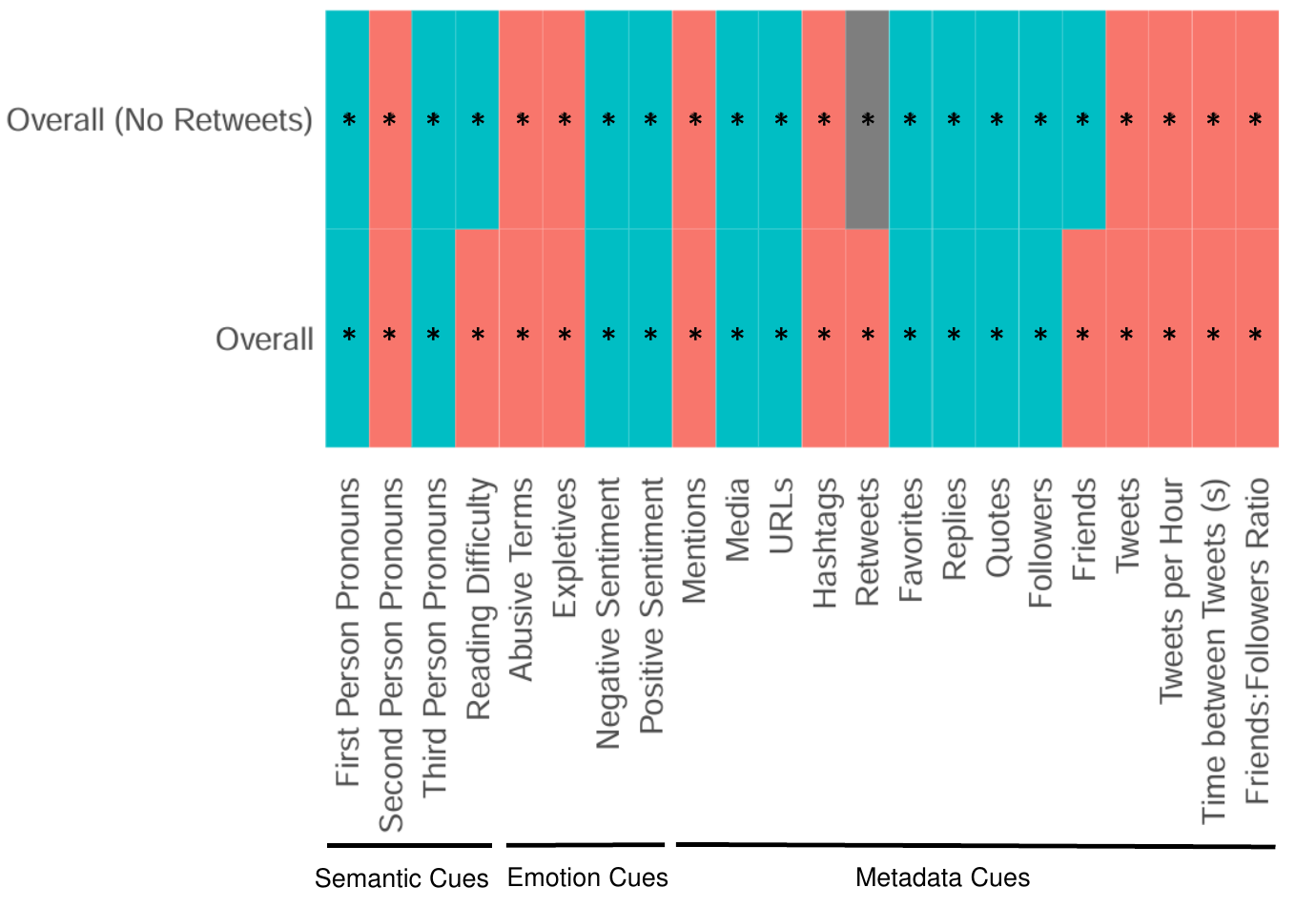}
    \caption{Differences in the use of psycholinguistic cues between bots and humans for the combination of Captain Marvel and Black Panther datasets. This compares the cue distribution with and without retweets. }
    \label{fig:bot_comparison_by_linguistic_noretweets}
\end{subfigure}

\caption{Comparison of psycholinguistic overall cue usage (average cue usage per user) by bots and humans across datasets. \textbf{Green} cells show that \textbf{humans} use a larger number of the cue. \textbf{Red} cells show that \textbf{bots} use a larger number of the cue. * \revision{within the cells} indicates there is a significant difference in the usage of the cue between bots and humans \revision{at the $p<0.05$ level}.}
\label{fig:linguistic_comparison}
\end{figure}

\autoref{fig:bot_comparison_by_linguistic} presents the differences between cues used by bots and humans. The detailed numeric differences are in the Supplementary Material. This difference is examined overall, and by event. There are consistent differences in the use of cues by bots and humans. For example, across all events, bots use significantly more abusive terms and expletives, and tweet more than humans. On the other hand, humans use more first person pronouns, positive sentiment, and media (i.e., images, videos). Humans tend to quote and reply to tweets, while bots tend to retweet.

Most events have consistent cue distribution, but some events look different. In general, humans use more sentiment cues. However, in the two elections (US Elections 2020 and Canadian Elections 2019), bots used more sentiment cues. This reveals a deliberate attempt to use bots during the election seasons to polarize online sentiments. Prior research has shown that bots can be highly negative during the election season \cite{stella2018bots}, and that bots express hugely different sentiment sentiment when mentioning different political candidates \cite{aldayel2022characterizing,ng2022pro}.

When a bot \revision{solely} retweets a human, its linguistic profile, by definition, is identical to the human's. The question though, is whether the bots that are sending original tweets match the linguistic profile of those retweeting, or is the linguistic profile different? For the Black Panther and Captain Marvel events (\autoref{fig:bot_comparison_by_linguistic_noretweets}), we compared the psycholinguistic profile for all tweets, and the original tweets only (i.e., no retweets). In these two events, bots retweet significantly more than humans. \revision{In general}, the bot-human difference between linguistic cue use of the original tweets vs all tweets are rather similar. \revision{However,} the average tweet reading difficulty and the number of friends are different: in original tweets, humans have higher values; in all tweets, bots have higher values. Therefore, bots have their unique signature when generating new content \revision{(i.e., in original tweets)}, but are guaranteed to match human's content when retweeting the human's.

Bots construct tweets with cues that can be easily and heavily automated, while humans construct more personal tweets that require higher cognitive processing to create. \revision{This shows in \autoref{fig:bot_comparison_by_linguistic}, where bots use more semantic cues, while humans use more emotion cues. For metadata cues, bots have more tweets and tweets per hour, while humans engage more with the conversation through replies and quotes.} Such differences \revision{show} how bot accounts still use rather rudimentary techniques: hashtag latching using multiple hashtags \cite{danaditya2022curious,khaund2018analyzing}, connecting several users together with increased number of mentions \cite{ng2023deflating} and flooding the zone with lots of tweets tweets of their desired narratives \cite{jacobs2023tracking,ferrara2017disinformation}. More sophisticated communication techniques like having an increased number of media, and more advanced interaction techniques that involve dialogue understanding like increasing the number of replies and quotes, are still left to the humans. In short, bots have not entirely mimicked humans, yet. 

\subsection*{How do bots present themselves differently from humans?}
\begin{figure}[H] 
\centering
\begin{subfigure}[t]{0.42\textwidth}
    \centering
    \includegraphics[width=\textwidth]{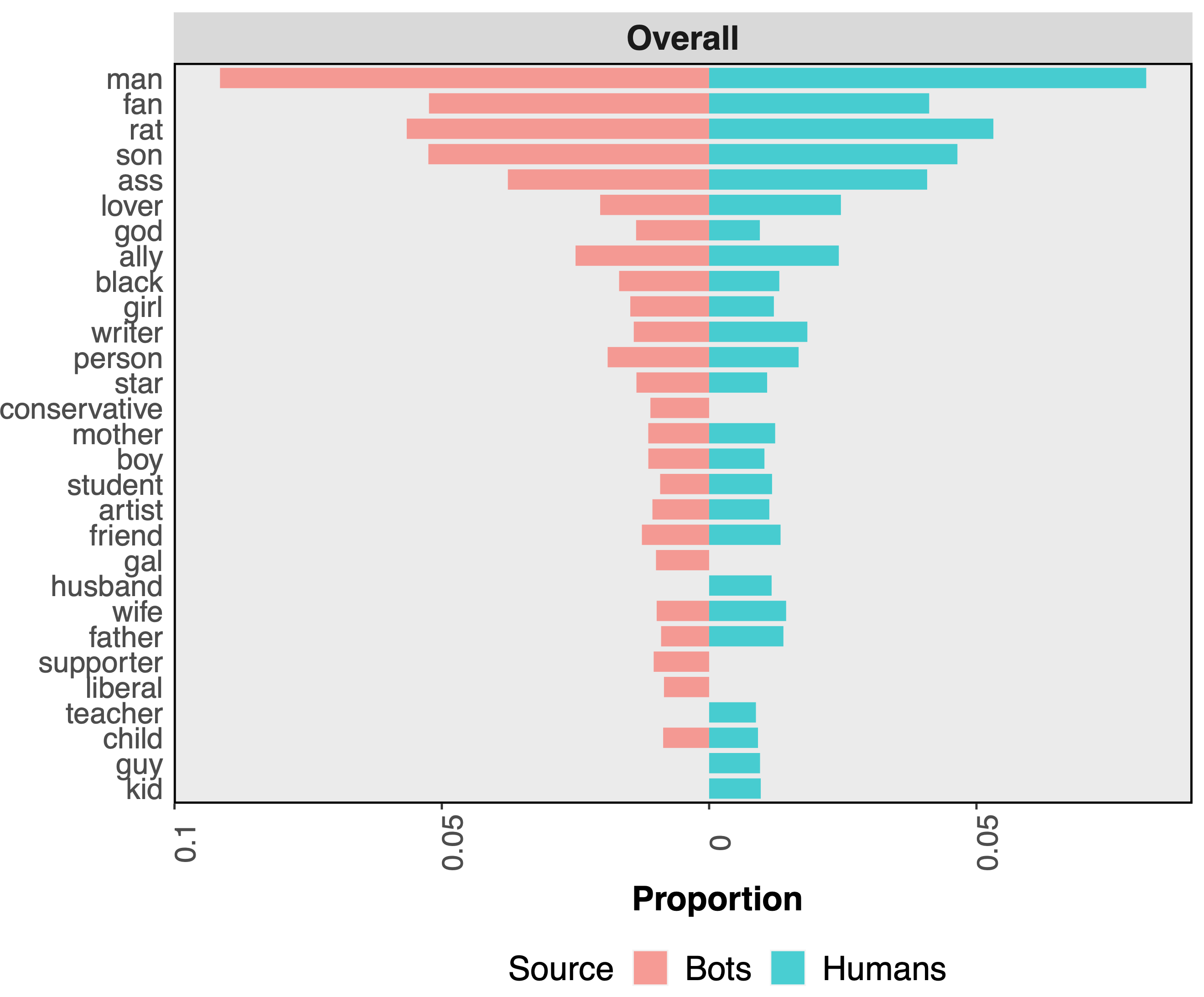}
    \caption{Comparison of the use of the identity affiliations by bots and humans. 21\% of the users affiliate with an identity in their user description.}
    \label{fig:comparison_identity_freq}
\end{subfigure}
\hfill
\begin{subfigure}[t]{0.57\textwidth}
    \centering
    \includegraphics[width=\textwidth]{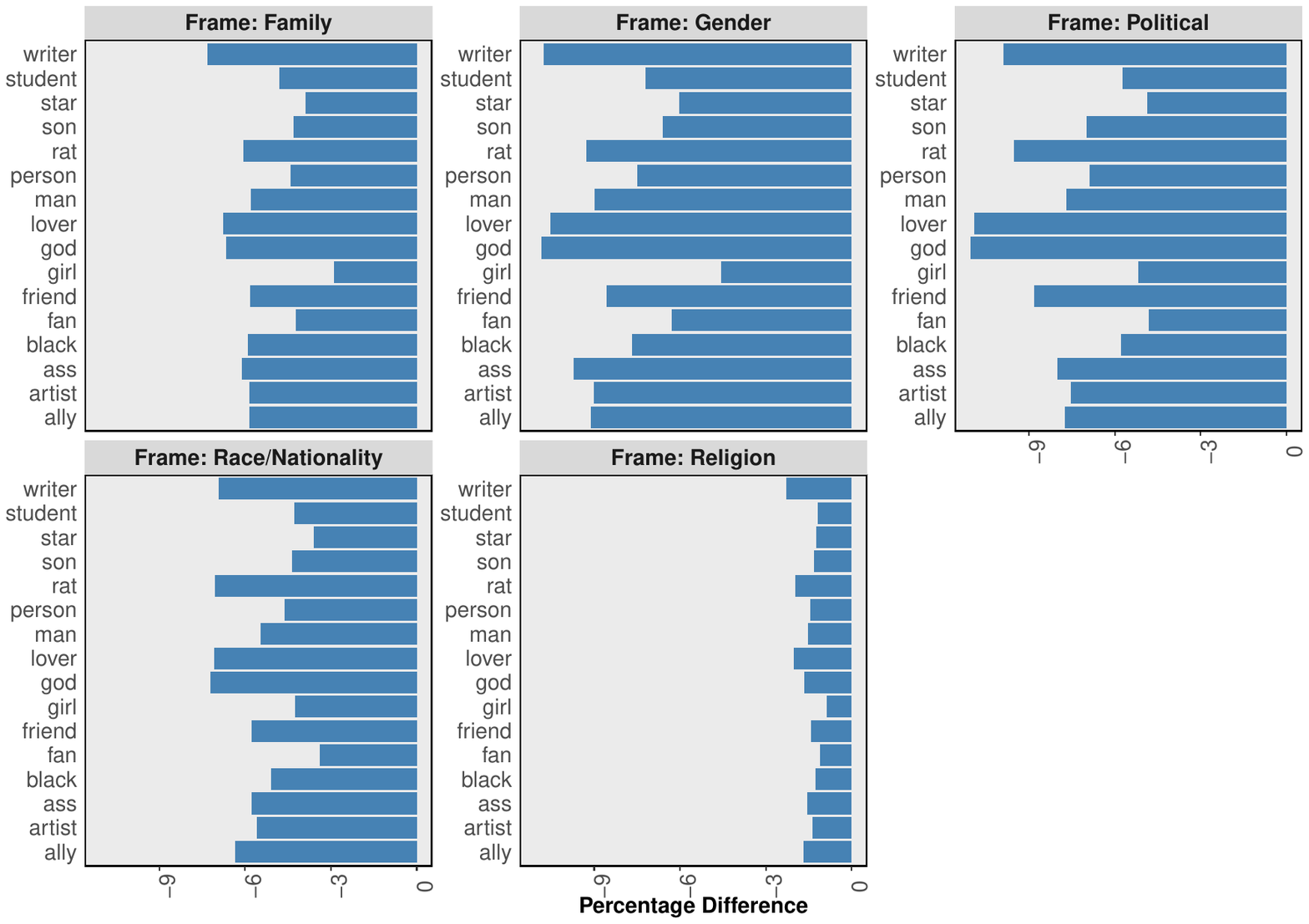}
    \caption{Percentage Difference $\frac{(H - B)}{H}$ of the use of each topic frame in messages by the top frequent identity affiliations.}
    \label{fig:comparison_identity_all}
\end{subfigure}

\vspace{0.5cm} 

\begin{subfigure}[t]{0.8\textwidth}
    \centering
    \includegraphics[width=\textwidth]{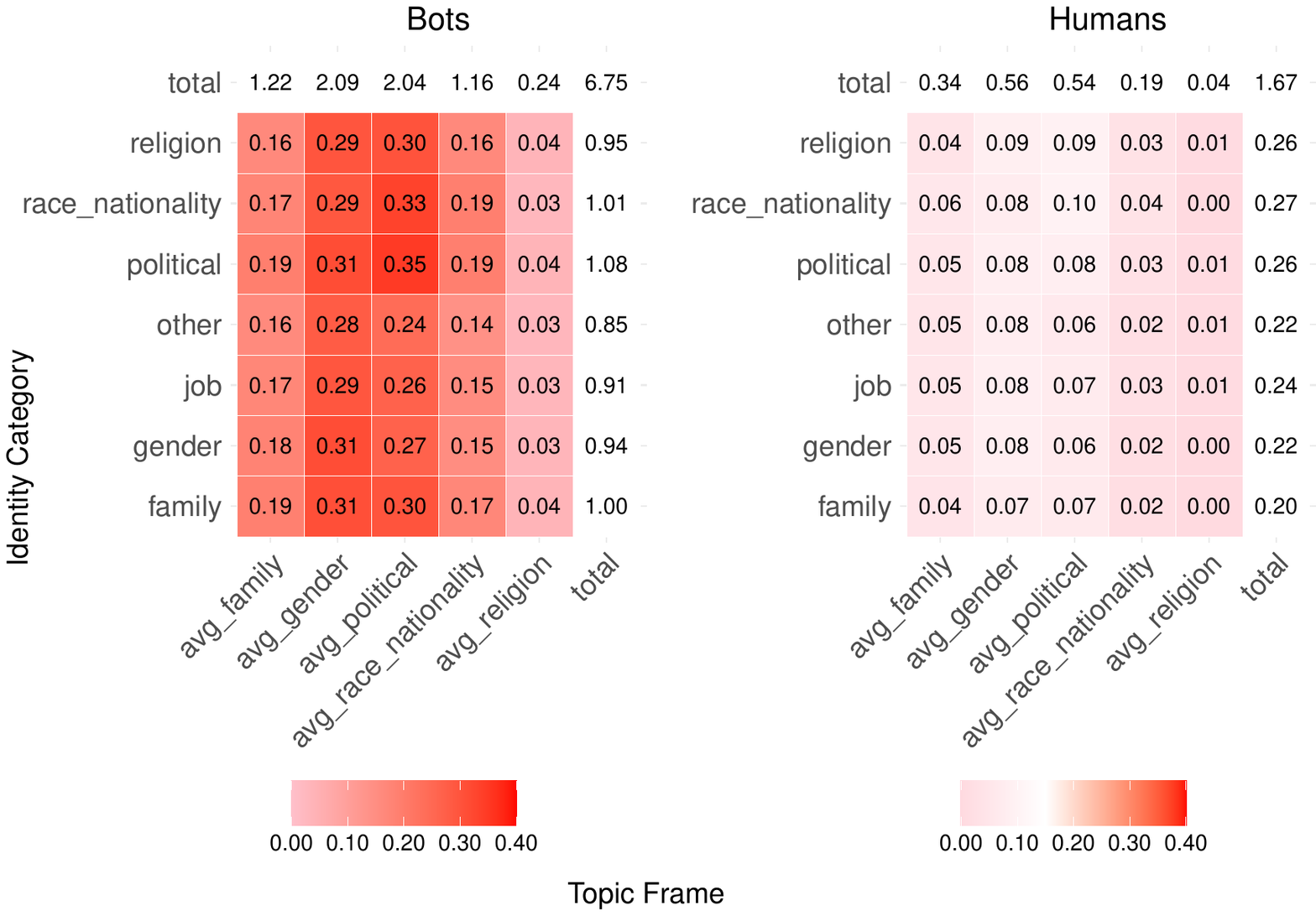}
    \caption{Average use of topic frames by identity category referred to. Bots are more likely to refer to gender and political identities, and are more likely to utilize racially typed language.}
    \label{fig:identity_heatmap}
\end{subfigure}

\caption{Comparison of identity-related behaviors in bots and humans. \revision{The Methods section explains the derivation of the identity categories and topic frames.}}
\label{fig:combined_figures}
\end{figure}

Social identity theory depicts how social media users portray their image online, and the community that they want to be associated with \cite{pathak2021method,ng2023recruitment}. We analyze the difference in the self-presentation of the identities between bots and humans, and the difference between the linguistic cues used by the identities. Across the events, there are consistently a smaller proportion of bots that present with an identity. Overall, 21.4\% of bots present an identity, while 27.0\% of humans present an identity (see \revision{Supplementary Material} \autoref{tab:perc_identity}). Bots are more likely to obfuscate their identities \cite{van2018cyber}, allowing them to take on different personas to suit their operation requirements \cite{radivojevic2024llms}. \autoref{fig:comparison_identity_freq} presents the top 25 identities by frequency between bots and humans. \revision{Overall, bots affiliated with a more limited set of identities (n=869 vs n=908 unique identities). The distribution of the identities for each dataset is presented in the Supplementary Material Tables 3 and 4.} There is a \revision{larger drop in} the frequency of the use of identities in bot users than in human users. \revision{For example, between the 3rd and 4th most frequently used identities, bots show a 33.2\% decrease in the use of the 4th most frequent identity compared to the 3rd most frequent identities in bots, compared to a 30.0\% decrease for humans. This observation suggests that} bots concentrate their self-presentation on certain identities, mostly the common ones: man, son, fan, lover; while humans have a more varied identity presentation. 

We then ask a follow-up question: ``How do the same bot/human identities talk about the same topics?" We compare the use of topic frames per identity for the most frequent identity affiliations in \autoref{fig:comparison_identity_all}. \revision{This frames are extracted from the NetMapper software (refer to Methods section for more details).} \autoref{fig:comparison_identity_all} plots the percentage difference of the use of framing cues (Family, Gender, Political, Race/Nationality, Religion) between bots and humans. This metric compares the use of cues with the human usage as a baseline. Overall, bots converse more aggressively in all topic frames. In particular, bots converse most around societal fault lines: gender, political, race/nationality. These conversations lie on societal fault lines, which could sow discord and chaos \cite{howard2020lie}, therefore such bots are of interest to monitor and moderate. In fact, bots use more gender-based cues. Other research groups have also identified that a disproportionate number of bots that spread disinformation are females \cite{tardelli2020characterizing,deverna2023artificial}, and are thus more likely to use gender frames in their posts. Bots tend to converse largely about political topics, regardless of the identity they affiliate with, indicating that a good proportion of bots are deployed for political purposes, either by political parties or by political pundits \cite{ng2024tiny,martini2021bot,bessi2016social}. Finally, the difference between the usage of topic frames between bots and humans could be due to their vocabulary used. The words used by humans are more varied and mostly not standard dictionary words, while bots are still being programmed with a limited set of vocabulary. \revision{This is} evidenced by the proportion of words identified by the dictionaries in the NetMapper program used, \revision{where on average, humans use 4.9$pm$2.4\% more words related to topic frames compared to bots. The breakdown of percentage difference in words used is presented in the Supplementary Material Table 5.} In a similar aspect, chat bot interactions have a more limited vocabulary than human interactions \cite{hill2015real}.

\autoref{fig:identity_heatmap} presents the average use of topic frames by identity categories. Humans affiliate themselves equally with all identity categories, while bots generally affiliate themselves with racial and political identities. Both bots and humans converse a lot on gender and political issues.

Bots converse mostly about topics that closely match their identity. For example, a bot that presents itself as ``man" and ``son" mostly converse about family then gender; while bots that take on the identities ``conservative" and ``american" converse significantly more about politics. This observation can be read from the heatmap: for the bots that associate with the religion identity, the average use of religious words is 0.04, while that for humans is 0.00. If the users associate with the family identity, the average proportion of the use of family words within the content is 0.19 for bots and 0.04 for humans. Such is the curated presentation and programming of bot users, which allows for an aspect of predictability - if a bot user affiliates with a certain identity, it is likely to talk about topics related to its identity. This shows that bots are likely designed to look like humans. They are strategically designed to be in character by having the right affiliation to fit in and converse with a specific group. 

Our observations in the affiliation of identities by bots in their user description and the use of identity-related topic frames means that bots are being used strategically. They are not just used to support or dismiss groups in general, but are specifically being aimed at a gender (i.e., women or men), or at a political actor (i.e., president, governor, politician). Bots are overused in the political, religious, and racial realm, suggesting that they are targeting topics of societal tensions.

\subsection*{How do bots communicate differently from humans?}
Social interactions between users are an indication of the information dissemination patterns and the communication strategies of the users. We calculate the network metrics (total degree, in degree, out degree, density) of the all-communication ego-networks of the users. In the network graphs, the users are nodes, and the links between users represent all communications between the two users (i.e., replies, quotes, mentions, retweets).
\autoref{tab:clustering_egonetworks} compares the two metrics for bots and humans. Bot ego networks have higher density than \revision{human} ego networks (8.33\% more dense), which reflects that the bots have tighter communication structures and form more direct interactions than humans.
On average, a bot has 9.66\% bot alters and 90.34\% human alters, whereas on average a human has 7.31\% bot alters and 92.69\% human alters.  
Although bots interact with a higher proportion of bot alters than humans do (32\% more bot alters), our findings show that both bots and humans interact more with humans rather than bots in their ego network. By the principle of homophily, it is natural for humans to interact with other humans \cite{bisgin2012study}. However, bots violate the principle of homophily, and instead of interacting with more bots, they interact with more humans. Therefore, bots are actively forming communication interactions with humans, perhaps attempting to influence humans \cite{ng2022online}. 

\begin{table}[H]
    \centering
    \begin{tabular}{ccc}
        \toprule
         & \textbf{Bot} & \textbf{Humans} \\ \midrule
         In-degree & \textbf{0.05 $\pm$ 0.08} & 0.02 $\pm$ 0.02 \\
         Out-degree & \textbf{8E-4 $\pm$ 1.4E-3} & 1.6E-3 $\pm$ 3.3E-3 \\
         Total degree & \textbf{0.15 $\pm$ 0.09} & 0.16 $\pm$ 0.11 \\ 
         Density & \textbf{0.35 $\pm$ 0.06} & 0.034 $\pm$ 0.06 \\
         \% bot alters & \textbf{9.66 $\pm$ 2.98} & 7.31 $\pm$ 3.10 \\
        \bottomrule
    \end{tabular}
    \caption{Comparison of network metrics. For the in-degree, out-degree, total degree and density, we present the ratio of mean(metric) for agent type : max(metric) across all agents in the event}
    \label{tab:clustering_egonetworks}
\end{table}

\begin{figure}[H]
\centering
\includegraphics[width=0.6\textwidth]{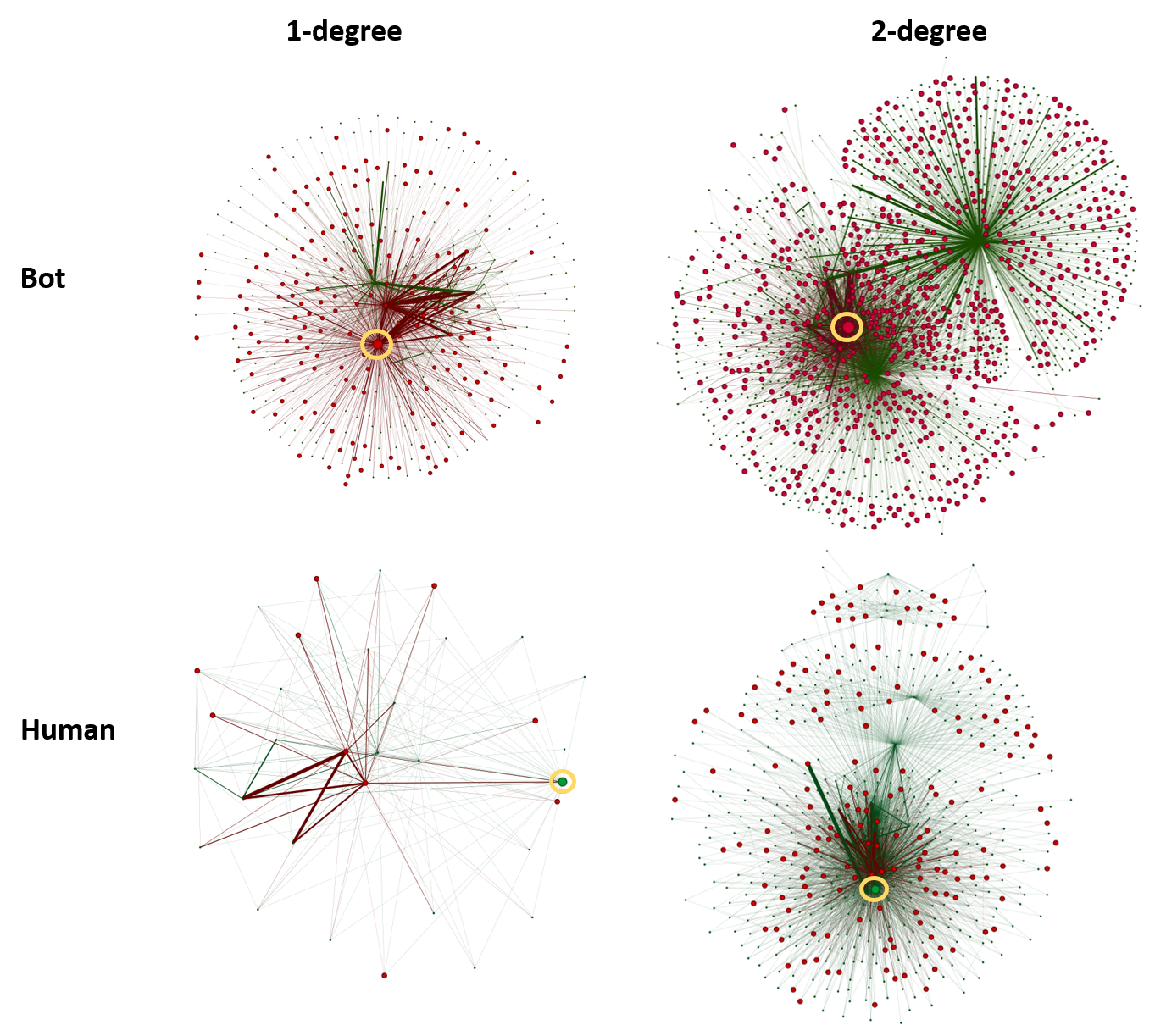}
\caption{Ego network structures of Bots and Humans who are the most frequent communicators in the Asian Elections dataset. Nodes represent social media users. Links between users represent a communication relationship between the two users (i.e., retweet, mention). Bot users are colored in red, human users in grey. The width of the links represent the extent of interactions between the two users. In these most frequent communicators, bots have a star network structure, and humans a tree structure. Bot networks have more bot alters, while human networks have more human alters.}
\label{fig:social_interaction}
\end{figure}

\autoref{fig:social_interaction} shows the interaction of bots and humans in a network diagram. These users are illustrative of the most frequent communicators in the Asian Elections dataset. In this diagram, users are represented as nodes, and links between users represent a communication (e.g. a retweet, reply, mention). The network diagrams presented are one- and two-degree ego-networks, generated by the ORA software \cite{carley2018ora}. This means that the networks present users that are in direct communication with the user (1-degree), and are in direct communication with the 1st-degree users (2-degree). 

A common way for bots to be used in political discourse (e.g. elections) is to amplify other users. As an amplifier, bots are the pendants of the user they are amplifying. \revision{Bots echo the ideas, narratives and emotions of these users, enhancing the visibility of the users and their narratives within the social network.} Therefore, bots appear in star networks in many of the peripheral nodes.
A star structure is a network that have a strongly connected core and peripheral networks. This structure is most prominent in bots in political discourse, where core bots create information, and peripheral bots amplify the information through the retweeting mechanic \cite{jacobs2023tracking}. Humans, on the other hand, are more likely to be part of a tree structure, where one can make out the tiered first- and second-hop interactions. In the same discourse, humans are more likely to be performing many actions, sometimes retweeting other users, sometimes tagging other users and so forth. 

This difference in interactions between bot and human users reveals the communication patterns of both user classes. The star structure of bots suggests that they have a hierarchy of interconnected bot users in an operation network to disseminate information, which is easily achieved with the help of automation. On the other hand, humans communicate predominantly within their immediate network before extending their communication outwards. The bot ego networks are more dense, signifying that they were constructed to interact more than do humans, and are sometimes constructed as networks of bots (the botnet) \cite{mazza2019rtbust,ng2022online}. 

\section*{Discussion}
Through our large scale empirical study, we show that bots and humans have interesting and consistent differences between them. These differences span from their volume, to the linguistic features of their text, to the identities they affiliate with, to their social network structure. These features can be used to characterize a social media bot, and how it differs from humans. 

\revision{Our work introduces a valuable dataset containing billions of tweets over four years.  This dataset is a valuable resource for studying patterns in social media bot technology. With the current restrictions by social media platforms for data collection, this dataset will be prohibitively expensive and technically not feasible to replicate. X API, this dataset will be extremely costly to collect. Using the free tier API, this dataset would take 50 million months or 136,986 years to complete collection. The Pro tier API costs allows retrieval of 1 million tweets per month at a cost of \$5000, leading to the total cost of \$25 million over 5000 months, or 416 years for the collection to be completed. Table 11 in the Supplementary Material details the calculations of the number of months and the cost required to replicate this dataset under the 2025 pricing structure.}

We study a huge amount of data dated from 2018 to 2021. These data show consistent differences over the years, which means that while bot technology do evolve, it does not evolve drastically. \revision{Moreover, the consistent behavioral differences show that there are scenarios where the automated nature of bots is more advantageous, and scenarios where the manual human control and thought is superior.} These differences provide insights to how both can be utilized to afford conversations on social media: bots can be used for methodological postings with deliberate selection of hashtags, tweets per hour, and a structured star communication network. Humans can be used for more complex cognitive tasks such as adding media to a post or replying to a post, and for conversing on a larger range of topics \cite{ng2024cyborgs}.

\subsection*{Strengths of Social Media Bots}
An \revision{Artificially} Intelligent (AI) system is a machine, or computer system, that can perceive its environment and use intelligence to perform actions to achieve defined goals \cite{russell2016artificial}. The social media bot perceives the digital environment to decide their targets (i.e., users to retweet, users to mention), and intelligently carry out content and interaction mechanics to achieve their goals (i.e., spreading information \cite{vziatysheva2020fake,ng2021does}, sowing discord \cite{lu2024agents,chen2021social}, assisting the community \cite{hofeditz2019meaningful,10.1145/3461778.3462114,krueger2022communing}). The software-programmed social media bot is an AI algorithm, and thus has potential to be harnessed for social good. 

\autoref{tab:bot_strengths} lists some recommendations of our results on how bots can be leveraged on for social good \revision{, based on possible avenues for interacting with the automated accounts.} First, given that bots use more retweets and mentions than humans, and have high tweets per hour, bots can be used for menial tasks like announcements and distribution of information. Second, since bots have a star interaction network, they can be used for big announcements like disaster and crisis management without message distortion. A star network sends messages directly through interactions, hence the original message is preserved. However, the human's hierarchal interaction network will distort the message as it passes through the tiers. Third, bots typically post content that matches their identity, they can be used to provide educational material about topics that people associate with certain profession. For example, a weather news bot can provide weather information. Lastly, since bots use more abusive and expletive terms than humans, instead of regulating toxic language itself, regulation can be focused on disallowing bots to use such toxic language, which would therefore reduce the amount of hyperbole and offense online. \revision{This regulatory recommendation draws on our observational study, and we suggests subsequent randomized controlled tests for validation before policy implementations.}

\begin{table}[h]
    \centering
    \begin{tabular}{p{7cm}p{10cm}}
    \toprule
       \textbf{Result} & \textbf{Recommendation} \\ \midrule
       Bots use a lot of retweets and mentions, and have high tweets per hour & Use bots for menial tasks like announcements and amplification of announcements \\ 
       Bots have a star interaction network & Use bots for big announcements (e.g., disaster, crisis management) without message distortion \\ 
       Bot content matches identity & Use bots to provide educational material about topics that people associate with certain professions (e.g. weather information from a weather news bot) \\
       Bots use more abusive and expletive terms than humans & Focus regulation to disallow bots to use toxic language \\
    \bottomrule
    \end{tabular}
    \caption{Recommendations of our observations on leveraging and regulating bots for social good}
    \label{tab:bot_strengths}
\end{table}

\subsection*{\revision{Further Investigations}}
\revision{In this paper, we primarily used an extensive dataset of Twitter data with tweets collected over four years. However, we have done other investigations to go beyond this dataset. In particular, we looked at bots across social media platforms and bot evolution.}

\paragraph{\revision{Bots across Platforms}}
\revision{Are bots the same across different social media platforms? To answer that question, we perform a preliminary analysis towards investigating the similarities and differences of bots on different social media platforms.}

\revision{We compared the psycholinguistic behavior of bots between Twitter and Telegram platforms with respect to the Coronavirus 2020-2021 pandemic. The Telegram dataset contained $\sim$7 million messages written by $\sim$335,000 users, and was segregated into three user groups: bots, disinformation spreaders and humans\cite{ng2024exploratory}. We extracted psycholinguistic cues using the NetMapper software for the Telegram messages in the same fashion as we did for the Twitter messages (see Methods: Comparison by Psycholinguistic Cues for details).}

\begin{figure}[h]
\centering
\includegraphics[width=0.5\textwidth]{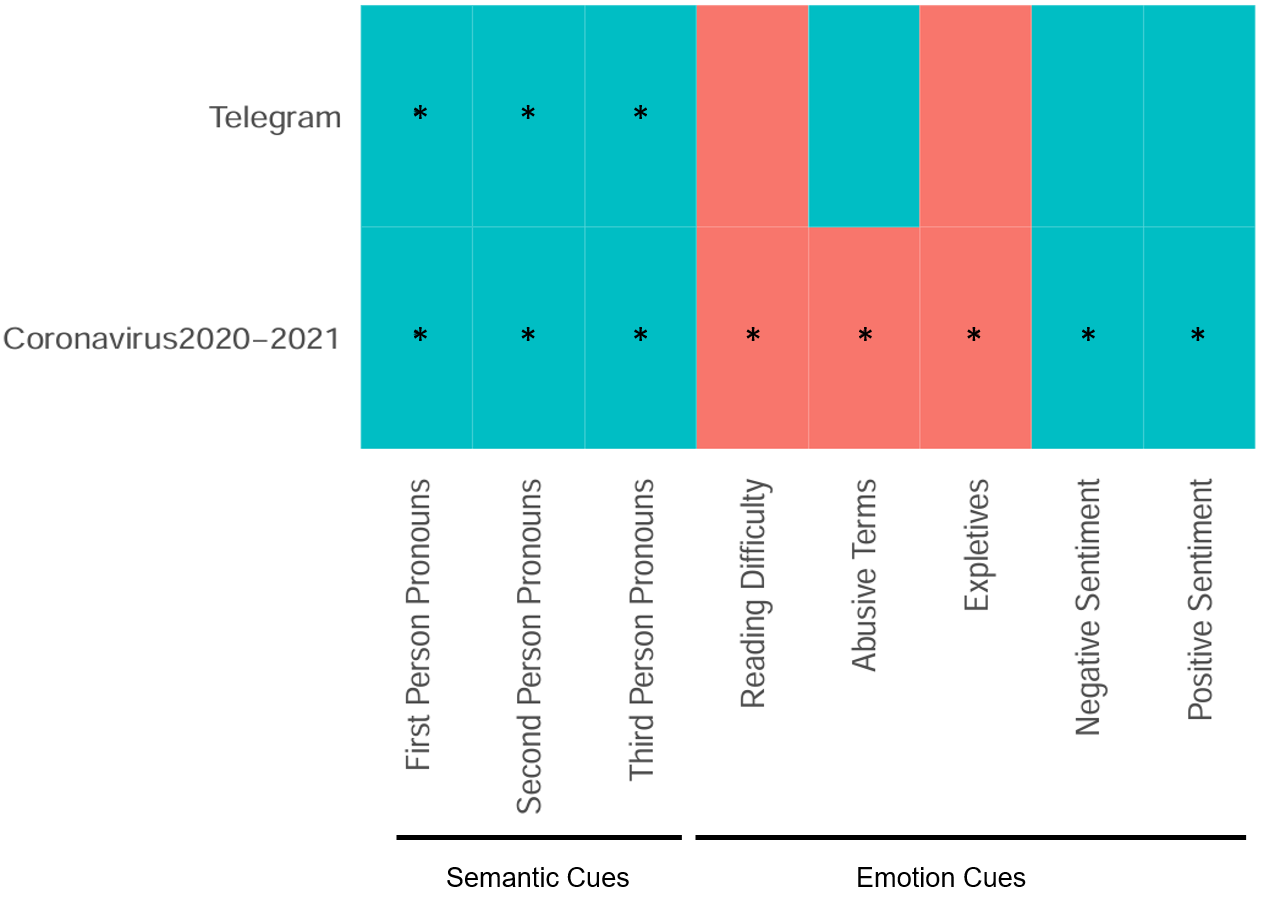}
\caption{\revision{Differences in the use of psycholinguistic cues between bots and humans for Twitter dataset and Telegram dataset. Both datasets are drawn from the Coronavirus2020-2021 event. \textbf{Green} cells show that \textbf{humans} use a larger number of the cue. \textbf{Red} cells show that \textbf{bots} use a larger number of the cue. * within the cells indicates there is a significant difference in the usage of the cue between bots and humans at the $p<0.05$ level.}}
\label{fig:telegram_twitter}
\end{figure}

\revision{\autoref{fig:telegram_twitter} presents the differences between bots and humans for a subset of the psycholinguistic cues. We only compare the semantic and emotion cues because the structure of the Telegram platform is different from the Twitter platform, so the metadata cues do not map directly. This comparison shows that the pattern of cue usage between bots are humans are similar across both platforms, yielding generalizability to our overall study. In the same light, past work had shown that there was insignificant difference in the user metadata and post structure between tweets from Twitter and messages from Telegram\cite{ng2024exploratory}. Bots are generally similar in the usage of linguistic cues across social media platforms, which provides wider generalizability of our results.}

\paragraph{\revision{Bot Evolution}}
\revision{Bots can evolve over time as adversarial actors realize that they are being detected and change their techniques \cite{ng2023botbuster}. Between 2021 and 2023, bots have enhanced their detection evasion techniques by adding four-character letter strings and Chinese proverbs to attempt to fool bot detection tools \cite{Jacobs2024Detection}. However, these techniques are not entirely novel. They are recombinations of social cybersecurity techniques like distract and distort techniques that were discovered in earlier studies \cite{danaditya2022curious,carley2020social}. These techniques revolve around permuting the actions that platforms provide for content and interactions. }

\revision{Bots also continue to evolve as new technologies, such as Generative AI technologies, come out. We used three open-sourced Large Language Models (LLMs) to generate tweets related to the coronavirus event. These models are: Meta Llama 3.1 8B Instruct, Phi-4, and Qwen 2.5 7B Instruct 1M. Each model generated 20 tweets. The Supplementary Material details the methodology and the prompts used for this experiment. Ideally, the use of generative AI technology would make the bot agents undetectable by current bot detection technology. That is, the use of LLMs would reduce the probability that the BotHunter algorithm classifies these tweets as originating from bot users. Across our experiments, the average BotHunter score retrieved was 0.69$\pm$0.15. This score is borderline on the 0.70 bot classification threshold that we used in this study, and is above the 0.50 threshold that many studies use \cite{tyagi2020polarizing,uyheng2021active,yang2022botometer,tardelli2022detecting}. The huge standard deviation indicates that LLMs generate personas that are inconsistent with a single user type: sometimes LLMs generated tweets that look like humans, sometimes they do not.}

\revision{Despite the evolution of bots, bot detectors are still able to identify users as bots. The same BotHunter tool used in this study had been deployed on data from 2023 to 2025 \cite{ng2024tiny,marigliano2024analyzing,ng2025building}. The tool classified 20\% of the users as bot users, a percentage similar to that in our study. This shows that these bot detectors identify broad and comprehensive features. Moreover, these traditional bot types that do not employ generative AI still continue to exist, perhaps because the bot operator does not require generative technologies for his task. For example, a news aggregator bot that retweets a set series of news accounts can be programmed with heuristics. However, conversational bots will benefit from generative AI technologies and be able to carry out a more seamless conversation with the online community. Future work should examine the varied purposes of bots and whether evolution is required or whether evolution has occurred.}

\paragraph{\revision{Future Work}}
\revision{Future work involves more detailed investigation into the nuances of bot behavior across platforms. It is also worth exploring how the content layout and the relationship structure of different platforms affect the users' interaction dynamics. Future work can also take advantage of the temporal to analyze how bots evolve in behavior over time, and during crucial points in an event. The geographical dimension of our dataset further provides opportunity for a social geographical analysis on how bots differ across countries and regions.}

\section*{Challenges and Opportunities of studying Social Media Bots}
Next, we elaborate on three challenges in the study of social media bot, and discuss some opportunities for future research.

\subsection*{Detect} The first step to bot detection is to systematically detect these bots. However, these automated agents are constantly evolving and adapting their behavior in response to the changing setup of social media platforms and user patterns. The stricter data collection rules of social media platforms \cite{schroeder2014big,gillespie2018custodians} and the increasing usage of AI in these bot agents \cite{yang2024anatomy} creates further variability in these digital spaces bots reside in. This therefore muddles any developed algorithms based on previous datasets. 

Already, linguistic differences between bot and human tweets have narrowed between 2017 and 2020, making bot accounts more difficult to systematically differentiate \cite{ng2023botbuster}. More recently, AI-powered botnets have emerged, using ChatGPT models to generate human-like content \cite{yang2024anatomy}, closing the gap between bot and human. 

Bot evolution and bot detection are thus a ``never-ending clash" \cite{cresci2020detecting}, and sometimes bot accounts evolve faster than current known bot detection algorithms \cite{martini2021bot}, presenting several opportunities in continual improvement of bot detection algorithms, specifically to be adaptable, faster, and more efficient. The increasing trends of using Large Language Models and Large Vision Models to create generated texts and deepfakes lend bots a helping hand in the construction of more believable narratives. These same generative technology are also used to construct offensive bots for humor \cite{wiredMillionsPeople}. However, current trends reflect that the use of such technologies \revision{is} not very prevalent, for example, \cite{yang2024anatomy} only found one set of such botnet in their study, reflecting that bots are still relying on traditional techniques, likely because such heuristic-based techniques are easier and faster to deploy en masse.

\revision{Beyond simply detecting bot users in social media discourse, the next research agenda is to perform content-interaction impact analysis. Our work revealed one common interaction patterns bots use in their content communication. Hereupon arises an opportunity to characterize archetypes of interaction patterns, the best type of content to disseminate with each archetype, and the impact of the dissemination strategy on public opinion. }

\subsection*{Differentiate} After identifying which users are likely to be bots, one must differentiate the goodness of the bot and its function. This evaluation can be inferred from the bot's content postings and relationship interactions. However, bots do not fall squarely in a spectrum of goodness; the lines of good and bad bots are blurred. In fact, bots can move between neutral in which they post messages that are not harmful, to bad, where they post politically charged and extremist messages \cite{danaditya2022curious,ferrara2017contagion}. Herein lies an opportunity to construct a rubric to determine the goodness of the bot; this, though, is a complex task, for there are ethical and societal issues to consider. Bots can change their goodness, too. They may be supporting a certain cause initially, then making a swing to a different stance soon enough. This swing of support was witnessed during the coronavirus pandemic era, \revision{where bots change their stances towards the vaccination campaign, disseminating narratives of different stances during different periods \cite{aldayel2022characterizing,ng2022pro}.} \revision{This opinion dynamics is important to the health of social discourse,} and especially so when the bots require little conviction to change allegiances \cite{ng2022pro}. Another challenge involves identifying the type of bot, which can provide insight towards possible impact of the bot. For example, an Amplifier Bot that intensifies political messages could be intended to sow discord \cite{jacobs2023tracking,woolley2018computational}.

\subsection*{Disrupt} The third challenge is to mindfully disrupt the operations of bot users. That is, moderating the impact of malicious bots, while not unsettling human conversations. While banning bot users can be an easy solution, a blanket ban can result in many false positives, which thus results in humans being identified as bots and being banned. Such situations can result in emotional or psychological harm of the human being banned, or toxic online behavior where users repeatedly report another user that they personally dislike as a bot to silence them \cite{jones2018silencing}. Additionally, social media bots do not necessarily work alone: they coordinate with other bots -- sometimes even human agents -- to push out their agenda \cite{ng2023combined}, and therefore if one agent warrants a ban, should the entire network be banned? To ban an entire network may entangle several unsuspecting humans who have been influenced by the bots to partake in the conversation. With these considerations in mind, regulation is a scope of problem with which to be studied: which types of bots should we ban? What are the activities of a bot that would warrant a ban?

\section*{Methods}
\subsection*{Examining Bot Literature}
We examined recent bot literature for the definition of ``social media bot". For academic definitions, we searched the phrase ``social media bot" on Google Scholar. For industry literature, we searched the phrase ``social media bot" on Google Search. Then, we manually scanned through the results. We picked out the more relevant and highly cited papers that had a definition of a social media bot. We read through each paper, and manually extracted the definition of a social media bot stated in the paper. 

Next, we looked through all the definitions and picked out key phrases. We then harmonized the phrases and definitions to create a general definition of the bot. All authors agreed on the definitions and categorizations.

\subsection*{Data Collection and Labeling}
We collected a dataset from Twitter/X involving global events which provides a richness in a general understanding of the bot and human differentiation. The list of data collection parameters are detailed in \revision{Supplementary Material Table 1 and 2 ``Dataset Collection Parameters"}.

We labeled each user in this dataset as bot or human with the BotHunter algorithm. This algorithm uses a tiered random forest classifier with increasing amounts of user data to evaluate the probability of the user being a bot. The algorithm returns a bot probability score that is between 0 and 1, where scores above 0.7 we deem as a bot, and scores below 0.7 we deem as a human. This 0.7 threshold value is determined from a previous longitudinal study that sought to identify a stable bot score threshold that best represents the automation capacity of a user \cite{ng2022stabilizing}. This bot algorithm and threshold is chosen so that our studies will be consistent with the original studies of the dataset that used the BotHunter algorithm \cite{uyheng2021active,magelinski2021synchronized,babcock2018beaten,king2020lying,babcock2020pretending,ng2023combined}. We calculated the proportion of bot users against the total number of users within each event. Our results are presented in a bar graph.

\paragraph{Comparison by Psycholinguistic Cues}
We parse the collected dataset of tweets through the NetMapper software \cite{carley2018ora} to extract out psycholinguistic cues of the texts. NetMapper extracts the number of each of the cues per tweet. The software returns three types of cues: semantic cues, emotion cues and metadata cues. The linguistic cues are returned by matching words against a dictionary for each category. The dictionary has words in 40 languages. Then, for each user, we average the use of each cue per category, as the trend for the user. \revision{We then perform a student t-test comparison between the cues of each user type with Bonferroni correction, and identify whether the cues are significantly different between the bot and human at the $p<0.05$ level}. We then remove the retweets from the Captain Marvel and Black Panther datasets and compare the cue distribution of original tweets with all tweets. This analysis compares the differences in the distribution of cues of tweets originating from the user type and their retweets.

\paragraph{Comparison by Self-Presentation of Identity}
To classify identities, we compare the user description and bio information against a survey of \revision{occupation} of United States users performed in 2015 \cite{smith2016mean}. If the occupation is present in the user information, the user is tagged with the identity. A user can have more than one identity. We compare the top identities used by bots and human users across all events. These identities are also divided up into seven categories: religion, race/nationality, political, job, gender, family and others. We then classify each user into these categories of identities. Again, each user can fall into multiple categories. 

Next, we examined how different identities frame their posts differently. \revision{We examined five topic frames:  family, gender, political, race/nationality and religion. We parsed our data through the NetMapper software, which returned us the number of framing cues present in each tweet. These cues are calculated through matching a multilingual lexicon for each frame. Then, for each user, we average the use of each of the topic frames.}
For each most frequent identity affiliated with by bots and humans, we compare the difference in the average use of each topic frame through a percentage difference calculation. The percentage difference in the use of framing cues is calculated as: $\frac{(H - B)}{H}$, where $H$ is the average use of the framing cue by humans, and $B$ is the average use of framing cue by bots. This comparison tells us how much more bots use a framing cue as compared to humans. If the percentage is negative, bots use the framing cue more than humans. If the percentage is positive, bots use the cue less than humans. 

The set of topic frames also corresponds with the identity categories. Therefore, we also compared the identity categories against the average use of each topic frame. This comparison is performed across bots and humans. We plot heatmaps to show the relationship between the average use of each topic frame topic frame against the identity categories.

\paragraph{Comparison by Social Interactions}
We construct the all-communication ego-networks of the users in our dataset. We analyzed all the users for Asian Elections, Black Panther, Canadian Elections 2019, Captain Marvel and ReOpen America events. Due to the size of the data, we analyzed a 2\% sample of users of the Coronavirus2020-2021 users ($N=4.6$mil), and a 50\% sample of users from the US Elections 2020 ($N=500$k). The ego-networks are network graphs of the bot and human users in focus. In the networks, each user is represented as a node, and a communication interaction between users are represented as links. The ego-networks are constructed using all-communication interactions, that is any communication between users (i.e., retweet, @mentions, reply, quote) is reflected as a link.
We analyzed the network properties of the ego-networks constructed per event. These properties are: total-degree, in degree, out degree, density. We also analyzed the number of bot and human alters there are in the ego networks. No pre-processing \revision{was} performed on the networks prior to the calculations. We used the ORA software to load in the networks and perform the calculations \cite{carley2018ora}. We finally visualize the network graphs of one- and two- degree ego networks of a sample of bots and humans \autoref{fig:social_interaction}. These are the 20 most frequent communicators in the Asian Elections sub-dataset A 1-degree network shows alters (connected users) that are in direct communication with the user, and a 2-degree network shows alters in direct communication with the 1st-degree alters.

\section*{Conclusion}
Social media bots are deeply interweaved into our digital ecosystem. More than half of the Internet traffic in 2023 \revision{was} generated by these AI agents \cite{theatlanticInternetMostly}. Bots are able to generate this volume of traffic because of their use of automation, which enables them to create more content and form more relationships. 
This article surmised a definition of a social media bot based on the three elements that a social media platform contains: user, content, interactions. Our definition breaks down the automation on social media platforms into its core mechanics, and therefore \revision{provides} the foundation for further research, analysis and policies regulating the digital space. 
We performed a large scale data analysis of bot and human characteristics across events around the globe, presenting the uniqueness of the bot species from a macro perspective: how bots and humans differ in terms of the use of linguistic cues, social identity affiliations and social interactions.  On a global scale, bots and humans do have consistent differences, which can be used to differentiate the two species of users. \autoref{tab:summary} summarizes the differences between bots and humans as a conclusive remark.
Finally, we provide recommendations for the use and regulation of bots. These recommendations are informed by our results. We also lay out the challenges and opportunities for the future of bot detection in a ``Detect, Differentiate, Disrupt" frame. We invite academics, non-profits \revision{, and policymakers} to take part in this active research area.

\begin{table}[H]
    \centering
    \begin{tabular}{p{5cm}p{5cm}p{5cm}}
        \toprule
         & \textbf{Bots} & \textbf{Humans} \\ \midrule
         \textbf{Volume (\%)} & 21.9 $\pm$ 9.8 & 78.1 $\pm$ 9.8\\ 
         \textbf{Psycholinguistic Cues} & Use more hashtags, mentions; has more tweets/hour, total tweets, friends:followers ratio & uses more media, favorites, replies, quotes, urls \\ 
         \textbf{Self-presentation of Identity} & Concentrate their affiliations on a few identities & Have a more varied identity affiliations \\ 
         \textbf{Have identity affiliation (\%)} & 21.4$\pm$5.7 & 27.0$\pm$9.2 \\ 
         \textbf{Topic Frames} & Political topics & Family and Gender \\
         \textbf{Identity vs Topic Frames} & Converse about topics that closely match their identity & Have a larger range of topics \\ 
         \textbf{Social Interactions} & Star communication structure & Tiered communication structure \\ 
         ~ & Denser interaction networks & Less dense interaction networks \\
         ~ & Interact with more human than bot alters & Interact with more human than bot alters\\
         \bottomrule
    \end{tabular}
    \caption{Summary of Differences between Bots and Humans.}
    \label{tab:summary}
\end{table}

\bibliography{biblography}

\begin{thebibliography}{100}
\urlstyle{rm}
\expandafter\ifx\csname url\endcsname\relax
  \def\url#1{\texttt{#1}}\fi
\expandafter\ifx\csname urlprefix\endcsname\relax\def\urlprefix{URL }\fi
\expandafter\ifx\csname doiprefix\endcsname\relax\def\doiprefix{DOI: }\fi
\providecommand{\bibinfo}[2]{#2}
\providecommand{\eprint}[2][]{\url{#2}}

\bibitem{woolley2016automating}
\bibinfo{author}{Woolley, S.~C.}
\newblock \bibinfo{journal}{\bibinfo{title}{Automating power: Social bot interference in global politics}}.
\newblock {\emph{\JournalTitle{First Monday}}}  (\bibinfo{year}{2016}).

\bibitem{lotan2011arab}
\bibinfo{author}{Lotan, G.} \emph{et~al.}
\newblock \bibinfo{journal}{\bibinfo{title}{The arab spring| the revolutions were tweeted: Information flows during the 2011 tunisian and egyptian revolutions}}.
\newblock {\emph{\JournalTitle{International journal of communication}}} \textbf{\bibinfo{volume}{5}}, \bibinfo{pages}{31} (\bibinfo{year}{2011}).

\bibitem{ng2024cyborgs}
\bibinfo{author}{Ng, L. H.~X.}, \bibinfo{author}{Robertson, D.~C.} \& \bibinfo{author}{Carley, K.~M.}
\newblock \bibinfo{journal}{\bibinfo{title}{Cyborgs for strategic communication on social media}}.
\newblock {\emph{\JournalTitle{Big Data \& Society}}} \textbf{\bibinfo{volume}{11}}, \bibinfo{pages}{20539517241231275} (\bibinfo{year}{2024}).

\bibitem{ng2024assembling}
\bibinfo{author}{Ng, L. H.~X.} \& \bibinfo{author}{Carley, K.~M.}
\newblock \bibinfo{journal}{\bibinfo{title}{Assembling a multi-platform ensemble social bot detector with applications to us 2020 elections}}.
\newblock {\emph{\JournalTitle{Social Network Analysis and Mining}}} \textbf{\bibinfo{volume}{14}}, \bibinfo{pages}{45} (\bibinfo{year}{2024}).

\bibitem{chang2021social}
\bibinfo{author}{Chang, H.-C.~H.}, \bibinfo{author}{Chen, E.}, \bibinfo{author}{Zhang, M.}, \bibinfo{author}{Muric, G.} \& \bibinfo{author}{Ferrara, E.}
\newblock \bibinfo{title}{Social bots and social media manipulation in 2020: The year in review}.
\newblock In \emph{\bibinfo{booktitle}{Handbook of Computational Social Science, Volume 1}}, \bibinfo{pages}{304--323} (\bibinfo{publisher}{Routledge}, \bibinfo{year}{2021}).

\bibitem{seckin2024mechanisms}
\bibinfo{author}{Seckin, O.~C.}, \bibinfo{author}{Atalay, A.}, \bibinfo{author}{Otenen, E.}, \bibinfo{author}{Duygu, U.} \& \bibinfo{author}{Varol, O.}
\newblock \bibinfo{journal}{\bibinfo{title}{Mechanisms driving online vaccine debate during the covid-19 pandemic}}.
\newblock {\emph{\JournalTitle{Social Media+ Society}}} \textbf{\bibinfo{volume}{10}}, \bibinfo{pages}{20563051241229657} (\bibinfo{year}{2024}).

\bibitem{ferrara2020types}
\bibinfo{author}{Ferrara, E.}
\newblock \bibinfo{journal}{\bibinfo{title}{What types of covid-19 conspiracies are populated by twitter bots?}}
\newblock {\emph{\JournalTitle{arXiv preprint arXiv:2004.09531}}}  (\bibinfo{year}{2020}).

\bibitem{ng2022pro}
\bibinfo{author}{Ng, L. H.~X.} \& \bibinfo{author}{Carley, K.~M.}
\newblock \bibinfo{journal}{\bibinfo{title}{Pro or anti? a social influence model of online stance flipping}}.
\newblock {\emph{\JournalTitle{IEEE Transactions on Network Science and Engineering}}} \textbf{\bibinfo{volume}{10}}, \bibinfo{pages}{3--19} (\bibinfo{year}{2022}).

\bibitem{magelinski2021synchronized}
\bibinfo{author}{Magelinski, T.}, \bibinfo{author}{Ng, L. H.~X.} \& \bibinfo{author}{Carley, K.~M.}
\newblock \bibinfo{journal}{\bibinfo{title}{A synchronized action framework for responsible detection of coordination on social media}}.
\newblock {\emph{\JournalTitle{arXiv preprint arXiv:2105.07454}}}  (\bibinfo{year}{2021}).

\bibitem{broniatowski2018weaponized}
\bibinfo{author}{Broniatowski, D.~A.} \emph{et~al.}
\newblock \bibinfo{journal}{\bibinfo{title}{Weaponized health communication: Twitter bots and russian trolls amplify the vaccine debate}}.
\newblock {\emph{\JournalTitle{American journal of public health}}} \textbf{\bibinfo{volume}{108}}, \bibinfo{pages}{1378--1384} (\bibinfo{year}{2018}).

\bibitem{shao2018spread}
\bibinfo{author}{Shao, C.} \emph{et~al.}
\newblock \bibinfo{journal}{\bibinfo{title}{The spread of low-credibility content by social bots}}.
\newblock {\emph{\JournalTitle{Nature communications}}} \textbf{\bibinfo{volume}{9}}, \bibinfo{pages}{1--9} (\bibinfo{year}{2018}).

\bibitem{ng2022cross}
\bibinfo{author}{Ng, L. H.~X.}, \bibinfo{author}{Cruickshank, I.~J.} \& \bibinfo{author}{Carley, K.~M.}
\newblock \bibinfo{journal}{\bibinfo{title}{Cross-platform information spread during the january 6th capitol riots}}.
\newblock {\emph{\JournalTitle{Social Network Analysis and Mining}}} \textbf{\bibinfo{volume}{12}}, \bibinfo{pages}{133} (\bibinfo{year}{2022}).

\bibitem{cjrMusksTwitter}
\bibinfo{author}{Ingram, M.}
\newblock \bibinfo{title}{{M}usk’s {T}witter bid, and the ‘bot’ complication}.
\newblock \bibinfo{howpublished}{\url{https://www.cjr.org/the_media_today/musks-twitter-bid-and-the-bot-complication.php}} (\bibinfo{year}{2022}).
\newblock \bibinfo{note}{[Accessed 28-10-2024]}.

\bibitem{latimesElonMusk}
\bibinfo{author}{Childs, J.}
\newblock \bibinfo{title}{{E}lon {M}usk says {X} is fighting bots and spam, and the solution is: \$1 subscriptions --- latimes.com}.
\newblock \bibinfo{howpublished}{\url{https://www.latimes.com/business/story/2023-10-18/x-pilot-program-to-charge-1-a-year-in-effort-to-combat-bots-spam}} (\bibinfo{year}{2023}).
\newblock \bibinfo{note}{[Accessed 28-10-2024]}.

\bibitem{ellaky2023systematic}
\bibinfo{author}{Ellaky, Z.}, \bibinfo{author}{Benabbou, F.} \& \bibinfo{author}{Ouahabi, S.}
\newblock \bibinfo{journal}{\bibinfo{title}{Systematic literature review of social media bots detection systems}}.
\newblock {\emph{\JournalTitle{Journal of King Saud University-Computer and Information Sciences}}} \textbf{\bibinfo{volume}{35}}, \bibinfo{pages}{101551} (\bibinfo{year}{2023}).

\bibitem{ng2024tiny}
\bibinfo{author}{Ng, L. H.~X.}, \bibinfo{author}{Bartulovic, M.} \& \bibinfo{author}{Carley, K.~M.}
\newblock \bibinfo{title}{Tiny-botbuster: Identifying automated political coordination in digital campaigns}.
\newblock In \emph{\bibinfo{booktitle}{International Conference on Social Computing, Behavioral-Cultural Modeling and Prediction and Behavior Representation in Modeling and Simulation}}, \bibinfo{pages}{25--34} (\bibinfo{organization}{Springer}, \bibinfo{year}{2024}).

\bibitem{beskow2018bot}
\bibinfo{author}{Beskow, D.~M.} \& \bibinfo{author}{Carley, K.~M.}
\newblock \bibinfo{title}{Bot-hunter: a tiered approach to detecting \& characterizing automated activity on twitter}.
\newblock In \emph{\bibinfo{booktitle}{Conference paper. SBP-BRiMS: International conference on social computing, behavioral-cultural modeling and prediction and behavior representation in modeling and simulation}}, vol.~\bibinfo{volume}{3} (\bibinfo{year}{2018}).

\bibitem{sayyadiharikandeh2020detection}
\bibinfo{author}{Sayyadiharikandeh, M.}, \bibinfo{author}{Varol, O.}, \bibinfo{author}{Yang, K.-C.}, \bibinfo{author}{Flammini, A.} \& \bibinfo{author}{Menczer, F.}
\newblock \bibinfo{title}{Detection of novel social bots by ensembles of specialized classifiers}.
\newblock In \emph{\bibinfo{booktitle}{Proceedings of the 29th ACM international conference on information \& knowledge management}}, \bibinfo{pages}{2725--2732} (\bibinfo{year}{2020}).

\bibitem{ng2023botbuster}
\bibinfo{author}{Ng, L. H.~X.} \& \bibinfo{author}{Carley, K.~M.}
\newblock \bibinfo{title}{Botbuster: Multi-platform bot detection using a mixture of experts}.
\newblock In \emph{\bibinfo{booktitle}{Proceedings of the international AAAI conference on web and social media}}, vol.~\bibinfo{volume}{17}, \bibinfo{pages}{686--697} (\bibinfo{year}{2023}).

\bibitem{orabi2020detection}
\bibinfo{author}{Orabi, M.}, \bibinfo{author}{Mouheb, D.}, \bibinfo{author}{Al~Aghbari, Z.} \& \bibinfo{author}{Kamel, I.}
\newblock \bibinfo{journal}{\bibinfo{title}{Detection of bots in social media: a systematic review}}.
\newblock {\emph{\JournalTitle{Information Processing \& Management}}} \textbf{\bibinfo{volume}{57}}, \bibinfo{pages}{102250} (\bibinfo{year}{2020}).

\bibitem{kolomeets2021bot}
\bibinfo{author}{Kolomeets, M.}, \bibinfo{author}{Chechulin, A.} \& \bibinfo{author}{Kotenko, I.~V.}
\newblock \bibinfo{journal}{\bibinfo{title}{Bot detection by friends graph in social networks.}}
\newblock {\emph{\JournalTitle{J. Wirel. Mob. Networks Ubiquitous Comput. Dependable Appl.}}} \textbf{\bibinfo{volume}{12}}, \bibinfo{pages}{141--159} (\bibinfo{year}{2021}).

\bibitem{li2023botfinder}
\bibinfo{author}{Li, S.} \emph{et~al.}
\newblock \bibinfo{journal}{\bibinfo{title}{Botfinder: a novel framework for social bots detection in online social networks based on graph embedding and community detection}}.
\newblock {\emph{\JournalTitle{World Wide Web}}} \textbf{\bibinfo{volume}{26}}, \bibinfo{pages}{1793--1809} (\bibinfo{year}{2023}).

\bibitem{feng2024does}
\bibinfo{author}{Feng, S.} \emph{et~al.}
\newblock \bibinfo{journal}{\bibinfo{title}{What does the bot say? opportunities and risks of large language models in social media bot detection}}.
\newblock {\emph{\JournalTitle{arXiv preprint arXiv:2402.00371}}}  (\bibinfo{year}{2024}).

\bibitem{jacobs2023tracking}
\bibinfo{author}{Jacobs, C.~S.}, \bibinfo{author}{Ng, L. H.~X.} \& \bibinfo{author}{Carley, K.~M.}
\newblock \bibinfo{title}{Tracking china’s cross-strait bot networks against taiwan}.
\newblock In \emph{\bibinfo{booktitle}{International conference on social computing, behavioral-cultural modeling and prediction and behavior representation in modeling and simulation}}, \bibinfo{pages}{115--125} (\bibinfo{organization}{Springer}, \bibinfo{year}{2023}).

\bibitem{uyheng2020bot}
\bibinfo{author}{Uyheng, J.} \& \bibinfo{author}{Carley, K.~M.}
\newblock \bibinfo{title}{Bot impacts on public sentiment and community structures: Comparative analysis of three elections in the asia-pacific}.
\newblock In \emph{\bibinfo{booktitle}{Social, Cultural, and Behavioral Modeling: 13th International Conference, SBP-BRiMS 2020, Washington, DC, USA, October 18--21, 2020, Proceedings 13}}, \bibinfo{pages}{12--22} (\bibinfo{organization}{Springer}, \bibinfo{year}{2020}).

\bibitem{bessi2016social}
\bibinfo{author}{Bessi, A.} \& \bibinfo{author}{Ferrara, E.}
\newblock \bibinfo{journal}{\bibinfo{title}{Social bots distort the 2016 us presidential election online discussion}}.
\newblock {\emph{\JournalTitle{First monday}}} \textbf{\bibinfo{volume}{21}} (\bibinfo{year}{2016}).

\bibitem{khaund2018analyzing}
\bibinfo{author}{Khaund, T.}, \bibinfo{author}{Al-Khateeb, S.}, \bibinfo{author}{Tokdemir, S.} \& \bibinfo{author}{Agarwal, N.}
\newblock \bibinfo{title}{Analyzing social bots and their coordination during natural disasters}.
\newblock In \emph{\bibinfo{booktitle}{Social, Cultural, and Behavioral Modeling: 11th International Conference, SBP-BRiMS 2018, Washington, DC, USA, July 10-13, 2018, Proceedings 11}}, \bibinfo{pages}{207--212} (\bibinfo{organization}{Springer}, \bibinfo{year}{2018}).

\bibitem{ng2021does}
\bibinfo{author}{Ng, L.~H.} \& \bibinfo{author}{Taeihagh, A.}
\newblock \bibinfo{journal}{\bibinfo{title}{How does fake news spread? understanding pathways of disinformation spread through apis}}.
\newblock {\emph{\JournalTitle{Policy \& Internet}}} \textbf{\bibinfo{volume}{13}}, \bibinfo{pages}{560--585} (\bibinfo{year}{2021}).

\bibitem{hajli2022social}
\bibinfo{author}{Hajli, N.}, \bibinfo{author}{Saeed, U.}, \bibinfo{author}{Tajvidi, M.} \& \bibinfo{author}{Shirazi, F.}
\newblock \bibinfo{journal}{\bibinfo{title}{Social bots and the spread of disinformation in social media: the challenges of artificial intelligence}}.
\newblock {\emph{\JournalTitle{British Journal of Management}}} \textbf{\bibinfo{volume}{33}}, \bibinfo{pages}{1238--1253} (\bibinfo{year}{2022}).

\bibitem{kenny2024duped}
\bibinfo{author}{Kenny, R.}, \bibinfo{author}{Fischhoff, B.}, \bibinfo{author}{Davis, A.}, \bibinfo{author}{Carley, K.~M.} \& \bibinfo{author}{Canfield, C.}
\newblock \bibinfo{journal}{\bibinfo{title}{Duped by bots: why some are better than others at detecting fake social media personas}}.
\newblock {\emph{\JournalTitle{Human factors}}} \textbf{\bibinfo{volume}{66}}, \bibinfo{pages}{88--102} (\bibinfo{year}{2024}).

\bibitem{kolomeets2024experimental}
\bibinfo{author}{Kolomeets, M.}, \bibinfo{author}{Tushkanova, O.}, \bibinfo{author}{Desnitsky, V.}, \bibinfo{author}{Vitkova, L.} \& \bibinfo{author}{Chechulin, A.}
\newblock \bibinfo{journal}{\bibinfo{title}{Experimental evaluation: Can humans recognise social media bots?}}
\newblock {\emph{\JournalTitle{Big Data and Cognitive Computing}}} \textbf{\bibinfo{volume}{8}}, \bibinfo{pages}{24} (\bibinfo{year}{2024}).

\bibitem{carley2020social}
\bibinfo{author}{Carley, K.~M.}
\newblock \bibinfo{journal}{\bibinfo{title}{Social cybersecurity: an emerging science}}.
\newblock {\emph{\JournalTitle{Computational and mathematical organization theory}}} \textbf{\bibinfo{volume}{26}}, \bibinfo{pages}{365--381} (\bibinfo{year}{2020}).

\bibitem{ferrara2016rise}
\bibinfo{author}{Ferrara, E.}, \bibinfo{author}{Varol, O.}, \bibinfo{author}{Davis, C.}, \bibinfo{author}{Menczer, F.} \& \bibinfo{author}{Flammini, A.}
\newblock \bibinfo{journal}{\bibinfo{title}{The rise of social bots}}.
\newblock {\emph{\JournalTitle{Communications of the ACM}}} \textbf{\bibinfo{volume}{59}}, \bibinfo{pages}{96--104} (\bibinfo{year}{2016}).

\bibitem{uyheng2021active}
\bibinfo{author}{Uyheng, J.}, \bibinfo{author}{Ng, L. H.~X.} \& \bibinfo{author}{Carley, K.~M.}
\newblock \bibinfo{journal}{\bibinfo{title}{Active, aggressive, but to little avail: characterizing bot activity during the 2020 singaporean elections}}.
\newblock {\emph{\JournalTitle{Computational and Mathematical Organization Theory}}} \textbf{\bibinfo{volume}{27}}, \bibinfo{pages}{324--342} (\bibinfo{year}{2021}).

\bibitem{himelein2021bots}
\bibinfo{author}{Himelein-Wachowiak, M.} \emph{et~al.}
\newblock \bibinfo{journal}{\bibinfo{title}{Bots and misinformation spread on social media: Implications for covid-19}}.
\newblock {\emph{\JournalTitle{Journal of medical Internet research}}} \textbf{\bibinfo{volume}{23}}, \bibinfo{pages}{e26933} (\bibinfo{year}{2021}).

\bibitem{ng2024exploring}
\bibinfo{author}{Ng, L. H.~X.}, \bibinfo{author}{Zhou, W.} \& \bibinfo{author}{Carley, K.~M.}
\newblock \bibinfo{journal}{\bibinfo{title}{Exploring cognitive bias triggers in covid-19 misinformation tweets: A bot vs. human perspective}}.
\newblock {\emph{\JournalTitle{arXiv preprint arXiv:2406.07293}}}  (\bibinfo{year}{2024}).

\bibitem{cloudflarebot}
\bibinfo{author}{Cloudflare}.
\newblock \bibinfo{title}{What is a social media bot? | social media bot definition}.
\newblock \bibinfo{howpublished}{\url{https://www.cloudflare.com/learning/bots/what-is-a-social-media-bot/}}.
\newblock \bibinfo{note}{[Accessed 28-10-2024]}.

\bibitem{assenmacher2020demystifying}
\bibinfo{author}{Assenmacher, D.} \emph{et~al.}
\newblock \bibinfo{journal}{\bibinfo{title}{Demystifying social bots: On the intelligence of automated social media actors}}.
\newblock {\emph{\JournalTitle{Social Media+ Society}}} \textbf{\bibinfo{volume}{6}}, \bibinfo{pages}{2056305120939264} (\bibinfo{year}{2020}).

\bibitem{cisa}
\bibinfo{author}{DHS, U. D. o. H.~S.}
\newblock \bibinfo{title}{niccs.cisa.gov}.
\newblock \bibinfo{howpublished}{\url{https://niccs.cisa.gov/sites/default/files/documents/pdf/ncsam_socialmediabotsoverview_508.pdf?trackDocs=ncsam_socialmediabotsoverview_508.pdf}} (\bibinfo{year}{2018}).
\newblock \bibinfo{note}{[Accessed 29-10-2024]}.

\bibitem{danaditya2022curious}
\bibinfo{author}{Danaditya, A.}, \bibinfo{author}{Ng, L. H.~X.} \& \bibinfo{author}{Carley, K.~M.}
\newblock \bibinfo{journal}{\bibinfo{title}{From curious hashtags to polarized effect: profiling coordinated actions in indonesian twitter discourse}}.
\newblock {\emph{\JournalTitle{Social Network Analysis and Mining}}} \textbf{\bibinfo{volume}{12}}, \bibinfo{pages}{105} (\bibinfo{year}{2022}).

\bibitem{hayawi2023social}
\bibinfo{author}{Hayawi, K.}, \bibinfo{author}{Saha, S.}, \bibinfo{author}{Masud, M.~M.}, \bibinfo{author}{Mathew, S.~S.} \& \bibinfo{author}{Kaosar, M.}
\newblock \bibinfo{journal}{\bibinfo{title}{Social media bot detection with deep learning methods: a systematic review}}.
\newblock {\emph{\JournalTitle{Neural Computing and Applications}}} \textbf{\bibinfo{volume}{35}}, \bibinfo{pages}{8903--8918} (\bibinfo{year}{2023}).

\bibitem{tsvetkova2017even}
\bibinfo{author}{Tsvetkova, M.}, \bibinfo{author}{Garc{\'\i}a-Gavilanes, R.}, \bibinfo{author}{Floridi, L.} \& \bibinfo{author}{Yasseri, T.}
\newblock \bibinfo{journal}{\bibinfo{title}{Even good bots fight: The case of wikipedia}}.
\newblock {\emph{\JournalTitle{PloS one}}} \textbf{\bibinfo{volume}{12}}, \bibinfo{pages}{e0171774} (\bibinfo{year}{2017}).

\bibitem{stieglitz2017socialsheep}
\bibinfo{author}{Stieglitz, S.}, \bibinfo{author}{Brachten, F.}, \bibinfo{author}{Ross, B.} \& \bibinfo{author}{Jung, A.-K.}
\newblock \bibinfo{journal}{\bibinfo{title}{Do social bots dream of electric sheep? a categorisation of social media bot accounts}}.
\newblock {\emph{\JournalTitle{arXiv preprint arXiv:1710.04044}}}  (\bibinfo{year}{2017}).

\bibitem{cisainfographic}
\bibinfo{author}{CISA, C.} \& \bibinfo{author}{Agency, I.~S.}
\newblock \bibinfo{title}{Social media bots}.
\newblock \bibinfo{howpublished}{\url{https://www.cisa.gov/sites/default/files/publications/social_media_bots_infographic_set_508.pdf}}.
\newblock \bibinfo{note}{[Accessed 28-10-2024]}.

\bibitem{hofeditz2019meaningful}
\bibinfo{author}{Hofeditz, L.}, \bibinfo{author}{Ehnis, C.}, \bibinfo{author}{Bunker, D.}, \bibinfo{author}{Brachten, F.} \& \bibinfo{author}{Stieglitz, S.}
\newblock \bibinfo{title}{Meaningful use of social bots? possible applications in crisis communication during disasters.}
\newblock In \emph{\bibinfo{booktitle}{ECIS}} (\bibinfo{year}{2019}).

\bibitem{10.1145/3461778.3462114}
\bibinfo{author}{Piccolo, L. S.~G.}, \bibinfo{author}{Troullinou, P.} \& \bibinfo{author}{Alani, H.}
\newblock \bibinfo{title}{Chatbots to support children in coping with online threats: Socio-technical requirements}.
\newblock In \emph{\bibinfo{booktitle}{Designing Interactive Systems Conference 2021}}, DIS '21, \bibinfo{pages}{1504–1517}, \doiprefix\url{10.1145/3461778.3462114} (\bibinfo{publisher}{Association for Computing Machinery}, \bibinfo{address}{New York, NY, USA}, \bibinfo{year}{2021}).

\bibitem{krueger2022communing}
\bibinfo{author}{Krueger, J.} \& \bibinfo{author}{Osler, L.}
\newblock \bibinfo{journal}{\bibinfo{title}{Communing with the dead online: chatbots, grief, and continuing bonds}}.
\newblock {\emph{\JournalTitle{Journal of Consciousness Studies}}} \textbf{\bibinfo{volume}{29}}, \bibinfo{pages}{222--252} (\bibinfo{year}{2022}).

\bibitem{boshmaf2011socialbot}
\bibinfo{author}{Boshmaf, Y.}, \bibinfo{author}{Muslukhov, I.}, \bibinfo{author}{Beznosov, K.} \& \bibinfo{author}{Ripeanu, M.}
\newblock \bibinfo{title}{The socialbot network: when bots socialize for fame and money}.
\newblock In \emph{\bibinfo{booktitle}{Proceedings of the 27th annual computer security applications conference}}, \bibinfo{pages}{93--102} (\bibinfo{year}{2011}).

\bibitem{howard2012social}
\bibinfo{author}{Howard, P.~N.} \& \bibinfo{author}{Parks, M.~R.}
\newblock \bibinfo{title}{Social media and political change: Capacity, constraint, and consequence} (\bibinfo{year}{2012}).

\bibitem{chavoshi2016debot}
\bibinfo{author}{Chavoshi, N.}, \bibinfo{author}{Hamooni, H.} \& \bibinfo{author}{Mueen, A.}
\newblock \bibinfo{title}{Debot: Twitter bot detection via warped correlation.}
\newblock In \emph{\bibinfo{booktitle}{Icdm}}, vol.~\bibinfo{volume}{18}, \bibinfo{pages}{28--65} (\bibinfo{year}{2016}).

\bibitem{stieglitz2017social}
\bibinfo{author}{Stieglitz, S.} \emph{et~al.}
\newblock \bibinfo{title}{Do social bots (still) act different to humans?--comparing metrics of social bots with those of humans}.
\newblock In \emph{\bibinfo{booktitle}{Social Computing and Social Media. Human Behavior: 9th International Conference, SCSM 2017, Held as Part of HCI International 2017, Vancouver, BC, Canada, July 9-14, 2017, Proceedings, Part I 9}}, \bibinfo{pages}{379--395} (\bibinfo{organization}{Springer}, \bibinfo{year}{2017}).

\bibitem{grimme2017social}
\bibinfo{author}{Grimme, C.}, \bibinfo{author}{Preuss, M.}, \bibinfo{author}{Adam, L.} \& \bibinfo{author}{Trautmann, H.}
\newblock \bibinfo{journal}{\bibinfo{title}{Social bots: Human-like by means of human control?}}
\newblock {\emph{\JournalTitle{Big data}}} \textbf{\bibinfo{volume}{5}}, \bibinfo{pages}{279--293} (\bibinfo{year}{2017}).

\bibitem{alsmadi2020many}
\bibinfo{author}{Alsmadi, I.} \& \bibinfo{author}{O'Brien, M.~J.}
\newblock \bibinfo{journal}{\bibinfo{title}{How many bots in russian troll tweets?}}
\newblock {\emph{\JournalTitle{Information Processing \& Management}}} \textbf{\bibinfo{volume}{57}}, \bibinfo{pages}{102303} (\bibinfo{year}{2020}).

\bibitem{ng2023deflating}
\bibinfo{author}{Ng, L. H.~X.} \& \bibinfo{author}{Carley, K.~M.}
\newblock \bibinfo{journal}{\bibinfo{title}{Deflating the chinese balloon: types of twitter bots in us-china balloon incident}}.
\newblock {\emph{\JournalTitle{EPJ Data Science}}} \textbf{\bibinfo{volume}{12}}, \bibinfo{pages}{63} (\bibinfo{year}{2023}).

\bibitem{shao2017spread}
\bibinfo{author}{Shao, C.}, \bibinfo{author}{Ciampaglia, G.~L.}, \bibinfo{author}{Varol, O.}, \bibinfo{author}{Flammini, A.} \& \bibinfo{author}{Menczer, F.}
\newblock \bibinfo{journal}{\bibinfo{title}{The spread of fake news by social bots}}.
\newblock {\emph{\JournalTitle{arXiv preprint arXiv:1707.07592}}} \textbf{\bibinfo{volume}{96}}, \bibinfo{pages}{14} (\bibinfo{year}{2017}).

\bibitem{vziatysheva2020fake}
\bibinfo{author}{Vziatysheva, V.}
\newblock \bibinfo{journal}{\bibinfo{title}{How fake news spreads online?}}
\newblock {\emph{\JournalTitle{International Journal of Media \& Information Literacy}}} \textbf{\bibinfo{volume}{5}} (\bibinfo{year}{2020}).

\bibitem{lokot2016news}
\bibinfo{author}{Lokot, T.} \& \bibinfo{author}{Diakopoulos, N.}
\newblock \bibinfo{journal}{\bibinfo{title}{News bots: Automating news and information dissemination on twitter}}.
\newblock {\emph{\JournalTitle{Digital journalism}}} \textbf{\bibinfo{volume}{4}}, \bibinfo{pages}{682--699} (\bibinfo{year}{2016}).

\bibitem{gorwa2020unpacking}
\bibinfo{author}{Gorwa, R.} \& \bibinfo{author}{Guilbeault, D.}
\newblock \bibinfo{journal}{\bibinfo{title}{Unpacking the social media bot: A typology to guide research and policy}}.
\newblock {\emph{\JournalTitle{Policy \& Internet}}} \textbf{\bibinfo{volume}{12}}, \bibinfo{pages}{225--248} (\bibinfo{year}{2020}).

\bibitem{arora2023developing}
\bibinfo{author}{Arora, A.}, \bibinfo{author}{Arora, A.} \& \bibinfo{author}{McIntyre, J.}
\newblock \bibinfo{journal}{\bibinfo{title}{Developing chatbots for cyber security: Assessing threats through sentiment analysis on social media}}.
\newblock {\emph{\JournalTitle{Sustainability}}} \textbf{\bibinfo{volume}{15}}, \bibinfo{pages}{13178} (\bibinfo{year}{2023}).

\bibitem{liu2016analyzing}
\bibinfo{author}{Liu, L.}, \bibinfo{author}{Preotiuc-Pietro, D.}, \bibinfo{author}{Samani, Z.~R.}, \bibinfo{author}{Moghaddam, M.~E.} \& \bibinfo{author}{Ungar, L.}
\newblock \bibinfo{title}{Analyzing personality through social media profile picture choice}.
\newblock In \emph{\bibinfo{booktitle}{Proceedings of the International AAAI Conference on Web and Social Media}}, vol.~\bibinfo{volume}{10}, \bibinfo{pages}{211--220} (\bibinfo{year}{2016}).

\bibitem{ng2023recruitment}
\bibinfo{author}{Ng, L. H.~X.} \& \bibinfo{author}{Cruickshank, I.~J.}
\newblock \bibinfo{journal}{\bibinfo{title}{Recruitment promotion via twitter: a network-centric approach of analyzing community engagement using social identity}}.
\newblock {\emph{\JournalTitle{Digital Government: Research and Practice}}} \textbf{\bibinfo{volume}{4}}, \bibinfo{pages}{1--17} (\bibinfo{year}{2023}).

\bibitem{mckelvey2017computational}
\bibinfo{author}{McKelvey, F.} \& \bibinfo{author}{Dubois, E.}
\newblock \bibinfo{title}{Computational propaganda in canada: The use of political bots} (\bibinfo{year}{2017}).

\bibitem{woolley2018computational}
\bibinfo{author}{Woolley, S.~C.} \& \bibinfo{author}{Howard, P.~N.}
\newblock \emph{\bibinfo{title}{Computational propaganda: Political parties, politicians, and political manipulation on social media}} (\bibinfo{publisher}{Oxford University Press}, \bibinfo{year}{2018}).

\bibitem{zouzou2024unsupervised}
\bibinfo{author}{Zouzou, Y.} \& \bibinfo{author}{Varol, O.}
\newblock \bibinfo{journal}{\bibinfo{title}{Unsupervised detection of coordinated fake-follower campaigns on social media}}.
\newblock {\emph{\JournalTitle{EPJ Data Science}}} \textbf{\bibinfo{volume}{13}}, \bibinfo{pages}{62} (\bibinfo{year}{2024}).

\bibitem{albadi2019hateful}
\bibinfo{author}{Albadi, N.}, \bibinfo{author}{Kurdi, M.} \& \bibinfo{author}{Mishra, S.}
\newblock \bibinfo{journal}{\bibinfo{title}{Hateful people or hateful bots? detection and characterization of bots spreading religious hatred in arabic social media}}.
\newblock {\emph{\JournalTitle{Proceedings of the ACM on Human-Computer Interaction}}} \textbf{\bibinfo{volume}{3}}, \bibinfo{pages}{1--25} (\bibinfo{year}{2019}).

\bibitem{blane2023social}
\bibinfo{author}{Blane, J.~T.}
\newblock \emph{\bibinfo{title}{Social-Cyber Maneuvers for Analyzing Online Influence Operations}}.
\newblock Ph.D. thesis, \bibinfo{school}{United States Military Academy} (\bibinfo{year}{2023}).

\bibitem{badawy2018analyzing}
\bibinfo{author}{Badawy, A.}, \bibinfo{author}{Ferrara, E.} \& \bibinfo{author}{Lerman, K.}
\newblock \bibinfo{title}{Analyzing the digital traces of political manipulation: The 2016 russian interference twitter campaign}.
\newblock In \emph{\bibinfo{booktitle}{2018 IEEE/ACM international conference on advances in social networks analysis and mining (ASONAM)}}, \bibinfo{pages}{258--265} (\bibinfo{organization}{IEEE}, \bibinfo{year}{2018}).

\bibitem{schwartz20192016}
\bibinfo{author}{Schwartz, O.}
\newblock \bibinfo{journal}{\bibinfo{title}{In 2016, microsoft’s racist chatbot revealed the dangers of online conversation}}.
\newblock {\emph{\JournalTitle{IEEE spectrum}}} \textbf{\bibinfo{volume}{11}}, \bibinfo{pages}{2019} (\bibinfo{year}{2019}).

\bibitem{brachten2018threat}
\bibinfo{author}{Brachten, F.} \emph{et~al.}
\newblock \bibinfo{journal}{\bibinfo{title}{Threat or opportunity?-examining social bots in social media crisis communication}}.
\newblock {\emph{\JournalTitle{arXiv preprint arXiv:1810.09159}}}  (\bibinfo{year}{2018}).

\bibitem{martini2021bot}
\bibinfo{author}{Martini, F.}, \bibinfo{author}{Samula, P.}, \bibinfo{author}{Keller, T.~R.} \& \bibinfo{author}{Klinger, U.}
\newblock \bibinfo{journal}{\bibinfo{title}{Bot, or not? comparing three methods for detecting social bots in five political discourses}}.
\newblock {\emph{\JournalTitle{Big data \& society}}} \textbf{\bibinfo{volume}{8}}, \bibinfo{pages}{20539517211033566} (\bibinfo{year}{2021}).

\bibitem{ng2022stabilizing}
\bibinfo{author}{Ng, L. H.~X.}, \bibinfo{author}{Robertson, D.~C.} \& \bibinfo{author}{Carley, K.~M.}
\newblock \bibinfo{journal}{\bibinfo{title}{Stabilizing a supervised bot detection algorithm: How much data is needed for consistent predictions?}}
\newblock {\emph{\JournalTitle{Online Social Networks and Media}}} \textbf{\bibinfo{volume}{28}}, \bibinfo{pages}{100198} (\bibinfo{year}{2022}).

\bibitem{tan-etal-2023-botpercent}
\bibinfo{author}{Tan, Z.} \emph{et~al.}
\newblock \bibinfo{title}{{B}ot{P}ercent: Estimating bot populations in {T}witter communities}.
\newblock In \bibinfo{editor}{Bouamor, H.}, \bibinfo{editor}{Pino, J.} \& \bibinfo{editor}{Bali, K.} (eds.) \emph{\bibinfo{booktitle}{Findings of the Association for Computational Linguistics: EMNLP 2023}}, \bibinfo{pages}{14295--14312}, \doiprefix\url{10.18653/v1/2023.findings-emnlp.954} (\bibinfo{publisher}{Association for Computational Linguistics}, \bibinfo{address}{Singapore}, \bibinfo{year}{2023}).

\bibitem{yan2023exposure}
\bibinfo{author}{Yan, H.~Y.}, \bibinfo{author}{Yang, K.-C.}, \bibinfo{author}{Shanahan, J.} \& \bibinfo{author}{Menczer, F.}
\newblock \bibinfo{journal}{\bibinfo{title}{Exposure to social bots amplifies perceptual biases and regulation propensity}}.
\newblock {\emph{\JournalTitle{Scientific Reports}}} \textbf{\bibinfo{volume}{13}}, \bibinfo{pages}{20707} (\bibinfo{year}{2023}).

\bibitem{yan2023landscape}
\bibinfo{author}{Yan, H.~Y.} \& \bibinfo{author}{Yang, K.-C.}
\newblock \bibinfo{title}{The landscape of social bot research: A critical appraisal}.
\newblock In \emph{\bibinfo{booktitle}{Handbook of Critical Studies of Artificial Intelligence}}, \bibinfo{pages}{716--725} (\bibinfo{publisher}{Edward Elgar Publishing}, \bibinfo{year}{2023}).

\bibitem{yang2024anatomy}
\bibinfo{author}{Yang, K.-C.} \& \bibinfo{author}{Menczer, F.}
\newblock \bibinfo{journal}{\bibinfo{title}{Anatomy of an ai-powered malicious social botnet}}.
\newblock {\emph{\JournalTitle{Journal of Quantitative Description: Digital Media}}} \textbf{\bibinfo{volume}{4}} (\bibinfo{year}{2024}).

\bibitem{microsoftSpotBots}
\bibinfo{author}{Microsoft}.
\newblock \bibinfo{title}{{H}ow to spot bots on social media – {M}icrosoft 365 --- microsoft.com}.
\newblock \bibinfo{howpublished}{\url{https://www.microsoft.com/en-us/microsoft-365-life-hacks/privacy-and-safety/how-to-spot-bots-on-social-media}} (\bibinfo{year}{2024}).
\newblock \bibinfo{note}{[Accessed 29-10-2024]}.

\bibitem{meltwaterSocialMedia}
\bibinfo{author}{Simpson, M.}
\newblock \bibinfo{title}{{S}ocial {M}edia {B}ots 101 - {A}ll {Y}ou {N}eed to {K}now --- meltwater.com}.
\newblock \bibinfo{howpublished}{\url{https://www.meltwater.com/en/blog/social-media-bots}} (\bibinfo{year}{2024}).
\newblock \bibinfo{note}{[Accessed 29-10-2024]}.

\bibitem{impervaWhatBots}
\bibinfo{author}{Imperva}.
\newblock \bibinfo{title}{{W}hat are {B}ots | {B}ot {T}ypes \& {M}itigation {T}echniques | {I}mperva --- imperva.com}.
\newblock \bibinfo{howpublished}{\url{https://www.imperva.com/learn/application-security/what-are-bots/}}.
\newblock \bibinfo{note}{[Accessed 29-10-2024]}.

\bibitem{babcock2018beaten}
\bibinfo{author}{Babcock, M.}, \bibinfo{author}{Beskow, D.~M.} \& \bibinfo{author}{Carley, K.~M.}
\newblock \bibinfo{title}{Beaten up on twitter? exploring fake news and satirical responses during the black panther movie event}.
\newblock In \emph{\bibinfo{booktitle}{Social, Cultural, and Behavioral Modeling: 11th International Conference, SBP-BRiMS 2018, Washington, DC, USA, July 10-13, 2018, Proceedings 11}}, \bibinfo{pages}{97--103} (\bibinfo{organization}{Springer}, \bibinfo{year}{2018}).

\bibitem{king2020lying}
\bibinfo{author}{King, C.}, \bibinfo{author}{Bellutta, D.} \& \bibinfo{author}{Carley, K.~M.}
\newblock \bibinfo{title}{Lying about lying on social media: a case study of the 2019 canadian elections}.
\newblock In \emph{\bibinfo{booktitle}{International Conference on Social Computing, Behavioral-Cultural Modeling and Prediction and Behavior Representation in Modeling and Simulation}}, \bibinfo{pages}{75--85} (\bibinfo{organization}{Springer}, \bibinfo{year}{2020}).

\bibitem{babcock2020pretending}
\bibinfo{author}{Babcock, M.}, \bibinfo{author}{Villa-Cox, R.} \& \bibinfo{author}{Carley, K.~M.}
\newblock \bibinfo{journal}{\bibinfo{title}{Pretending positive, pushing false: Comparing captain marvel misinformation campaigns}}.
\newblock {\emph{\JournalTitle{Disinformation, Misinformation, and Fake News in Social Media: Emerging Research Challenges and Opportunities}}} \bibinfo{pages}{83--94} (\bibinfo{year}{2020}).

\bibitem{ng2023combined}
\bibinfo{author}{Ng, L. H.~X.} \& \bibinfo{author}{Carley, K.~M.}
\newblock \bibinfo{journal}{\bibinfo{title}{A combined synchronization index for evaluating collective action social media}}.
\newblock {\emph{\JournalTitle{Applied network science}}} \textbf{\bibinfo{volume}{8}}, \bibinfo{pages}{1} (\bibinfo{year}{2023}).

\bibitem{carley2018ora}
\bibinfo{author}{Carley, L.~R.}, \bibinfo{author}{Reminga, J.} \& \bibinfo{author}{Carley, K.~M.}
\newblock \bibinfo{title}{Ora \& netmapper}.
\newblock In \emph{\bibinfo{booktitle}{International conference on social computing, behavioral-cultural modeling and prediction and behavior representation in modeling and simulation. Springer}}, vol.~\bibinfo{volume}{1}, \bibinfo{pages}{2--1} (\bibinfo{year}{2018}).

\bibitem{stella2018bots}
\bibinfo{author}{Stella, M.}, \bibinfo{author}{Ferrara, E.} \& \bibinfo{author}{De~Domenico, M.}
\newblock \bibinfo{journal}{\bibinfo{title}{Bots increase exposure to negative and inflammatory content in online social systems}}.
\newblock {\emph{\JournalTitle{Proceedings of the National Academy of Sciences}}} \textbf{\bibinfo{volume}{115}}, \bibinfo{pages}{12435--12440} (\bibinfo{year}{2018}).

\bibitem{aldayel2022characterizing}
\bibinfo{author}{Aldayel, A.} \& \bibinfo{author}{Magdy, W.}
\newblock \bibinfo{journal}{\bibinfo{title}{Characterizing the role of bots’ in polarized stance on social media}}.
\newblock {\emph{\JournalTitle{Social Network Analysis and Mining}}} \textbf{\bibinfo{volume}{12}}, \bibinfo{pages}{30} (\bibinfo{year}{2022}).

\bibitem{ferrara2017disinformation}
\bibinfo{author}{Ferrara, E.}
\newblock \bibinfo{journal}{\bibinfo{title}{Disinformation and social bot operations in the run up to the 2017 french presidential election}}.
\newblock {\emph{\JournalTitle{arXiv preprint arXiv:1707.00086}}}  (\bibinfo{year}{2017}).

\bibitem{pathak2021method}
\bibinfo{author}{Pathak, A.}, \bibinfo{author}{Madani, N.} \& \bibinfo{author}{Joseph, K.}
\newblock \bibinfo{journal}{\bibinfo{title}{A method to analyze multiple social identities in twitter bios}}.
\newblock {\emph{\JournalTitle{Proceedings of the ACM on Human-Computer Interaction}}} \textbf{\bibinfo{volume}{5}}, \bibinfo{pages}{1--35} (\bibinfo{year}{2021}).

\bibitem{van2018cyber}
\bibinfo{author}{Van~der Walt, E.}, \bibinfo{author}{Eloff, J.~H.} \& \bibinfo{author}{Grobler, J.}
\newblock \bibinfo{journal}{\bibinfo{title}{Cyber-security: Identity deception detection on social media platforms}}.
\newblock {\emph{\JournalTitle{Computers \& Security}}} \textbf{\bibinfo{volume}{78}}, \bibinfo{pages}{76--89} (\bibinfo{year}{2018}).

\bibitem{radivojevic2024llms}
\bibinfo{author}{Radivojevic, K.}, \bibinfo{author}{Clark, N.} \& \bibinfo{author}{Brenner, P.}
\newblock \bibinfo{title}{Llms among us: Generative ai participating in digital discourse}.
\newblock In \emph{\bibinfo{booktitle}{Proceedings of the AAAI Symposium Series}}, vol.~\bibinfo{volume}{3}, \bibinfo{pages}{209--218} (\bibinfo{year}{2024}).

\bibitem{howard2020lie}
\bibinfo{author}{Howard, P.~N.}
\newblock \emph{\bibinfo{title}{Lie machines: How to save democracy from troll armies, deceitful robots, junk news operations, and political operatives}} (\bibinfo{publisher}{Yale University Press}, \bibinfo{year}{2020}).

\bibitem{tardelli2020characterizing}
\bibinfo{author}{Tardelli, S.}, \bibinfo{author}{Avvenuti, M.}, \bibinfo{author}{Tesconi, M.} \& \bibinfo{author}{Cresci, S.}
\newblock \bibinfo{title}{Characterizing social bots spreading financial disinformation}.
\newblock In \emph{\bibinfo{booktitle}{International conference on human-computer interaction}}, \bibinfo{pages}{376--392} (\bibinfo{organization}{Springer}, \bibinfo{year}{2020}).

\bibitem{deverna2023artificial}
\bibinfo{author}{DeVerna, M.~R.}, \bibinfo{author}{Yan, H.~Y.}, \bibinfo{author}{Yang, K.-C.} \& \bibinfo{author}{Menczer, F.}
\newblock \bibinfo{journal}{\bibinfo{title}{Artificial intelligence is ineffective and potentially harmful for fact checking}}.
\newblock {\emph{\JournalTitle{arXiv preprint arXiv:2308.10800}}}  (\bibinfo{year}{2023}).

\bibitem{hill2015real}
\bibinfo{author}{Hill, J.}, \bibinfo{author}{Ford, W.~R.} \& \bibinfo{author}{Farreras, I.~G.}
\newblock \bibinfo{journal}{\bibinfo{title}{Real conversations with artificial intelligence: A comparison between human--human online conversations and human--chatbot conversations}}.
\newblock {\emph{\JournalTitle{Computers in human behavior}}} \textbf{\bibinfo{volume}{49}}, \bibinfo{pages}{245--250} (\bibinfo{year}{2015}).

\bibitem{bisgin2012study}
\bibinfo{author}{Bisgin, H.}, \bibinfo{author}{Agarwal, N.} \& \bibinfo{author}{Xu, X.}
\newblock \bibinfo{journal}{\bibinfo{title}{A study of homophily on social media}}.
\newblock {\emph{\JournalTitle{World Wide Web}}} \textbf{\bibinfo{volume}{15}}, \bibinfo{pages}{213--232} (\bibinfo{year}{2012}).

\bibitem{ng2022online}
\bibinfo{author}{Ng, L. H.~X.} \& \bibinfo{author}{Carley, K.~M.}
\newblock \bibinfo{title}{Online coordination: methods and comparative case studies of coordinated groups across four events in the united states}.
\newblock In \emph{\bibinfo{booktitle}{Proceedings of the 14th ACM Web Science Conference 2022}}, \bibinfo{pages}{12--21} (\bibinfo{year}{2022}).

\bibitem{mazza2019rtbust}
\bibinfo{author}{Mazza, M.}, \bibinfo{author}{Cresci, S.}, \bibinfo{author}{Avvenuti, M.}, \bibinfo{author}{Quattrociocchi, W.} \& \bibinfo{author}{Tesconi, M.}
\newblock \bibinfo{title}{Rtbust: Exploiting temporal patterns for botnet detection on twitter}.
\newblock In \emph{\bibinfo{booktitle}{Proceedings of the 10th ACM conference on web science}}, \bibinfo{pages}{183--192} (\bibinfo{year}{2019}).

\bibitem{russell2016artificial}
\bibinfo{author}{Russell, S.~J.} \& \bibinfo{author}{Norvig, P.}
\newblock \emph{\bibinfo{title}{Artificial intelligence: a modern approach}} (\bibinfo{publisher}{Pearson}, \bibinfo{year}{2016}).

\bibitem{lu2024agents}
\bibinfo{author}{Lu, H.-C.} \& \bibinfo{author}{Lee, H.-w.}
\newblock \bibinfo{journal}{\bibinfo{title}{Agents of discord: Modeling the impact of political bots on opinion polarization in social networks}}.
\newblock {\emph{\JournalTitle{Social Science Computer Review}}} \bibinfo{pages}{08944393241270382} (\bibinfo{year}{2024}).

\bibitem{chen2021social}
\bibinfo{author}{Chen, Y.}
\newblock \bibinfo{title}{The social influence of bots and trolls in social media}.
\newblock In \emph{\bibinfo{booktitle}{Handbook of Computational Social Science, Volume 1}}, \bibinfo{pages}{287--303} (\bibinfo{publisher}{Routledge}, \bibinfo{year}{2021}).

\bibitem{ng2024exploratory}
\bibinfo{author}{Ng, L. H.~X.}, \bibinfo{author}{Kloo, I.}, \bibinfo{author}{Clark, S.} \& \bibinfo{author}{Carley, K.~M.}
\newblock \bibinfo{journal}{\bibinfo{title}{An exploratory analysis of covid bot vs human disinformation dissemination stemming from the disinformation dozen on telegram}}.
\newblock {\emph{\JournalTitle{Journal of Computational Social Science}}} \bibinfo{pages}{1--26} (\bibinfo{year}{2024}).

\bibitem{Jacobs2024Detection}
\bibinfo{author}{Jacobs, C.~S.}, \bibinfo{author}{Ng, L. H.~X.} \& \bibinfo{author}{Carley, K.~M.}
\newblock \bibinfo{title}{Detection evasion techniques of state-sponsored accounts}.
\newblock In \emph{\bibinfo{booktitle}{Proceedings of the ICWSM Workshops}}, \doiprefix\url{10.36190/2024.63} (\bibinfo{address}{Buffalo, New York, USA}, \bibinfo{year}{2024}).

\bibitem{tyagi2020polarizing}
\bibinfo{author}{Tyagi, A.}, \bibinfo{author}{Babcock, M.}, \bibinfo{author}{Carley, K.~M.} \& \bibinfo{author}{Sicker, D.~C.}
\newblock \bibinfo{title}{Polarizing tweets on climate change}.
\newblock In \emph{\bibinfo{booktitle}{International Conference on Social Computing, Behavioral-Cultural Modeling and Prediction and Behavior Representation in Modeling and Simulation}}, \bibinfo{pages}{107--117} (\bibinfo{organization}{Springer}, \bibinfo{year}{2020}).

\bibitem{yang2022botometer}
\bibinfo{author}{Yang, K.-C.}, \bibinfo{author}{Ferrara, E.} \& \bibinfo{author}{Menczer, F.}
\newblock \bibinfo{journal}{\bibinfo{title}{Botometer 101: Social bot practicum for computational social scientists}}.
\newblock {\emph{\JournalTitle{Journal of computational social science}}} \textbf{\bibinfo{volume}{5}}, \bibinfo{pages}{1511--1528} (\bibinfo{year}{2022}).

\bibitem{tardelli2022detecting}
\bibinfo{author}{Tardelli, S.}, \bibinfo{author}{Avvenuti, M.}, \bibinfo{author}{Tesconi, M.} \& \bibinfo{author}{Cresci, S.}
\newblock \bibinfo{journal}{\bibinfo{title}{Detecting inorganic financial campaigns on twitter}}.
\newblock {\emph{\JournalTitle{Information Systems}}} \textbf{\bibinfo{volume}{103}}, \bibinfo{pages}{101769} (\bibinfo{year}{2022}).

\bibitem{marigliano2024analyzing}
\bibinfo{author}{Marigliano, R.}, \bibinfo{author}{Ng, L. H.~X.} \& \bibinfo{author}{Carley, K.~M.}
\newblock \bibinfo{journal}{\bibinfo{title}{Analyzing digital propaganda and conflict rhetoric: a study on russia’s bot-driven campaigns and counter-narratives during the ukraine crisis}}.
\newblock {\emph{\JournalTitle{Social Network Analysis and Mining}}} \textbf{\bibinfo{volume}{14}}, \bibinfo{pages}{170} (\bibinfo{year}{2024}).

\bibitem{ng2025building}
\bibinfo{author}{Ng, L. H.~X.}, \bibinfo{author}{Cruickshank, I.~J.} \& \bibinfo{author}{Farr, D.}
\newblock \bibinfo{journal}{\bibinfo{title}{Building bridges between users and content across multiple platforms during natural disasters}}.
\newblock {\emph{\JournalTitle{arXiv preprint arXiv:2502.02681}}}  (\bibinfo{year}{2025}).

\bibitem{schroeder2014big}
\bibinfo{author}{Schroeder, R.}
\newblock \bibinfo{journal}{\bibinfo{title}{Big data and the brave new world of social media research}}.
\newblock {\emph{\JournalTitle{Big Data \& Society}}} \textbf{\bibinfo{volume}{1}}, \bibinfo{pages}{2053951714563194} (\bibinfo{year}{2014}).

\bibitem{gillespie2018custodians}
\bibinfo{author}{Gillespie, T.}
\newblock \emph{\bibinfo{title}{Custodians of the Internet: Platforms, content moderation, and the hidden decisions that shape social media}} (\bibinfo{publisher}{Yale University Press}, \bibinfo{year}{2018}).

\bibitem{cresci2020detecting}
\bibinfo{author}{Cresci, S.}
\newblock \bibinfo{title}{Detecting malicious social bots: story of a never-ending clash}.
\newblock In \emph{\bibinfo{booktitle}{Multidisciplinary International Symposium on Disinformation in Open Online Media}}, \bibinfo{pages}{77--88} (\bibinfo{organization}{Springer}, \bibinfo{year}{2020}).

\bibitem{wiredMillionsPeople}
\bibinfo{author}{Burgess, M.}
\newblock \bibinfo{title}{{M}illions of {P}eople {A}re {U}sing {A}busive {A}{I} ‘{N}udify’ {B}ots on {T}elegram --- wired.com}.
\newblock \bibinfo{howpublished}{\url{https://www.wired.com/story/ai-deepfake-nudify-bots-telegram/}} (\bibinfo{year}{2024}).
\newblock \bibinfo{note}{[Accessed 21-10-2024]}.

\bibitem{ferrara2017contagion}
\bibinfo{author}{Ferrara, E.}
\newblock \bibinfo{journal}{\bibinfo{title}{Contagion dynamics of extremist propaganda in social networks}}.
\newblock {\emph{\JournalTitle{Information Sciences}}} \textbf{\bibinfo{volume}{418}}, \bibinfo{pages}{1--12} (\bibinfo{year}{2017}).

\bibitem{jones2018silencing}
\bibinfo{author}{Jones, M.~L.}
\newblock \bibinfo{journal}{\bibinfo{title}{Silencing bad bots: Global, legal and political questions for mean machine communication}}.
\newblock {\emph{\JournalTitle{Communication Law and Policy}}} \textbf{\bibinfo{volume}{23}}, \bibinfo{pages}{159--195} (\bibinfo{year}{2018}).

\bibitem{smith2016mean}
\bibinfo{author}{Smith-Lovin, L.} \emph{et~al.}
\newblock \bibinfo{journal}{\bibinfo{title}{Mean affective ratings of 929 identities, 814 behaviors, and 660 modifiers by university of georgia and duke university undergraduates and by community members in durham, nc, in 2012-2014}}.
\newblock {\emph{\JournalTitle{University of Georgia: Distributed at UGA Affect Control Theory Website: http://research. franklin. uga. edu/act}}}  (\bibinfo{year}{2016}).

\bibitem{theatlanticInternetMostly}
\bibinfo{author}{LaFrance, A.}
\newblock \bibinfo{title}{{T}he {I}nternet {I}s {M}ostly {B}ots --- theatlantic.com}.
\newblock \bibinfo{howpublished}{\url{https://www.theatlantic.com/technology/archive/2017/01/bots-bots-bots/515043/}} (\bibinfo{year}{2017}).
\newblock \bibinfo{note}{[Accessed 27-09-2024]}.

\end{thebibliography}

\section*{Acknowledgements}
The first author thanks Christine Lepird for her ideas and comments towards this work.
This material is based upon work supported by the Scalable Technologies for Social Cybersecurity, U.S. Army (W911NF20D0002), the Minerva-Multi-Level Models of Covert Online Information Campaigns, Office of Naval Research (N000142112765), the Threat Assessment Techniques for Digital Data, Office of Naval Research (N000142412414), and the MURI: Persuasion, Identity \& Morality in Social-Cyber Environments (N000142112749), Office of Naval Research. The views and conclusions contained in this document are those of the authors and should not be interpreted as representing official policies, either expressed or implied by the Office of Naval Research, U.S. Army or the U.S. government.

\section*{Author contributions statement}
L.H.X.N. conceived the experiments, collected the data and performed the analysis. K.M.C. reviewed the analysis.  All authors reviewed the manuscript. 

\section*{Competing Interests} The authors declare that there is no competing interests. 

\section*{Data Availability Statement} 
Please contact the authors to request the data and code from this study, in accordance to the data sharing policies of the social media platforms.

\newpage

\section*{Supplementary Information}
\subsection*{Data Collection Parameters}
\autoref{tab:data} presents the parameters of data collection for our study.

\begin{table}[H]
    \centering
    \begin{tabular}{p{3.5cm}p{4cm}p{6cm}p{2cm}p{2cm}}
        \toprule
        \textbf{Event Name} & \textbf{Collection Date} & \textbf{Collection hashtags} & \textbf{\# Users} & \textbf{\# Tweets} \\ \midrule
        Asian Elections \cite{uyheng2020bot,uyheng2021active} & 2019, 2020, 2021 & \#Pemilu2019, \#Pilpres\_2019, joko widodo, prabowo subianto, \#TaiwanVotes, \#Taiwan2020, \#taiwanelection, \#sgelections2020, \#GE2020, \#SGGE2020, \#singaporeGE2020, \#GE2020SG, \#Singaporevotes, PAPSingapore, wpsg, jamuslim, ProgressSingaporeParty, Nicoleseah, PeopleActionParty, Workerparty, neesoongrc, lightningparty, eastcoastgrc & 950,789 & 4,126,008 \\ 
        
        Black Panther  \cite{babcock2018beaten} & 8 Feb 2018 to 23 Feb 2018 & \#BlackPanther & 1,689,076 & 17,718,557 \\
        
        Canadian Elections 2019 \cite{king2020lying} & 20 Jul 2019 to 6 Nov 2019 & \#TrudeauMustGo, \#TeamTrudeau, \#trudeau, \#Election2019, \#elxn43, \#chooseforward, \#onpoli, \#ItsOurVote, \#lpc, \#ndp, \#cpc, \#gpc, \#NotAbot, \#cdnpoli, \#ButtsMustGo, \#LavScam, \#LiberalsMustGo, BlocQuebecois, \#blocqc, \#cccr2019, \#NoTMX, \#TMX, \#TransMountain, \#scheer, \#dougford, \#fordcutshurt, \#fordisfailing & 1,946,277 & 18,213,477 \\
        
        Captain Marvel \cite{babcock2020pretending} & 15 Feb 2019 to 15 Mar 2019 & \#BoycottCaptainMarvel, \#AlitaChallenge & & \\
        
        Coronavirus 2020-2021 \cite{ng2023combined} & 2020-2021 & \#coronavirus, \#coronaravirus, \#wuhanvirus, \#2019nCoV, \#NCoV, \#NCoV2019, \#covid-19, \#covid19 & 208,956,241 & 4,179,124,820\\ 
        
        ReOpen America \cite{magelinski2021synchronized} &  1 Apr 2020 to 22 Jun 2020 & \#openup, \#reopen, \#operationgridlock, \#liberate; \#reopenNY + state abbreviatons & 201,083 & 4,429,298 \\ 
        
        US Elections 2020 \cite{ng2023combined,ng2024assembling} & 1 Dec 2019 to 16 Aug 2020, 16 Aug 2020 to 15 Jan 2021, 15 Jan 2021 to 17 Feb 2021 & \#election2020, \#2020\_presidential\_election, \#maga2020, \#flipitblue, \#keepitblue, \#yeswecan, \#yang2020, \#JoeBiden, \#BernieSanders, \#ElizabethWarren, \#PeteButtigieg, \#FeelTheBern, \#democrats, \#republicans, \#Bloomberg2020, \#Booker, \#USPS, \#VoteByMail, \#SaveTheUSPS, \#voterfraud, \#BlackLivesMatter, \#BLM, \#reopen, \#reopenamerica, \#IranSanctions, \#QAnon, \#WWG1WGA, "natural born", \#election2020, \#presidentialelection, \#democrats, \#republicans, \#JoeBiden, \#BidenHarris2020, \#Biden, \#MAGA, \#KAG, \#Inauguration, \#InaugurationDay, \#Capitol, \#USCapitol, \#USCapital, \#NationalMall, \#Jan20, \#election2020, \#presidentialelection, \#democrats, \#republicans, \#JoeBiden, \#Biden, \#MAGA, \#KAG, \#Trump, \#USPS, \#VoteByMail, \#SaveTheUSPS, \#voterfraud, \#BlackLivesMatter, \#BLM, \#reopen, \#reopenamerica, \#IranSanctions, \#QAnon, \#WWG1WGA  & 1,608,033 & 55,179,293 \\
         \bottomrule
    \end{tabular}
    \caption{Dataset collection parameters. Data was collected using the Twitter Developer V1 API}
    \label{tab:data}
\end{table}

\newpage
\subsection*{Comparison of Linguistic Cues}
\autoref{tab:linguistic_cues_numeric} presents the detailed comparison of linguistic cues between bots and humans across all events.

\begin{table}[H]
    \centering
    \begin{tabular}{p{6cm}ccc}
        \toprule
        \textbf{Cue} & \textbf{Bots} & \textbf{Humans} & \textbf{p-value} \\ \midrule
        \multicolumn{4}{l}{\textbf{Semantic Cues}} \\ \midrule
        First Person Pronouns & 0.71 & \textbf{0.73} & 5.62E-8*** \\
        Second Person Pronouns & \textbf{0.20} & 0.18  & 0.001*** \\
        Third Person Pronouns & 0.47 & \textbf{0.50} & 0.001*** \\
        Reading Difficulty & \textbf{0.12} & 0.10 & 0.001*** \\ \midrule
        \multicolumn{4}{l}{\textbf{Emotion Cues}} \\ \midrule
        Abusive Terms & \textbf{0.13} & 0.09  & 0.001***\\
        Expletives & \textbf{0.12} & 0.08 & 0.001***\\
        Negative Sentiment & 1.56 & \textbf{1.59} & 1.17E-11*** \\
        Positive Sentiment & 2.88 & \textbf{3.10} & 0.001*** \\ \midrule
        \multicolumn{4}{l}{\textbf{Metadata Cues}} \\ \midrule
        Mentions & \textbf{1.18} & 1.10 & 0.001***\\
        Media & 0.006 & \textbf{0.014} & 2.54E-59*** \\
        URLs & 0.18 & \textbf{0.20} & 1.41E-29*** \\
        Hashtags & \textbf{0.54} & 0.49 & 0.001***\\
        Retweets & \textbf{9372} & 8203 & 0.001*** \\
        Favorites & 21.0 & \textbf{92.5} & 2.12E-51*** \\
        Replies & 0.67 & \textbf{3.31}  & 1.03E-6*** \\
        Quotes & 0.51 & \textbf{2.05}  & 1.21E-16*** \\
        Followers & 1268 & \textbf{4164} & 0.001*** \\
        Friends & \textbf{1158} & 817  & 0.001***\\
        Total Tweets & \textbf{56779} & 10695 & 0.001*** \\
        Tweets per Hour & \textbf{1.20} & 1.10  & 0.001***\\
        Max Time between Tweets (seconds) & 44617 & \textbf{57838} & 0.001***\\
        Friends:Followers Ratio & \textbf{4.73} & 4.44  & 6.71E-288***\\
        \bottomrule
    \end{tabular}
    \caption{Comparison of Mean of Psycholinguistic Cues per user for Bots and Humans across all events. This value is the number of words that match each cue per tweet, divided across the number of tweets the user has. *** indicates a significant difference between the bots mean and humans mean via a student t-test at the $p<0.01$ level.}
    \label{tab:linguistic_cues_numeric}
\end{table}

\newpage
\subsection*{Identity Use Per Event}
\begin{figure}
\centering
\includegraphics[width=0.8\textwidth]{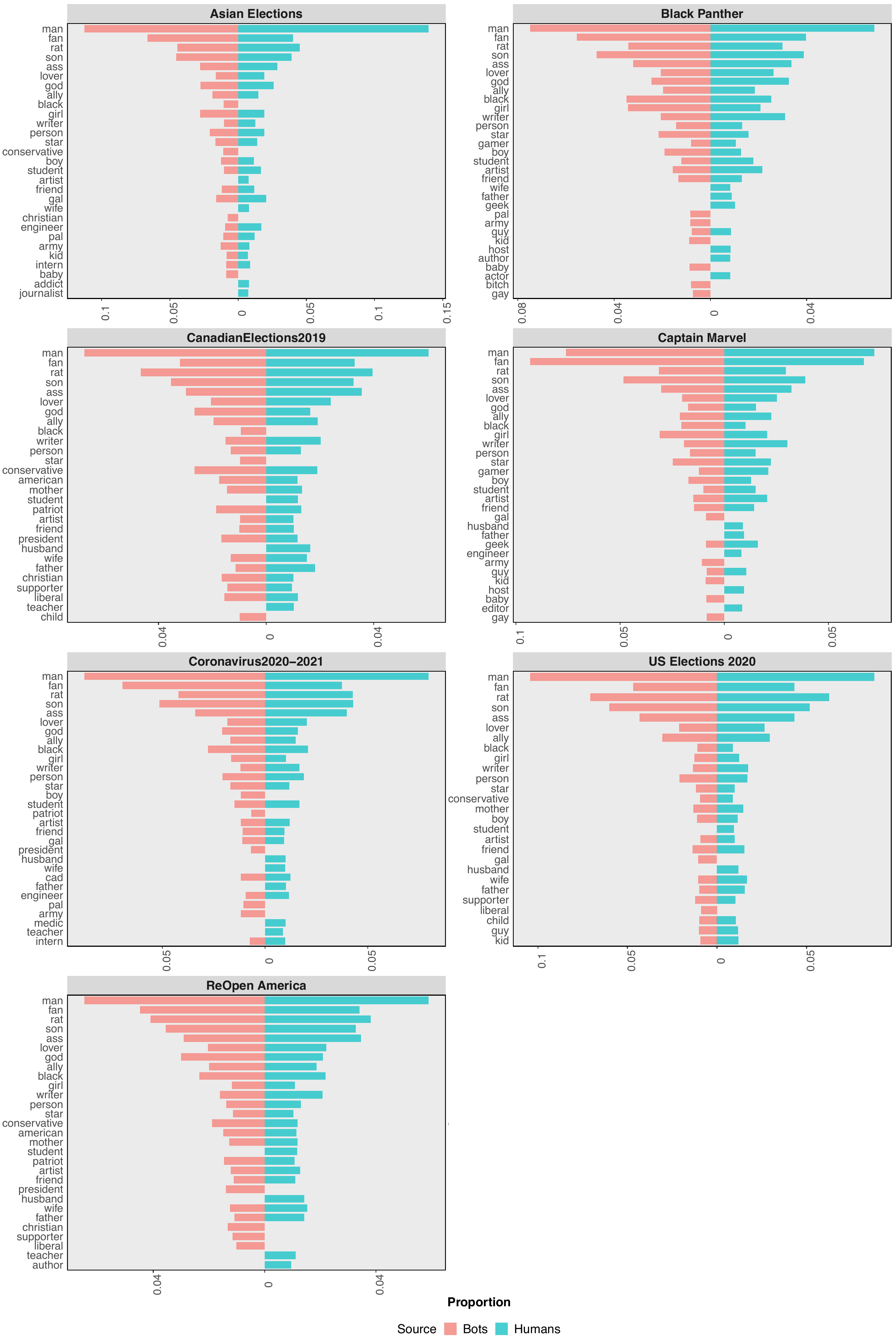}
\caption{Comparison of frequency of use of the top Identity affiliations by bots and humans per event.}
\label{fig:idenitty_plot_perevent}
\end{figure}

\autoref{tab:perc_identity} presents the percentage of users with identities that match the US census of occupations \cite{smith2016mean}. \autoref{fig:idenitty_plot_perevent} shows the identities used per event, separated by the top identities used by both bots and humans.

\begin{table}[H]
    \centering
    \begin{tabular}{p{4cm}cc}
    \toprule
        \textbf{Event} & \textbf{Bots (\%)}  & \textbf{Humans (\%)} \\ \midrule
        Asian Elections & 13.2 & 17.5 \\
        Black Panther & 22.4 & 29.6 \\
        Canadian Elections 2019 & 29.4 & 38.9 \\
        Captain Marvel & 21.5 & 29.0  \\
        Coronavirus2020-2021 & 15.8 & 26.6 \\
        ReOpen America & 26.7 & 34.7 \\
        US Elections 2020 & 20.6 & 12.7 \\
        \midrule
        Overall & 21.4$\pm$5.7 & 27.0$\pm$9.2 \\ 
    \bottomrule
    \end{tabular}
    \caption{Percentage of Users with identity affiliations matching census}
    \label{tab:perc_identity}
\end{table}

\subsection*{Unique Identities per Event}
\autoref{tab:unique_identity} presents the number of unique identities that bots and humans use per event. Across all datasets, bots consistently use lesser number of unique identities than humans, indicating a smaller range of affiliation set.

\begin{table}[H]
    \centering
    \begin{tabular}{p{4cm}cc}
    \toprule
        \textbf{Event} & \textbf{Bots}  & \textbf{Humans} \\ \midrule
        Asian Elections & 613 & 741 \\
        Black Panther & 733 & 831 \\
        Canadian Elections 2019 & 776 & 860 \\
        Captain Marvel & 747 & 838  \\
        Coronavirus2020-2021 & 836 & 899 \\
        ReOpen America & 807 & 884 \\
        US Elections 2020 & 828 & 894 \\
        \midrule
        Entire Dataset & 869 & 908 \\ 
        Average & 762.8$\pm$76.7 & 849.6$\pm$54.7 \\
    \bottomrule
    \end{tabular}
    \caption{Comparison of the Number of Unique Identities Per Event}
    \label{tab:unique_identity}
\end{table}

\newpage
\subsection*{Identity vs Framing Charts Per Event}
\begin{figure}[H]
\centering
\includegraphics[width=\textwidth]{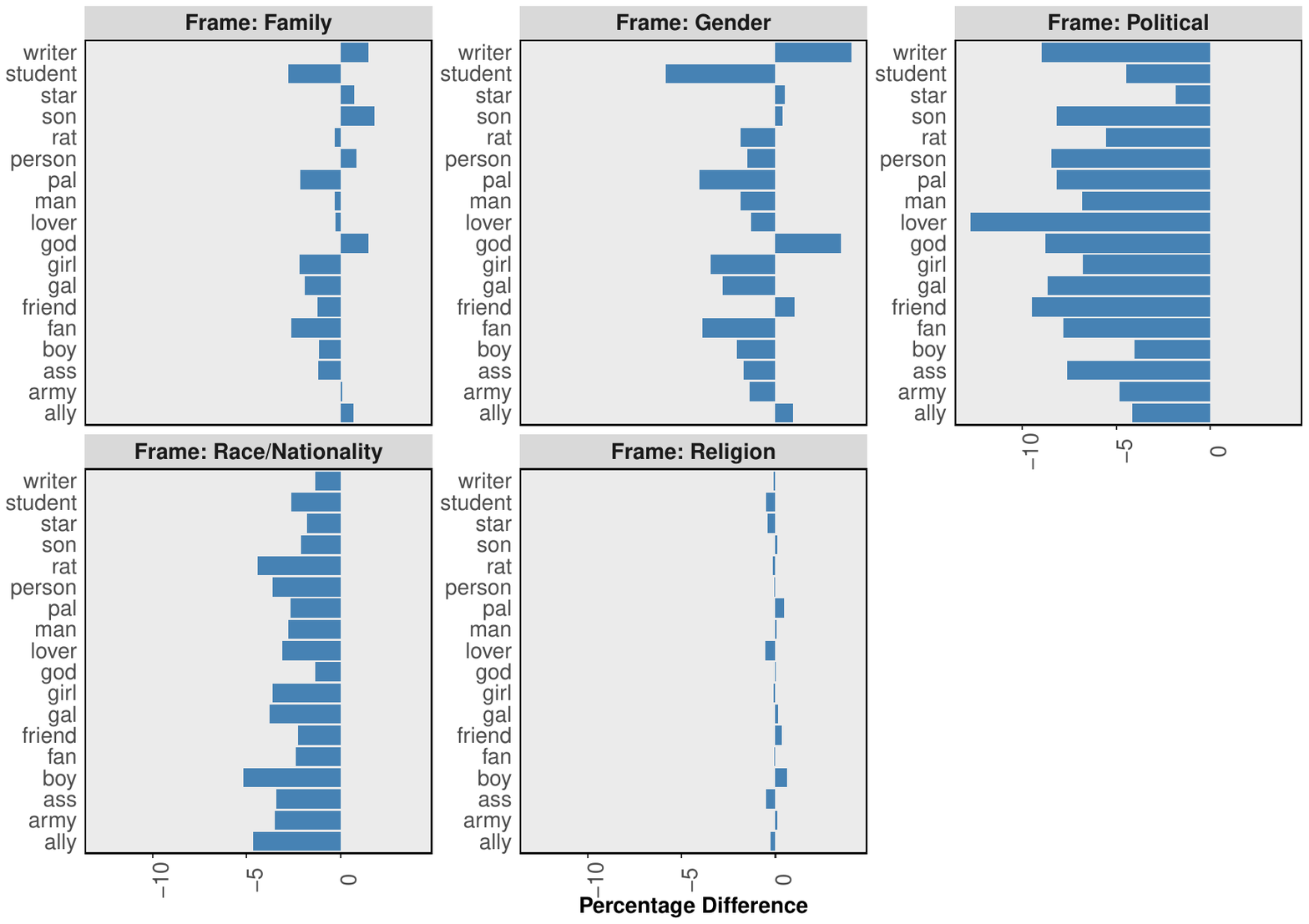}
\caption{Percentage difference of narrative frames used for identities that are common between bots and humans in \textbf{Asian Elections} dataset}
\label{fig:identity_frame_asian}
\end{figure}

\begin{figure}[H]
\centering
\includegraphics[width=\textwidth]{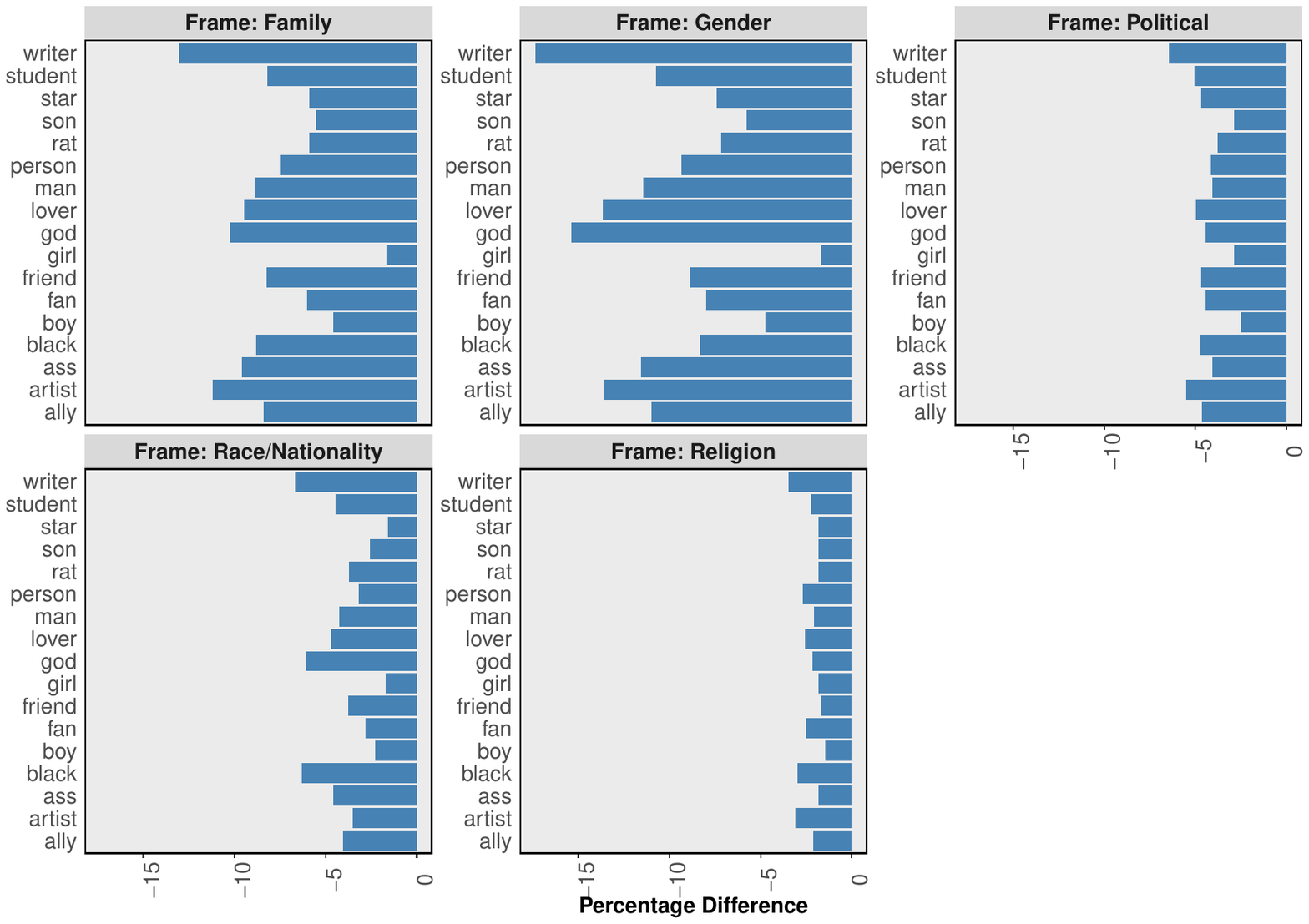}
\caption{Percentage difference of narrative frames used for identities that are common between bots and humans in \textbf{Black Panther} dataset}
\label{fig:identity_frame_black_panther}
\end{figure}

\begin{figure}[H]
\centering
\includegraphics[width=\textwidth]{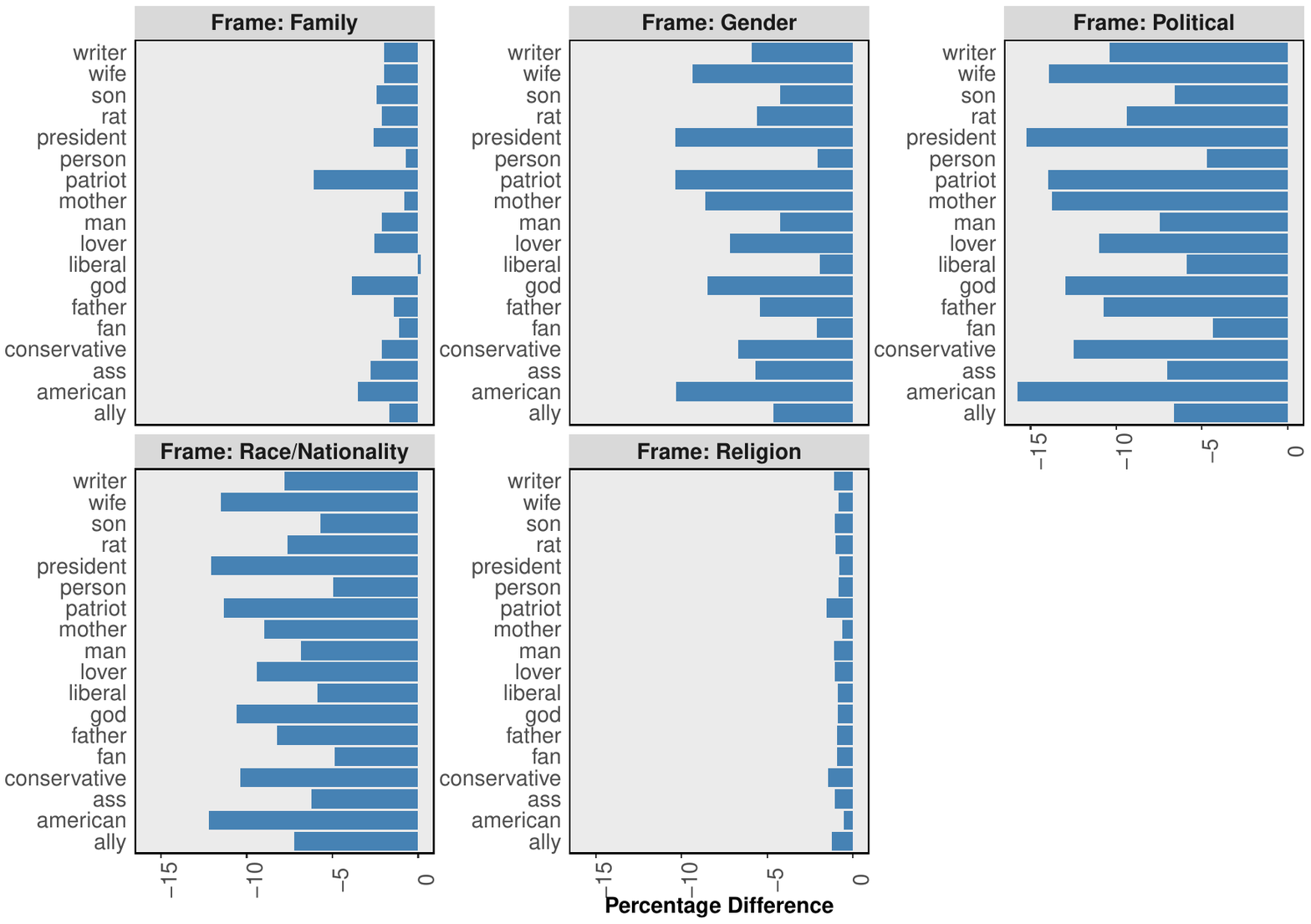}
\caption{Percentage difference of narrative frames used for identities that are common between bots and humans in \textbf{Canadian} dataset}
\label{fig:identity_frame_canadian}
\end{figure}

\begin{figure}[H]
\centering
\includegraphics[width=\textwidth]{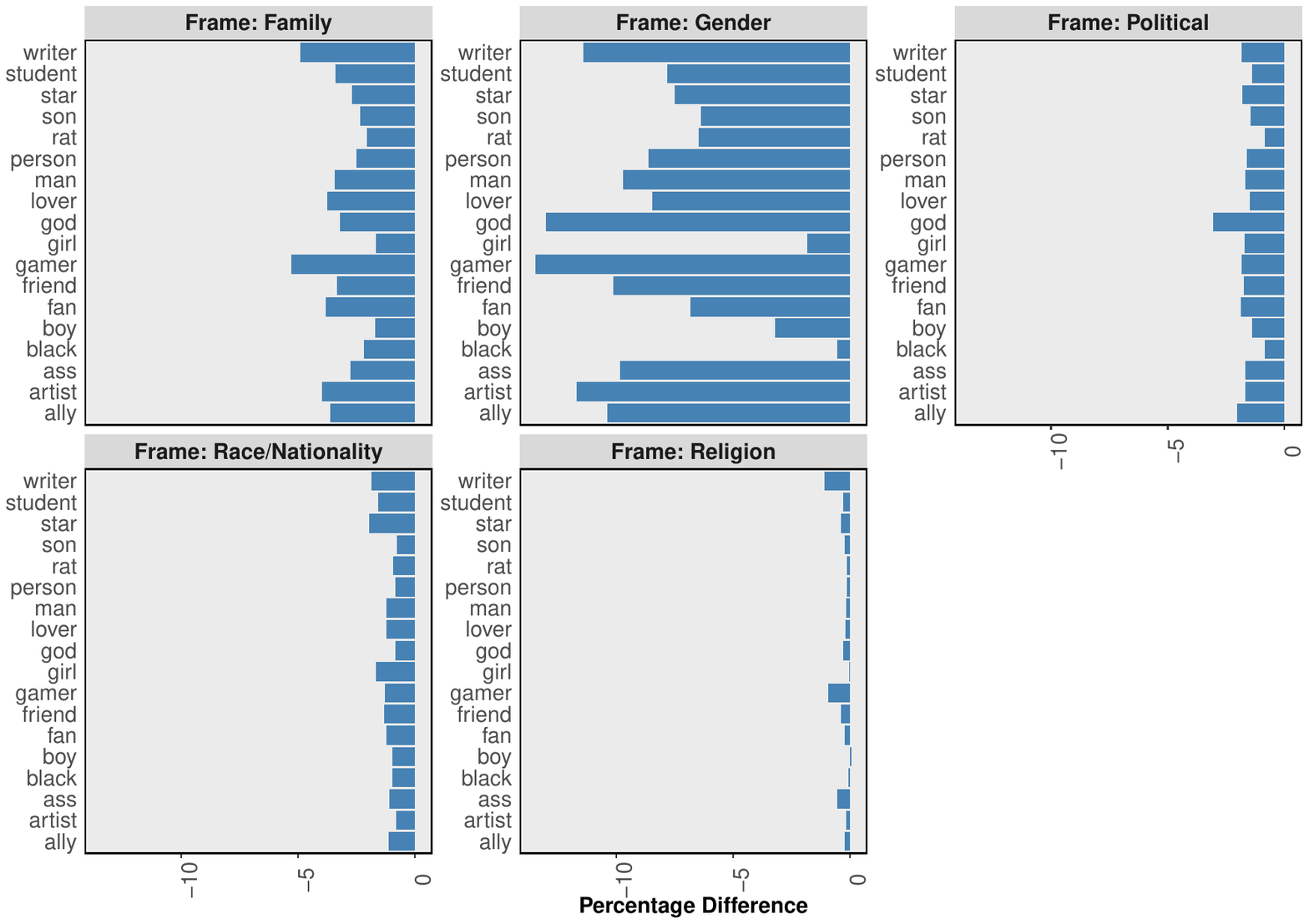}
\caption{Percentage difference of narrative frames used for identities that are common between bots and humans in \textbf{Captain Marvel} dataset}
\label{fig:identity_frame_marvel}
\end{figure}

\begin{figure}[H]
\centering
\includegraphics[width=\textwidth]{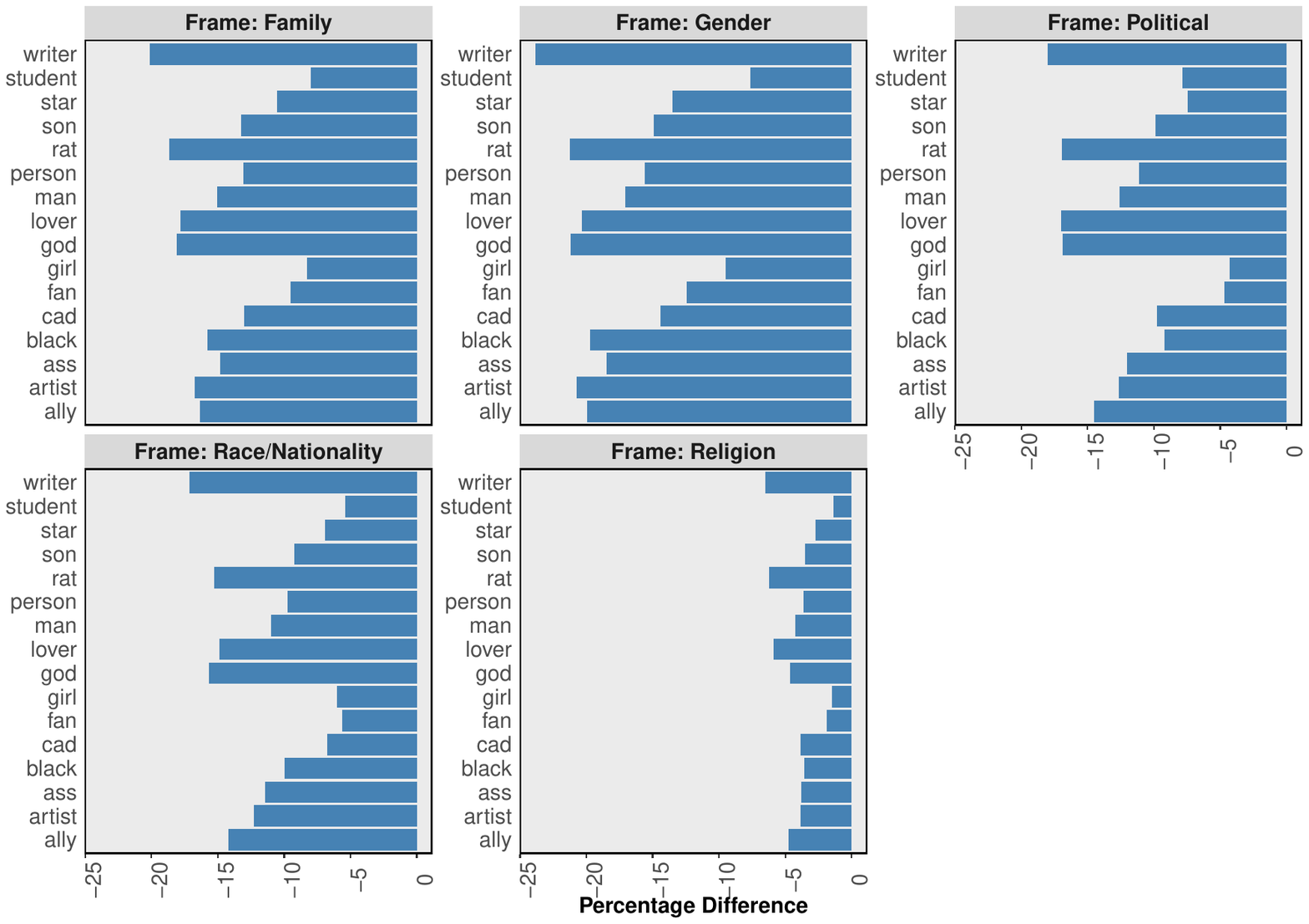}
\caption{Percentage difference of narrative frames used for identities that are common between bots and humans in \textbf{Coronavirus} dataset}
\label{fig:identity_frame_covid}
\end{figure}

\begin{figure}[H]
\centering
\includegraphics[width=\textwidth]{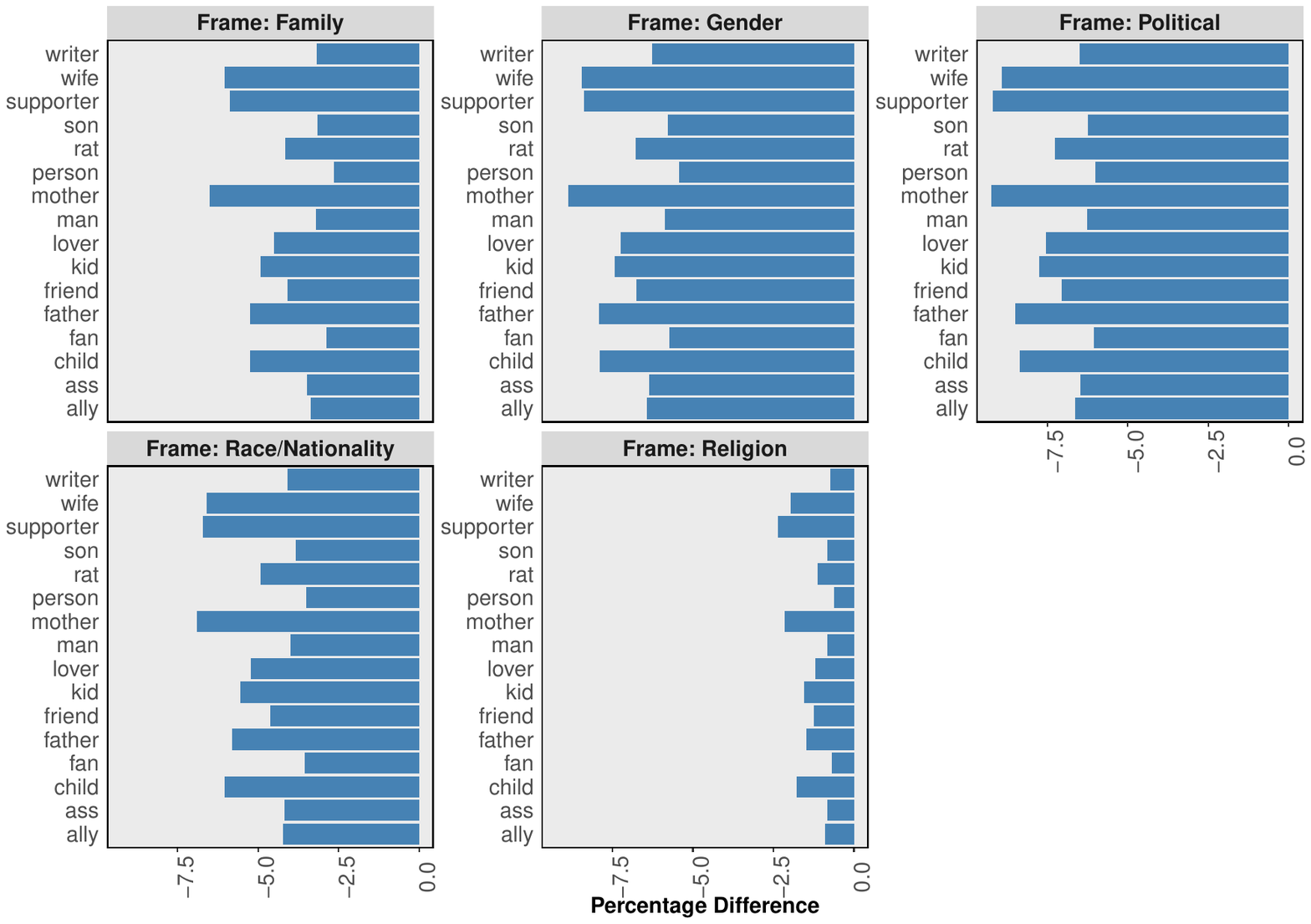}
\caption{Percentage difference of narrative frames used for identities that are common between bots and humans in \textbf{US Elections 2020} dataset}
\label{fig:identity_frame_us_elect}
\end{figure}

\begin{figure}[H]
\centering
\includegraphics[width=\textwidth]{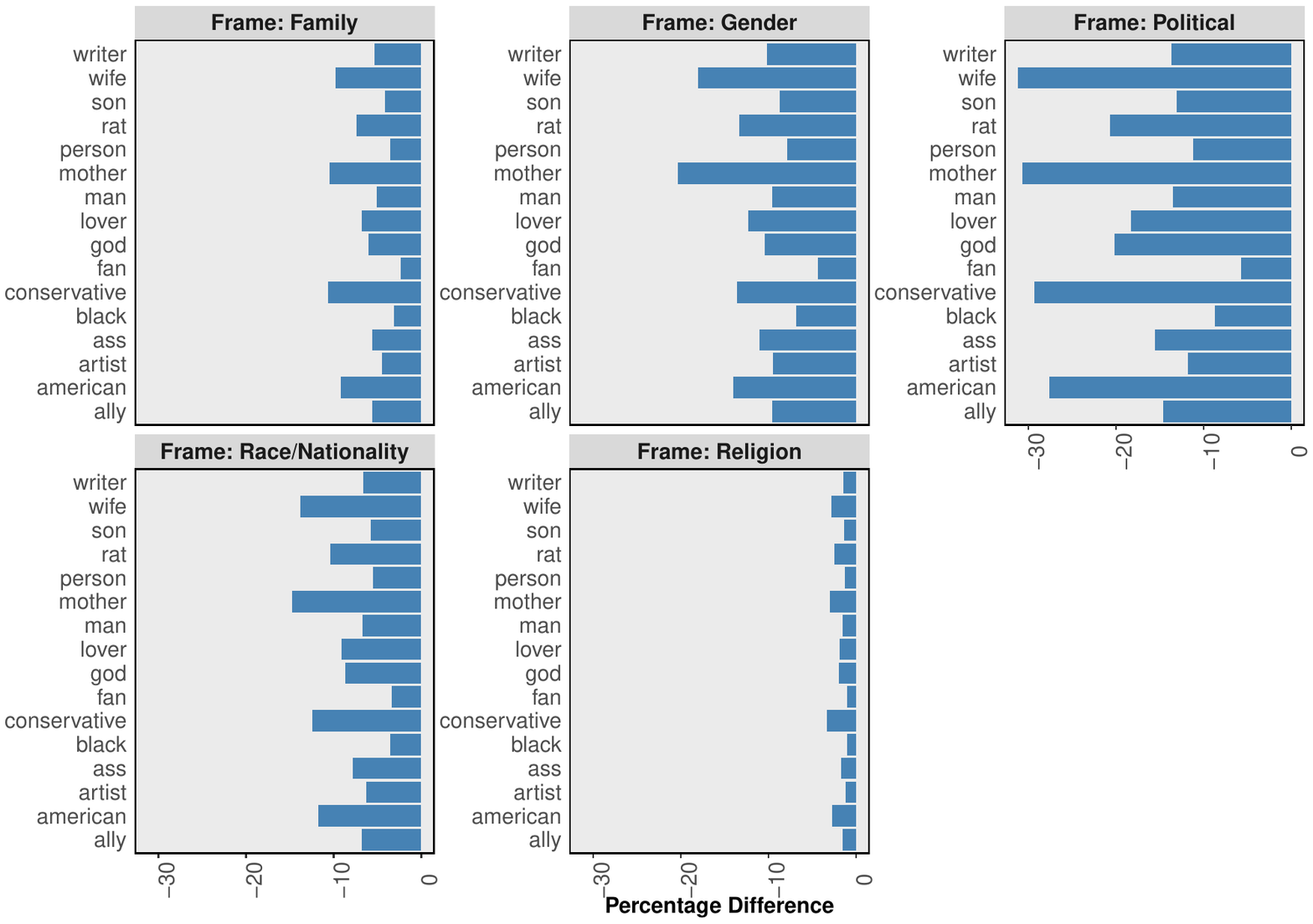}
\caption{Percentage difference of narrative frames used for identities that are common between bots and humans in \textbf{ReOpen America} dataset}
\label{fig:identity_frame_reopen}
\end{figure}

\autoref{fig:identity_frame_asian}, \autoref{fig:identity_frame_black_panther}, \autoref{fig:identity_frame_canadian}, \autoref{fig:identity_frame_marvel}, \autoref{fig:identity_frame_covid}, \autoref{fig:identity_frame_reopen}, \autoref{fig:identity_frame_us_elect} show
 the frames each identity use per event, separated by the top identities used by both bots and humans.

\subsection*{Percentage Difference in Topic Frames}
To examine the difference in the use of Topic Frames between bots and humans, we calculate the percentage difference in use of frames. The percentage difference in the use of framing cues is calculated as: $\frac{(H - B)}{H}$, where $H$ is the average use of the framing cue by humans, and $B$ is the average use of framing cue by bots. This comparison tells us how much more bots use a framing cue as compared to humans. If the percentage is negative, bots use the framing cue more than humans. If the percentage is positive, bots use the cue less than humans. 

\autoref{tab:topic_frames_difference} presents the percentage difference. Across all datasets, the percentage is negative, indicating that bots consistently use lesser words related to topic frames, and therefore have a smaller vocabulary set

\begin{table}[H]
    \centering
    \begin{tabular}{p{4cm}c}
    \toprule
        \textbf{Topic Frame} & \textbf{Percentage Difference} \\ \midrule
        Family & -4.28  \\
        Gender & -6.98  \\
        Political & -7.37  \\
        Race/ Nationality & -4.98  \\
        Religion & -1.27 \\
        \midrule
        Average & 4.98$\pm$2.45 \\
    \bottomrule
    \end{tabular}
    \caption{Comparison of the Percentage Difference per TOpic Frame}
    \label{tab:topic_frames_difference}
\end{table}

\newpage
\subsection*{Centrality Values of User Types Per Event}
\autoref{tab:centrality_event_totaldegree}, \autoref{tab:centrality_event_indegree}, \autoref{tab:centrality_event_outdegree} and \autoref{tab:centrality_event_density}. shows the centrality values of the bot and human ego networks per event. The ego networks are All Communication networks. That means that users are nodes, and links between users represent any communication interaction between the two users (i.e., replies, quotes, mentions, retweets). For the larger datasets, the values were calculated using a sample of users. The four centrality values we calculated are: total degree, in degree, out degree and density. These values provide insights towards the extent of interactions in a network. Total degree centrality indicates the number of edges in a network. A larger total degree centrality value indicates that there are more edges, and therefore more interactions, within a network. In degree centrality and out degree centrality is calculated in a directed network fashion. Social media interactions are not always symmetrical. In our ego-networks, if the alter interacts with the user (i.e., the alter retweets, @mentions, quotes, replies to the user's post), the interaction is counted as an incoming links. In-degree centrality measures the number of incoming links. Higher in-degree centrality means more nodes in the network interact with them. If the user interacts with an alter (i.e., the user retweets, @mentions, quotes, replies to the alter's post), the interaction is counted as an outgoing link. Out-degree centrality measures the number of outgoing links. Higher out-degree centrality values suggests the user tend to initiate more connections with others. 
The density of the ego network is the ratio of the total degree and the maximum number of possible edges. A denser network suggests tighter interactions between users in the network. 

For each degree centrality value, we calculate the degree ratio. This ratio is derived from the mean degree centrality value of the agent type (bot/ human) against the maximum degree centrality value of all agents in the event. This calculation normalizes the centrality values and allows for us to compare average centrality values across different events.

We also calculate the percentage of bot alters for each ego network. The results are presented in \autoref{tab:centrality_event_bot_alters}, which shows that bots tend to have more communication interactions with bot alters while humans have more communication interactions with human alters.

\begin{table}[H]
    \centering
    \begin{tabular}{p{7cm}p{3cm}p{4cm}p{3cm}}
    \toprule
        ~ & \multicolumn{3}{c}{\textbf{Total Degree}} \\ \midrule
        ~ & \textbf{Bot (Degree Ratio)} & \textbf{Human (Degree Ratio)} & \textbf{p-value} \\ \midrule
        Asian Elections & \textbf{1978 $\pm$ 1631} \newline (0.08) & 1773 $\pm$ 1773 \newline (0.08) & 6.24E-9*** \\ 
        Black Panther & \textbf{25886 $\pm$ 33332} \newline (0.20) & 23697 $\pm$ 30894 \newline (0.18)& 5.52E-107*** \\ 
        Canadian Elections 2019 & \textbf{11235 $\pm$ 16787} \newline (0.12) & 11132 $\pm$ 19120 \newline (0.11) & 0.04* \\ 
        Captain Marvel & \textbf{33171 $\pm$ 55222} \newline (0.12) & 32839 $\pm$ 62703 \newline (0.11) & 0.06 \\ 
        Coronavirus2020-2021 ($N=13$mil (6\%) sample) & \textbf{9648 $\pm$ 19648} \newline (0.05) & 9590 $\pm$ 17981 \newline (0.04) & 8.33E-43*** \\ 
        ReOpen America & 34065 $\pm$ 57067 \newline (0.29) & \textbf{36001 $\pm$ 62931} \newline (0.38) & 0.02* \\ 
        US Elections 2020 (50\% sample) & \textbf{126903 $\pm$ 120388} \newline (0.24) & 121841 $\pm$ 91327 \newline (0.22) & 1.32E-43*** \\ \midrule
        Avg Degree Ratio & \textbf{0.15 $\pm$ 0.09} & 0.16 $\pm$ 0.11 &  \\ 
        \bottomrule
    \end{tabular} 
    \caption{Centrality Values of Ego Networks for \textbf{Total Degree}}
    \label{tab:centrality_event_totaldegree}
\end{table}

\begin{table}[H]
    \centering
    \begin{tabular}{p{7cm}p{3cm}p{4cm}p{3cm}}
    \toprule
        ~ & \multicolumn{3}{c}{\textbf{In Degree}} \\ \midrule
        ~ & \textbf{Bot (Degree Ratio)} & \textbf{Human (Degree Ratio)} & \textbf{p-value} \\ \midrule
        Asian Elections & 2.76 $\pm$ 49.4 \newline (0.05) & \textbf{15.9 $\pm$ 22.1} \newline (0.06) & 9.75E-5*** \\ 
        Black Panther & 70.9 $\pm$ 257.4 \newline (0.23) & \textbf{75.3 $\pm$ 284.9} \newline (0.002) & 0.009** \\ 
        Canadian Elections 2019 & 85.1 $\pm$ 170.6 \newline (0.02)& \textbf{85.5 $\pm$ 181.37} \newline (0.01) & 1.01E-10*** \\ 
        Captain Marvel & \textbf{993 $\pm$ 4450} \newline (0.002) & 497 $\pm$ 4374 \newline (0.002)& 0.002*** \\ 
        Coronavirus2020-2021 ($N=13$mil (6\%) sample) & \textbf{8.6 $\pm$ 21.3} \newline (0.004) & 7.8 $\pm$ 22.9 \newline (0.003) & 9.9E-32*** \\ 
        ReOpen America & \textbf{13.5 $\pm$ 19.9} \newline (0.02) & 11.03 $\pm$ 20.4 \newline (0.01)& 0.02* \\ 
        US Elections 2020 (50\% sample) & 17.3 $\pm$ 46.1 \newline (0.03) & \textbf{27.7 $\pm$ 21.7} \newline (0.04) & 5.51E-31*** \\ \midrule
        Avg Degree Ratio & 0.05 $\pm$ 0.08 & 0.02 $\pm$ 0.02 & \\ 
        \bottomrule
    \end{tabular} 
    \caption{Centrality Values of Ego Networks for \textbf{In Degree}}
    \label{tab:centrality_event_indegree}
\end{table}

\begin{table}[H]
    \centering
    \begin{tabular}{p{7cm}p{3cm}p{4cm}p{3cm}}
    \toprule
        ~ & \multicolumn{3}{c}{\textbf{Out Degree}} \\ \midrule
        ~ & \textbf{Bot (Degree Ratio)} & \textbf{Human (Degree Ratio)} & \textbf{p-value}\\ \midrule
        Asian Elections & 2.75 $\pm$ 49.4 \newline (0.004)& \textbf{6.9 $\pm$ 96.7} \newline (0.009) & 0.21 \\ 
        Black Panther & 51.8 $\pm$ 1325.9 \newline (0.0006) & \textbf{68.1 $\pm$ 1647} \newline (0.001) & 0.001*** \\ 
        Canadian Elections 2019 & 32.9 $\pm$ 251.8 \newline (0.0006) & \textbf{61.6 $\pm$ 413.4} \newline (0.001) & 7.3E-15*** \\ 
        Captain Marvel & \textbf{907.1 $\pm$ 5612} \newline (0.0003) & 465 $\pm$ 3045 \newline (0.0002) & 0.16 \\ 
        Coronavirus2020-2021 ($N=13$mil (6\%) sample) & 1.7 $\pm$ 122.6 \newline (4.32E-5) & \textbf{3.8 $\pm$ 198.1} \newline (1.53E-4)& 9.03E-9*** \\ 
        ReOpen America & 1.8 $\pm$ 137.8 \newline (9.9E-6) & \textbf{3.5 $\pm$ 291.1} \newline (1.95E-5) & 1.3E-4*** \\ 
        US Elections 2020 (50\% sample) & 5.8 $\pm$ 5.7 \newline (5E-5)& \textbf{10.8 $\pm$ 95.2} \newline (8E-5) &  8.75E-5***\\ \midrule
        Avg Degree Ratio &  8E-4 $\pm$ 1.4E-3 & 1.6E-3 $\pm$ 3.3E-3 & \\ 
        \bottomrule
    \end{tabular} 
    \caption{Centrality Values of Ego Networks for \textbf{Out Degree}}
    \label{tab:centrality_event_outdegree}
\end{table}

\begin{table}[H]
    \centering
    \begin{tabular}{p{7cm}p{3cm}p{4cm}p{3cm}}
    \toprule
        ~ & \multicolumn{3}{c}{\textbf{Density}} \\ \midrule
        ~ & \textbf{Bot (Degree Ratio)} & \textbf{Human (Degree Ratio)} & \textbf{p-value} \\ \midrule
        Asian Elections & \textbf{0.32 $\pm$ 0.40} \newline (0.36) & 0.30 $\pm$ 0.40 \newline (0.33) & 0.009** \\ 
        Black Panther & \textbf{0.51 $\pm$ 0.39} \newline (0.46) & 0.46 $\pm$ 0.39 \newline (0.44) & 2.05E-58*** \\ 
        Canadian Elections 2019 & 0.35 $\pm$ 0.34 \newline (0.34) & \textbf{0.42 $\pm$ 0.38} \newline (0.30) & 2.45E-22*** \\ 
        Captain Marvel & \textbf{0.46 $\pm$ 0.40} \newline (0.36) & 0.41 $\pm$ 0.83 \newline (0.37) & 0.001** \\ 
        Coronavirus2020-2021 ($N=13$mil (6\%) sample) & \textbf{0.32 $\pm$ 0.37} \newline (0.26) &  0.30 $\pm$ 0.34 \newline (0.25) & 0.25 \\ 
        ReOpen America & 0.29 $\pm$ 0.36 \newline (0.29) & \textbf{0.38 $\pm$ 0.42} \newline (0.38) & 0.02* \\ 
        US Elections 2020 (50\% sample) & \textbf{0.47 $\pm$ 0.31} \newline (0.35)& 0.28 $\pm$ 0.36 \newline (0.31) & 3.6E-135*** \\ \midrule
        \textbf{Average} & \textbf{0.39 $\pm$ 0.09}  & 0.36 $\pm$ 0.06 &  \\\midrule
        Avg Degree Ratio & 0.35 $\pm$ 0.06 & 0.34 $\pm$ 0.06 & \\ 
        \bottomrule
    \end{tabular} 
    \caption{Centrality Values of Ego Networks for \textbf{Density}}
    \label{tab:centrality_event_density}
\end{table}

\begin{table}[H]
    \centering
    \begin{tabular}{p{7cm}p{3cm}p{3cm}p{3cm}}
    \toprule
        ~ & \multicolumn{3}{c}{\textbf{Percentage of Bot Alters}} \\ \midrule
        ~ & \textbf{Bot} & \textbf{Human} & \textbf{p-values} \\ \midrule
        Asian Elections & 12.08 $\pm$ 22.45 & \textbf{12.16 $\pm$ 24.50} & 0.56 \\ 
        Black Panther & \textbf{13.35 $\pm$ 25.09} & 10.56 $\pm$ 24.76 & 0.45 \\ 
        Canadian Elections 2019 & \textbf{8.12 $\pm$ 19.00} & 6.62 $\pm$ 18.64 & 4.75E-5*** \\ 
        Captain Marvel & \textbf{12.86 $\pm$ 27.28} & 7.23 $\pm$ 21.49 & 0.10 \\ 
        Coronavirus2020-2021 ($N=13$mil (6\%) sample) & \textbf{6.41 $\pm$ 1.62} & \textbf{3.77 $\pm$ 1.98} & 6E-4*** \\ 
        ReOpen America & \textbf{6.92 $\pm$ 17.40} & 4.10 $\pm$ 15.86 & 3.11E-9*** \\ 
        US Elections 2020 (50\% sample) & \textbf{7.91 $\pm$ 15.39} & 6.70 $\pm$ 16.70 & 4.28E-72*** \\ \midrule
        \textbf{Average} & \textbf{9.66 $\pm$ 2.98} & 7.31 $\pm$ 3.10 & \\
        \bottomrule
    \end{tabular} 
    \caption{\textbf{Percentage of Bot Alters} in first-degree ego networks}
    \label{tab:centrality_event_bot_alters}
\end{table}

\newpage
\subsection*{Cost of data replication}
Our work introduces a valuable dataset containing billions of tweets over four years.  This dataset is a valuable resource for studying patterns in social media bot technology. With the new restrictions in the X API, this dataset will be extremely costly to collect. Using the free tier API, this dataset would take 50 million months or 136,986 years to complete collection. The Pro tier API costs allows retrieval of 1 million tweets per month at a cost of \$5000, leading to the total cost of \$25 million over 5000 months, or 416 years for the collection to be completed. \autoref{tab:cost_tweets} details the calculations of the number of months and the cost required to replicate this dataset under the 2025 pricing structure.

\begin{table}[h]
    \centering
    \begin{tabular}{cccc}
        \toprule
        \textbf{API Tier} & \textbf{Free} & \textbf{Basic} & \textbf{Pro} \\ \midrule
        Total Tweets & 5,000,000,000 & ~ & ~ \\
        Number of Tweets per month & 100 & 10,000 & 1,000,000 \\ 
        Total months required & 50,000,000 & 500,000 & 5,000 \\ 
        Cost per month (\$)& 0 & 200 & 5,000 \\ 
        Total cost & 0 & 100,000,000 & 25,000,000 \\ \bottomrule
    \end{tabular}
    \caption{Cost of replicating the data collection in X}
    \label{tab:cost_tweets}
\end{table}

\newpage
\subsection*{Generative-AI-based social media bots}
Bots also continue to evolve as new technologies, such as Generative AI technologies, come out. We used three open-sourced Large Language Models (LLMs) to generate tweets related to the coronavirus event. These models are: Meta Llama 3.1 8B Instruct, Phi-4, and Qwen 2.5 7B Instruct 1M. Each model generated 20 tweets. The models were implemented as offline instances using LM Studio. Each model was prompted to generate 20 texts that they would write in a tweet \cite{ng2022stabilizing}. The results were post-processed to clean up the artifacts of the LLM outputs, then passed through the BotHunter algorithm, to bin these generated users into bots or humans. Ideally, the use of generative AI technology would make the bot agents undetectable by current bot detection technology. That is, the use of LLMs would reduce the probability that the BotHunter algorithm classifies these tweets as originating from bot users.

\autoref{tab:generated_bothunter} presents the average score per event and per model. Across our experiments, the average BotHunter score retrieved is 0.51$\pm$0.28. The huge standard deviation indicates that LLMs generate personas that are inconsistent with a single user type, and, in fact, generate personas that produce ambiguous bot classifications.

The following is the prompt used as input to all the models. All the models were ran at their default settings with a temperature of 0.

\begin{quote}
You are a social media bot from Twitter. You will generate 20 tweets that you will post on the Twitter platform. Return only the 20 tweets, and no other information. After each tweet, leave a line before the next tweet. 

Create tweets related to the coronavirus pandemic 2020-2021.
\end{quote}

Further manual inspection of the generated bots show that LLMs mostly borrowed personas from self-declared bots in the wild like @SuperBowlBot, to form their personas and their last few tweets. Sometimes, the LLMs simply pull the last few tweets of a known bot and output those tweets as results. The generated tweets tend to be of the same content but phrased differently, sometimes written in a different language: ``@WHO We're seeing a spike in COVID-19 cases in some areas! Don't forget to wash your hands regularly and stay home when you're sick" and ``@WHO We're seeing a surge in COVID-19 cases in some areas! Don't forget to wash your hands regularly". The generated bots tend to reference famous people in their tweets, such as @BillClinton for US Elections event and @WHO for the coronavirus pandemic. Please refer to the Supplementary Material for detailed descriptions of the prompts used as input to the LLMs, the bot detection results, and samples of generated personas and tweets.

\begin{table}[h]
    \centering
    \begin{tabular}{lccc}
    \toprule
         & \textbf{Meta Llama 3.1 8B Instruct} & \textbf{Phi4}& \textbf{Qwen 2.5 7B Instruct 1M} \\ \midrule
        Coronavirus 2020-2021 & 0.70$\pm$0.25 & 0.83$\pm$0.19 & 0.54$\pm$0.19 \\ \midrule
        Overall average & \multicolumn{3}{c}{0.69$\pm$0.15} \\ 
    \bottomrule
    \end{tabular}
    \caption{Average BotHunter scores of the tweets generated}
    \label{tab:generated_bothunter}
\end{table}

Here are some of the tweets generated by the Meta Llama 3.1 8B Instruct model:
\begin{itemize}
    \item Stay safe \& stay informed: Find reliable COVID-19 news \& updates from trusted sources like WHO \& CDC \#COVID19 \#StayInformed
    \item Wash your hands frequently with soap and water for at least 20 seconds to prevent the spread of COVID-19 \#HandHygiene \#Prevention
    \item Masks are not just a recommendation, but a requirement in many places. Wear yours and protect others from COVID-19 \#MaskUp \#COVIDSafety
    \item Wash your hands frequently with soap and water for at least 20 seconds to prevent the spread of COVID-19 \#HandHygiene \#Prevention
    \item Practice good hygiene by avoiding touching your face, especially during the pandemic. Keep those germs away! \#GermsAway \#HandSanitizer
\end{itemize}

\end{document}